\documentclass[twocolumn,traditabstract]{aa} 
\usepackage{graphicx}
\usepackage{txfonts}
\usepackage{natbib}       
\usepackage{multirow}
\usepackage{aalongtable}
\usepackage{xtab,booktabs}  
\usepackage[draft]{hyperref}

\def\etal{{et\,al.}}
\def\msun{M$_{\odot}$}
\def\h{$^{\rm h}$}\def\m{$^{\rm m}$}

\def\degs{\ifmmode ^{\circ}\else$^{\circ}$\fi}
\def\fss{\hbox{$.\!\!^{\rm s}$}}        
\def\fdg{\hbox{$.\!\!^\circ$}}          
\def\farcs{\hbox{$.\!\!^{\prime\prime}$}}  
\def\farcm{\hbox{$.\!\!^{\prime}$}}
\def\amin{\ifmmode ^{\prime}\else$^{\prime}$\fi}
\def\asec{\ifmmode ^{\prime\prime}\else$^{\prime\prime}$\fi}
\newbox\grsign \setbox\grsign=\hbox{$>$}
\newdimen\grdimen \grdimen=\ht\grsign
\newbox\laxbox \newbox\gaxbox
\setbox\gaxbox=\hbox{\raise.5ex\hbox{$>$}\llap
     {\lower.5ex\hbox{$\sim$}}}\ht1=\grdimen\dp1=0pt
\setbox\laxbox=\hbox{\raise.5ex\hbox{$<$}\llap
     {\lower.5ex\hbox{$\sim$}}}\ht2=\grdimen\dp2=0pt
\def\gax{$\mathrel{\copy\gaxbox}$}
\def\lax{$\mathrel{\copy\laxbox}$}

\newcommand{\gp}{\mbox{$g^\prime$}}
\newcommand{\rp}{\mbox{$r^\prime$}}
\newcommand{\ip}{\mbox{$i^\prime$}}
\newcommand{\zp}{\mbox{$z^\prime$}}
\newcommand{\griz}{\gp\rp\ip\zp}
\begin{document}

   \title{The GROND GRB sample: I. Overview and Statistics}

   \author{J. Greiner\inst{1} \and
     T. Kr\"uhler\inst{1}\thanks{\emph{Present address:} 82008 M\"unchen, Fasanenstr. 31} \and
     J. Bolmer\inst{1}\thanks{\emph{Present address:} 80335 M\"unchen, Erzgiessereistr. 22} \and
     S. Klose\inst{2} \and
     P.M.J. Afonso\inst{3} \and
     J. Elliott \inst{4} \and
     R. Filgas\inst{5} \and
     J.F. Graham\inst{6} \and
     D.A.~Kann\inst{7}\thanks{\emph{deceased}} \and
     F. Knust\inst{1}\thanks{\emph{Present address:} 80337 M\"unchen, Augsburgerstr. 5} \and
     A. K\"{u}pc\"{u} Yolda\c{s}\inst{8} \and
     M. Nardini\inst{1}\thanks{\emph{Present address:} Via Dante Alighieri 49, 20092, Cinisello Balsamo (MI), Italy} \and
     A.M. Nicuesa Guelbenzu\inst{2} \and
     F. Olivares Estay\inst{9} \and
     A.~Rossi\inst{10} \and
     P. Schady\inst{11} \and
     T. Schweyer\inst{12} \and
     V. Sudilovsky\inst{13} \and
     K. Varela\inst{14} \and
     P. Wiseman\inst{15}
   }   
   \institute{Max-Planck Institut f\"ur extraterrestrische Physik,
              Giessenbachstr. 1, 85748 Garching, Germany\\
         \email{jcg@mpe.mpg.de}
         \and
             Th\"uringer Landessternwarte Tautenburg, Sternwarte 5,
             07778 Tautenburg,  Germany 
         \and
           American River College, Physics \& Astronomy Dpt., 4700 College Oak Drive, Sacramento, CA 95841, U.S.A.
         \and
             StubHub Inc., 120 Broadway, Suite 200, Santa Monica, CA 90401, U.S.A
         \and
           Institute of Experimental and Applied Physics, Czech Technical University in Prague, Husova 240/5, 11000 Prague, Czech Republic
           \and
           Visitor, University of Hawaii at Manoa, 2505 Correa Rd., Honolulu, HI 96822, U.S.A.
           \and
           Hessian Research Cluster ELEMENTS, Goethe University Frankfurt,
           Max-von-Laue-Str. 12, D-60438 Frankfurt a.M., Germany
           \and
           European Molecular Biology Laboratory, European Bioinformatics
           Institute, Hinxton CB10 1SD, United Kingdom
           \and
           Instituto de Astronom\'{\i}a y Ciencias Planetarias,
            Universidad de Atacama, Copayapu 485, Copiap\'o, Chile
           \and
         INAF - Osservatorio di Astrofisica e Scienza dello Spazio, via
         Piero Gobetti 93/3, 40129 Bologna, Italy
         \and
           Department of Physics, University of Bath, Building 3 West,
           Bath BA2 7AY, United Kingdom
         \and
          Department of Astronomy, The Oskar Klein Centre, Stockholm University,
          AlbaNova, 10691 Stockholm, Sweden
          \and
          Amazon Web Services (AWS), 55 Pier 4 Blvd., Boston, MA 02210,  U.S.A. 
          \and
          Influur Corporation, 1111 Brickell Ave, Miami, FL 33131, U.S.A.
         \and
         School of Physics and Astronomy, University of Southampton, Southampton
           SO17 1BJ, United Kingdom
             }

   \date{Received  2024; accepted ...}


   \abstract{
   With GROND, a 7-channel optical and near-infrared imager at the 
   2.2m telescope of the Max-Planck Society at ESO/La Silla, a dedicated
   GRB afterglow observing program was performed between 2007 and 2016.
   In this first of a series of papers, we describe the GRB observing plan,
   provide first readings of all so far unpublished GRB afterglow measurements
   and some observing statistics.
   In total, we observed 514 GRBs with GROND, among those 434
   Swift-detected GRBs, representing 81\% of the observable Swift sample.
   For GROND-observations within 30 min of the GRB trigger, the optical/NIR
   afterglow detection rate is 81\% for long- and 57\% for short-duration GRBs.
   We report the discovery of 10 new GRB
   afterglows plus one candidate, redshift estimates (partly improved)
   for 4 GRBs, new host detections for 7 GRBs.
   We identify the (already known) afterglow of GRB 140209A
     as the sixth GRB exhibiting a 2175 \AA\ dust feature.
   As a side result, we identify
   two blazars, one of these at a redshift of z=3.8 (in the GRB 131209A field). 
   }
   
   \keywords{Gamma rays: bursts --
     Techniques: photometric
   \vspace*{3.cm}
    }
   \begin{@twocolumnfalse}
    \maketitle
     \setlength{\fboxrule}{0.8pt}
     \vspace*{-3.6cm}\framebox[0.943\textwidth][t]{
      \begin{minipage}[t][1.7cm][c]{16.85cm}
      \it This publication is dedicated to David Alexander Kann who passed
       away in 2023 at the age of 46, just a day after some of us had the chance
       to discuss recent GRB news with him. Being a member of the GROND team
       since 2011, Alex was our encyclopedic dictionary for nearly
       every GRB mentioned herein. He was eagerly anticipating this
       manuscript, but we were too late. We deeply miss him and will always
       remember him.
      \end{minipage}
    }
   \end{@twocolumnfalse}

\section{Introduction}

Gamma-ray bursts  (GRBs)  are  the  high-energy
signatures of the  death of massive stars (long-duration sub-class)
or the merger of two neutron stars (short-duration sub-class),
though outliers to these associations have been reported.
While the prompt high-energy emission informs about the activity
of the central engine and the energy dissipation mechanism,
the ensuing afterglow emission carries information about the
GRB surrounding, and moreover can be used to probe the host galaxy
and the Universe at large. Due to this versatility of the GRB afterglows,
they have been the focus of GRB research
for more than a decade, before Fermi observations have put the prompt
gamma-ray emission again in the centre of attention
and the gravitational-wave detection of GRB 170817A has given
a boost to multi-wavelength observations of short GRBs.

Afterglow emission is detectable
in high-energy gamma-rays \citep{Mirzoyan+19} for a few hours,
in the X-ray to the optical/near-infrared for days up to weeks,
and in the radio band up to years after the gamma-ray burst. The
Neil-Gehrels Swift observatory \citep{gcg04} has pioneered systematic
X-ray/UV/optical afterglow observations. An early surprise of these
observations was the finding that the X-ray afterglow does not just
fade exponentially, as the early predictions and observations suggested.
Instead, the 'canonical' X-ray light curve consists of at least 5 different
segments \citep{nkg06}. While at X-rays basically every GRB afterglow
was detected whenever Swift slewed immediately, the detection rate
in the optical/UV was only about 25\% \citep{Roming+2009}, and thus Swift's 
optical afterglow light curve data set is less extensive. Despite this,
Swift/UVOT as well as many ground-based observations have shown that the
optical light does not trace the behaviour of the X-ray afterglow
during the first hours of observations which has spawned a
number of investigations \citep{pan06, gng09}.
As of today, a large part of this mismatch provides a challenge in
our understanding of the GRB afterglow phenomenon.
The latest Swift/UVOT afterglow catalogue \citep{Roming+2017}
is the largest coherent optical/UV sample of afterglows, with
538 GRBs observed. Due to its short-wavelength coverage, it is
biased against dust extinction and high-redshift.
Thus, higher sensitivity and near-infrared (NIR) coverage was in demand.
Both of these improvements could be provided by GROND@2.2m telescope.

Here we describe and summarise the dedicated optical/near-infrared GRB afterglow
observing program executed with GROND between May 2007 (commissioning) and
Sep. 30, 2016 (when the MPE directorate terminated this program).

\section{GROND and the afterglow observing strategy}

GROND, a simultaneous 7-channel optical/near-infrared ima\-ger \citep{gbc08} 
moun\-ted at  the 2.2\,m  MPI/ESO telescope at La Silla (ESO, Chile),
was designed and developed to rapidly identify GRB afterglows and measure their 
redshift via the drop-out technique, using the 7 filter bands between
0.4--2.4 $\mu$m (\gp\rp\ip\zp$JHK_s$).
GROND was commissioned at the MPG 2.2m telescope in April/May 2007,
and the first gamma-ray burst followed up was GRB 070521 \citep{gck07a}.
For the first few months (until end of September 2007), follow-up observations
depended on the willingness of the scheduled observers to share observing time.
Thereafter, a general override permission and a 15\% share
of the total telescope time allowed us to follow every well-localised
GRB which was visible from La Silla, weather permitting, with only
few exceptions (see statistics in the next section).

GROND observations of GRBs within the first 24 hrs were fully automated
(see \citealt{gbc08} for more details). Usually, each GRB was scheduled
to be observed for its full visibility period in the first night.
For subsequent nights, a human
decision process kicked in, based on the results of the previous night.
In general, nightly observations were performed for the first 2--4 nights,
after which the coverage decreased depending on the brightness of the 
afterglow. In general, we attempted to follow each GRB with detected afterglow
until it was no longer detectable with GROND in a 2-hr exposure.
For nearby GRBs, the coverage was enhanced again after 10$\times$(1+z)
days to search for the GRB-supernova.
In many cases, single late-time observations (after months to years)
were performed, motivated by the search for the underlying host galaxy.

GROND data have been reduced in the standard manner \citep{kkg08} using 
pyraf/IRAF \citep{Tody1993, kkg08b}.
The optical/NIR imaging was calibrated against the Sloan Digital Sky Survey
(SDSS)\footnote{\url{http://www.sdss.org}} \citep{Eisenstein+2011},
Pan-STARRS1\footnote{\url{https://www2.ifa.hawaii.edu/research/Pan-STARRS.shtml}}
\citep{Chambers+2016}
or the SkyMapper (SM) Survey\footnote{\url{http://skymapper.anu.edu.au}}
\citep{Wolf+2018}
catalogues for $g^\prime r^\prime i^\prime z^\prime$, 
and the 2MASS catalogue \citep{Skrutskie+2006} for the $JHK_{\rm s}$ bands.
This leads to typical
absolute accuracies  of $\pm$0.03~mag in $g^\prime r^\prime i^\prime 
z^\prime$ and $\pm$0.05~mag in $JHK_{\rm s}$.
Since the GROND dichroics were designed to match the Sloan and 2MASS
filter systems \citep{gbc08}, the colour  terms are very small,
below 0.01 mag, except for the $i^\prime$ band \citep[for details see][]{gby21}.

Photometric redshifts are derived with the publicly available hyperZ code
\citep{Bolzonella+2000},
which minimises the $\chi^2$ from synthetic photometry of a template
spectrum against the observed data.
In addition to the built-in handling of the Lyman absorption according to
\cite{Madau1995} and several default reddening templates, the following
features have been added:
(i) a number of powerlaw spectra with different spectral index, to account
for the synchrotron afterglow spectra of GRB afterglows,
(ii) neutral hydrogen absorption following the description of
\cite{Totani+2006} to account for damped Ly-$\alpha$ absorbers (DLA),
(iii) a reddening law according to \cite{Maiolino+2004},
(iv) the filter responses of GROND and Swift/UVOT including
  all optical components between the primary telescope mirror and the detector,
  and its quantum eﬃciency \citep{gbc08, Poole+2008}.
Very similar to the nominal hyperZ usage, we fit for 4 variables simultaneously:
the powerlaw slope, the redshift, the host extinction and the normalisation
(fixing the Galactic foreground extinction).
Simulations show that photometric redshifts of GRB afterglows with the
12 filters of Swift/UVOT and GROND result at $\Delta z / (1+z)$ \lax\ 0.1,
substantially more accurate than galaxies/AGN due to the simple powerlaw shape
of afterglow SEDs  \citep{ksg11}.
A full description of the procedure, these simulations of the redshift-dependent
error, comparison to spectroscopic redshifts and application to the first
sample of UVOT/GROND detected afterglows is given in \cite{ksg11}.

\section{Observations}

A total of 1018 GRB triggers with localisation errors smaller than
20\amin\ occurred between May 2007 and September 2016\footnote{see 
  \url{https://www.mpe.mpg.de/~jcg/grbgen.html} for a complete list}.
Out of these, 255 GRBs were at declinations north of +37\fd5
which is the northern-most declination reachable with GROND due to
a minimum 22\degr\ horizon distance requirement of the 2.2m telescope
(the northern-most burst observed with GROND is GRB 101008A at +37\fd06).
Another 104 and 4 were too close to the Sun and Moon, respectively,
to be observable.
Out of these 655 observable GRBs, 519 (79\%)
were actually observed with GROND.
The reasons for the 21\% (136 GRBs) non-observations
split into the following sub-groups:
8.5\% (56 GRBs, =41\% of the non-observed) bad weather,
5.3\% (35, 26\%) GROND being off due to instrument or telescope (M1 coating) maintenance,
2.7\% (18, 13\%) missing override permission (during Chilean or MPIA time), 
1.4\% (9, 7\%) technical problems of the telescope, and
1.2\% (8, 6\%) being purposely not observed (see Tab. \ref{ignored}),
and 1.5\% (10, 7\%) due to positions available more than 48 hrs after the
event (mostly Swift/XRT-follow-up of IPN, AGILE, MAXI or Fermi/LAT positions).

For the subset of 829 Swift-detected GRBs with immediately (up to few hours) 
well-localised {\it Swift}/XRT afterglow positions during the period
considered here, 533 were observable
for GROND, and 434 were actually observed. This implies a follow-up
efficiency of these well-localised {\it Swift}-GRBs of 81\%.

Out of the 519 observed triggers, 2 were later re-classified as
galactic X-ray transients and are thus not covered here any further.
For 3 further GRBs (100225A, 110426A and 130310A), the original error boxes
were much larger than the FOV of GROND, but nevertheless observed with
multiple pointings in the hope that follow-up X-ray observations would
better constrain the afterglow position. This didn't happen for those
three GRBs, so we continue with a sample of 514 GRBs.

\begin{figure}[bht]
\includegraphics[angle=270,width=9.0cm, viewport=28 18 490 750, clip]{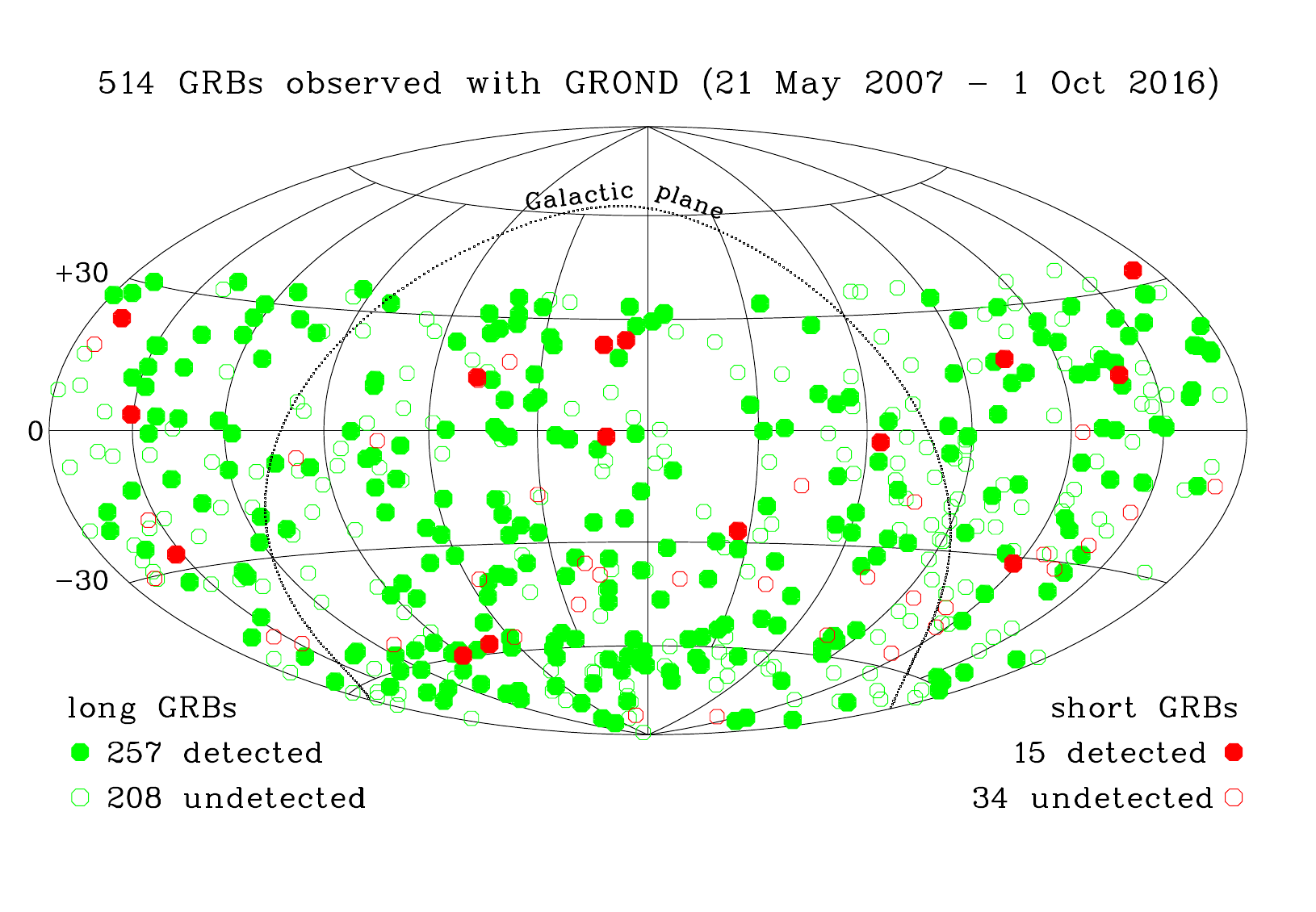}
\caption{Sky distribution in equatorial coordinates of the GROND GRB
  sample. Note the lack of extinction bias towards the Galactic plane,
  visible through the ratio of detected/non-detected afterglows.
  The apparent 'concentration' of observed GRBs towards the South Pole
    is due to a 2$\sigma$ lack of triggers in the
    $-40$\degs\ $ < Decl. < -50$\degs\ range: 32 vs. a mean of 50 (between
    -70\degs\ and +20\degs).
    [Updated from \cite{jcg19}]}
\label{grondGRBs}
\end{figure}

The distribution of these 514 GRBs in equatorial coordinates is shown in
Fig. \ref{grondGRBs}, separated into long-/short-duration (465/49) GRBs and
labelled differently for detected (257 long and 15 short) and
non-detected afterglows.

For 363 of these 514 observed GRBs, preliminary results of the
first observing sequences were quickly reported via the GRB Circular
Network (GCN), while 29 were published 
in refereed journals or conference proceedings without a GCN.
In total, GROND afterglow measurements for 123 GRBs have so far
been published in refereed journals,
i.e. 94 GCN-notified afterglows have a follow-up paper.

For the remaining 122 previously unpublished GRBs,
Tab. \ref{unpub} gives the detections or upper limits in all 7 bands,
with two changes: (i) two GRBs are omitted:
 GRB 100707A (GROND follow-up was on two PTF-detected candidates)
and GRB 120522A (IPN error box covered with 3 GROND pointings), for which
the lack of a second GROND epoch prohibits any statement on objects
fainter than the SM and 2MASS catalogues;
(ii) three GRBs (080703, 130211A, 140719A) are added
clarifying/improving earlier GCN notifications (see text in
sect. \ref{sect:130211A}).
Table \ref{unpub} includes 10 new afterglow identifications, namely GRBs
081127, 101017A, 120215A, 120302A, 130211A, 130725A, 140412A, 140619A, 140710B, 150518A
(marked in boldface in Tab. \ref{unpub}) and one candidate
(GRB 120328A),
as well as new host detections and new or updated photometric redshifts.
The majority of the other 27 GRBs with GROND-detected afterglows
(Tab. \ref{unpub}) are those where GROND observations were either late or not
deeper (or more informative) than measurements already published in GCNs.
The upper limits are not necessarily the very earliest GROND
observations, but compromised for depth (either higher photometric
accuracy, or deeper limits), and often therefore are stacks of
multiple OBs where single outliers in seeing or ellipticity
have been removed.

For selected GRBs with afterglow detection and/or interesting upper limits,
the next section provides the context and more GROND observing details.
Table \ref{summary} lists the GRBs with new afterglow detections,
new host identification, and new or corrected redshift estimates.

\begin{table}[th]
  \caption{8 GRBs purposely ignored for GROND observations}
   \vspace{-0.2cm}
     \begin{tabular}{ll}
     \hline
     \noalign{\smallskip}
     GRB & Reason \\
      \noalign{\smallskip}
      \hline
      \noalign{\smallskip}
       080409 & skipped in favour of 080408 \\
       080607 & skipped in favour of 080605 \\
       080727 & skipped in favour of 080727B \\
       081003 & skipped in favour of 081003B \\
       090515 & not observable in first night, then faint limits \\
              &  known from others\\
       100703A& alert only 2 days later; ignored \\
       120121A& ignored by observer \\
       131026A& not observable in first night, then retracted by \\
              & BAT team \\
      \noalign{\smallskip}
      \hline
      \noalign{\smallskip}
     \end{tabular}
   \label{ignored}
\end{table}

\small
\addtolength{\tabcolsep}{-1pt}   
\begin{longtable}{lcccccccccc}
  \caption{Measurements of either unpublished GROND-observed GRBs or those with updates relative to published GCNs. All magnitudes are in the AB system, and are not corrected for the galactic foreground extinction. GRB names in boldface are those for which we report hitherto unknown optical afterglows. The 2nd and 3rd columns are the mid-time of the data used (T$_o$ denotes the trigger time of the gamma-ray instrument), for which column 4 gives the exposure times for the optical and NIR channels, respectively. Single or stacked multiple observation blocks are used; for the latter, the exposure times deviate from the nominal single-OB exposure times. The long/short classification is given in Tab. \ref{allgrbs}.}\\
     \noalign{\vspace{-0.2cm}}
     \hline
     \noalign{\smallskip}
     GRB & \multicolumn{2}{c}{Mid-Time} & Expo & \gp & \rp & \ip & \zp & $J$ & $H$ & $K$ \\
         & $\!\!\!$\scriptsize{MMDD-hh:mm} & $\!\!$T-T$_o$ (hrs)$\!\!$ & min  & mag & mag & mag & mag & mag & mag & mag \\    
      \noalign{\smallskip}
      \hline
      \noalign{\smallskip}
           \endfirsthead
      \caption{continued.}\\
      \hline
      \noalign{\smallskip}
     GRB & \multicolumn{2}{c}{Mid-Time} & Expo & \gp & \rp & \ip & \zp & $J$ & $H$ & $K$ \\
     & $\!\!\!$\scriptsize{MMDD-hh:mm}&  $\!\!$T-T$_o$ (hrs)$\!\!$ & min & mag & mag & mag & mag & mag & mag & mag \\
      \noalign{\smallskip}
      \hline
      \noalign{\smallskip}
      \endhead
      \endfoot
 070621  & 0622-05:41 & ~~6.39 & 18/20 & $>$24.4 & $>$24.8 & $>$23.8 & $>$23.3 & $>$21.0 &$>$20.2  &$>$18.4  \\
 070917\footnotemark[1] & 0918-00:57 & 17.38 & 73/64 & $>$24.0 & 23.66$\pm$0.30 & 22.84$\pm$0.17 & 23.32$\pm$0.27 & $>$20.9 &$>$19.5  &$>$18.0  \\ 
 071021\footnotemark[1]  & 1022-01:55 & 16.22 & 148/123 & $>$24.3 & $>$24.4 & $>$23.9 & $>$23.8 & $>$21.4 &$>$20.5  & 21.08$\pm$0.32   \\
 071117\footnotemark[1]  & 1118-01:17 &10.45 & 20/15 & $>$23.8 & 22.92$\pm$0.14 & 22.42$\pm$0.15 & 22.37$\pm$0.23 & $>$20.6 &$>$19.9  &$>$19.0 \\
 080130  & 0131-08:25 & 21.20 & 11/11 & $>$23.6 & $>$22.9 & $>$22.5 & $>$22.4 & $>$19.8 &$>$19.2  &$>$18.8  \\
 080319B & 0325-07:45 & 145.54~ & 59/45 & $>$23.5 & $>$23.3 & $>$23.0 & $>$22.8 & $>$20.2 &$>$19.5  &$>$18.9   \\ 
 080405\footnotemark[1]  & 0407-00:25 & 39.10 & 90/76 & $>$25.9 & $>$25.1 & $>$24.4 & $>$24.3 & $>$21.6 &$>$21.0  &$>$20.2    \\
 080414\footnotemark[1]  & 0415-05:04 & ~~6.51 & 40/36 & $>$23.1 & $>$23.6 & $>$22.6 & $>$22.7 & $>$18.3 &$>$17.4  &$>$16.9    \\
 080523\footnotemark[1]  & 0524-09:28 & 12.10 & 40/40 & 23.89$\pm$0.29 & 23.41$\pm$0.12 & 22.90$\pm$0.22 & 22.58$\pm$0.27 & $>$20.6 &$>$19.9  &$>$19.1   \\
 080604  & 0605-01:19 & 17.87 & 26/22 & 22.87$\pm$0.08 & 22.69$\pm$0.09 & 22.55$\pm$0.21& 22.32$\pm$0.28& $>$21.2 &$>$20.1  &$>$19.1  \\
 080703  & 0703-23:12 & ~~4.20 & 4/6 & $>$22.5 & 21.98$\pm$0.16 & 21.87$\pm$0.39 & 21.01$\pm$0.28 & $>$19.2 & $>$18.6  &$>$17.9 \\
 080723  & 0724-02:49 & 22.50 & 6/5 & $>$22.4 & $>$21.1 & $>$20.7 & $>$20.5 & $>$19.4 &$>$18.9  &$>$18.3  \\
 080802  & 0802-23:08 & ~~7.93 &  8/8 & $>$21.5 & $>$21.4 & $>$21.4 & $>$21.5 & $>$18.4 &$>$17.9  &$>$17.4    \\ 
 080810  & 0811-06:52 & 17.70 & 5/6 & $>$20.3 & $>$20.0 & $>$19.5 & $>$19.2 & $>$17.3 &$>$16.3 &$>$15.7 \\ 
 080822B\footnotemark[2] & 0823-08:45 & 11.70 & 4/4  & $>$21.3 & $>$21.8 & $>$21.3 & $>$21.2 & $>$19.4 &$>$18.9  &$>$18.3   \\ 
 080828  & 0830-00:08 & 44.89 & 25/20 & $>$24.1 & $>$23.2 & $>$22.8 & $>$22.3 & $>$19.1 &$>$18.8  &$>$18.4 \\ 
 080905B\footnotemark[1] & 0906-07:45 &14.82 &6/6& $>$23.1 & $>$22.8 & $>$22.1 & $>$21.6 & $>$19.1 &$>$18.8  &$>$17.6  \\
 080915B & 0915-23:52 & ~~7.97&  6/6  & $>$21.5 & $>$22.1 & $>$21.9 & $>$21.5 & $>$19.1 &$>$18.0  &$>$17.4      \\ 
 080922  & 0923-00:08 & 13.07 & 25/20 & $>$24.8 & $>$24.3 & $>$23.7 & $>$23.2 & $>$18.6 &$>$17.8  &$>$16.9      \\
 081003B & 1004-01:46 & ~~4.96 & 25/20 & $>$24.0 & $>$23.5 & $>$23.0 & $>$22.5 & $>$19.5 &$>$19.0  &$>$18.6      \\ 
 081016  & 1017-00:24 & 17.53 & 25/20 & $>$24.0 & $>$23.7 & $>$22.8 & $>$22.4 & $>$19.7 &$>$18.6  &$>$17.9      \\ 
 081016B & 1017-01:27 & ~~5.66 & 8/8 & $>$23.6 & $>$23.9 & $>$23.0 & $>$22.6 & $>$20.1 &$>$19.7  &$>$18.1        \\
 {\bf 081127}\footnotemark[1]  & 1128-01:20 & 18.25 & 55/52 & 23.26$\pm$0.11 & 22.98$\pm$0.15 & 22.76$\pm$0.17 & 22.03$\pm$0.14 & $>$20.5 &$>$20.0  &$>$19.1      \\
 090129  & 0130-08:56 & 11.81 & 8/8 & $>$22.1 & $>$21.5 & $>$20.8 & $>$20.4 & $>$18.2 &$>$17.7  &$>$16.7     \\ 
 090201  & 0202-00:55 & ~~7.13 & 8/8 & $>$23.5 & $>$24.0 & $>$23.1 & $>$22.8 & $>$19.8 &$>$19.5  &$>$18.4    \\
 090324  & 0324-09:18 & ~~6.50 & 13/12 & $>$23.9 & $>$24.0 & $>$23.6 & $>$23.2 & $>$19.4 &$>$19.1  &$>$18.8  \\ 
 090401B & 0402-01:06 & 16.51 & 23/24 & 23.77$\pm$0.19 & 22.60$\pm$0.06 & 22.10$\pm$0.08 & 21.65$\pm$0.07 & 20.93$\pm$0.13 & 20.46$\pm$0.16 & 20.29$\pm$0.26 \\
 090418  & 0419-09:31 & 22.39 & 13/12 & $>$23.5 & 22.81$\pm$0.18 & 22.62$\pm$0.39 & 22.11$\pm$0.19 & $>$20.0 &$>$19.5  &$>$18.6 \\
 090529  & 0531-04:05 & 37.87 & 50/40 & 23.45$\pm$0.15 & 22.83$\pm$0.09 & 23.02$\pm$0.22 &23.02$\pm$0.34 & $>$21.7 &$>$21.2  &$>$20.5     \\
 090809\footnotemark[3]  & 0810-02:07 & ~~8.60 & 15/16 & 22.16$\pm$0.17 & 21.52$\pm$0.09 & 20.99$\pm$0.10 & 21.12$\pm$0.17 & 20.23$\pm$0.24 & $>$20.8  &$>$19.9  \\ 
 090827\footnotemark[1] & 0829-02:16 & 31.16& 18/15 & $>$23.4 & $>$23.9 & $>$23.2 & $>$22.8 & $>$20.0 &$>$18.8  & --          \\
 090926B & 0927-04:43 & ~~6.79 & 75/60 & 23.16$\pm$0.11 & 23.15$\pm$0.10 & 22.82$\pm$0.12 & 22.31$\pm$0.11 & 21.89$\pm$0.32 & 20.65$\pm$0.17 & 20.21$\pm$0.26   \\ 
 090929  & 0930-03:42 & 23.15 & 24/15 & $>$22.9 & $>$22.8 & $>$22.3 & $>$22.3 & $>$19.9 &$>$19.3  &$>$18.6        \\ 
 090929B & 0930-07:44 & 21.58 & 4/4 & $>$19.7 & 19.45$\pm$0.10 & 19.40$\pm$0.20 & $>$19.4 & $>$19.3 &$>$19.1  &$>$18.2  \\ 
 091015  & 1017-01:27 & 26.45 & 25/20 & $>$24.9 & $>$24.2 & $>$23.7 & $>$23.2 & $>$20.5 &$>$19.8  &$>$19.2 \\
 100115A\footnotemark[1] & 0116-01:16 & 14.01 & 9/15 & $>$22.4 & 22.31$\pm$0.12 & 22.43$\pm$0.26 & 22.03$\pm$0.22 & $>$20.0 &$>$19.1  &$>$17.4  \\ 
 100203A & 0205-02:32 & 32.01 & 19/15 &$>$22.7 & $>$22.3 & $>$21.4 & $>$20.9 & $>$18.1 &$>$17.5  &$>$16.4  \\
 100526A\footnotemark[4] & 0527-06:04 & 13.63 & 0/4 & -- & -- & -- & -- & $>$20.7 &$>$20.0  &$>$19.4   \\
 100704A & 0704-23:41 & 20.10 & 25/20 &$>$23.7 & $>$23.5 & $>$22.8 & $>$22.6 & $>$20.9 &$>$20.2  &$>$19.2   \\ 
 100816A\footnotemark[1] & 0818-06:49 & 54.19 & 25/20 & $>$25.1 & $>$25.0 & $>$24.1 & $>$23.6 & $>$20.8 &$>$20.1  &$>$19.2   \\ 
 100823A\footnotemark[1] & 0824-05:02 & 11.61 & 25/20 & $>$22.4 & $>$22.9 & $>$22.6 & $>$22.3 & $>$20.3 &$>$19.7  &$>$18.4   \\ 
 100901A & 0904-06:56 & 65.36 & 25/20 & 20.85$\pm$0.03 & 20.54$\pm$0.03 & 20.31$\pm$0.03 & 20.01$\pm$0.03 & 19.75$\pm$0.07 &19.49$\pm$0.09 &19.45$\pm$0.12  \\ 
 100915B & 0915-09:06 & ~~3.27 & 25/20 & $>$24.4 & $>$24.3 & $>$23.5 & $>$23.0 & $>$20.4 &$>$19.6  &$>$18.9      \\ 
 100917A & 0917-23:43 & 18.66& 4/4 & $>$22.6 & $>$22.8 & $>$21.8 & $>$21.8 & $>$19.6 &$>$18.9  &$>$18.1     \\ 
 100928A & 0928-23:40 & 21.34 &  4/4 & $>$21.7 & $>$22.3 & $>$21.3 & $>$21.1 & $>$18.9 &$>$18.3  &$>$17.9    \\ 
 {\bf 101017A}\footnotemark[1] & 1018-01:10 & 14.62 &  36/44 & 23.30$\pm$0.30 & 23.25$\pm$0.14 & 22.67$\pm$0.12 & 23.04$\pm$0.24 & $>$21.0 &$>$20.5  &$>$18.9        \\
 110128A & 0129-08:13 & 30.47 & 75/60 &24.66$\pm$0.35 & 23.82$\pm$0.13 & $>$23.4 & $>$23.3 & $>$20.9 &$>$20.0  &$>$18.1        \\ 
 110207A & 0208-00:46 & 13.48 & 12/13 & $>$22.5 & $>$22.4 & $>$21.6 & $>$21.4 & $>$19.3 &$>$18.9  &$>$17.9        \\
 110519A & 0520-04:18 & 26.10 & 7/11 & $>$21.3 & $>$21.3 & $>$20.8 & $>$20.7 & $>$18.5 &$>$16.9  &$>$16.0     \\ 
 110719A & 0720-09:07 & 26.96 & 8/8 & $>$21.7 & $>$22.1 & $>$21.7 & $>$21.5 & $>$19.3 &$>$18.7  &$>$18.4      \\ 
 110818A\footnotemark[1] & 0820-04:20 & 31.70 & 37/30 & $>$24.1 & 24.29$\pm$0.29 & 23.43$\pm$0.26 & 23.05$\pm$0.29 & $>$20.8 &$>$20.1  &$>$19.4   \\
 110921A & 0921-23:45 & ~~9.89 &  3.3/3 & $>$23.1 & $>$23.0 & $>$22.0 & $>$21.3 & $>$19.0 &$>$18.5  &$>$17.8      \\
 111020A\footnotemark[1] & 1021-00:26 & 17.87 & 50/35& $>$24.8 & $>$25.0 & $>$24.1 & $>$23.6 & $>$20.5 &$>$19.7  &$>$18.7  \\ 
 111022A & 1023-00:10 & ~~8.05 &  9/8 & $>$22.7 & $>$21.9 & $>$21.5 & $>$21.0 & $>$16.9 &$>$16.6  &$>$15.9  \\ 
 111026A\footnotemark[1] & 1027-00:08 & 17.34& 10/12 & $>$21.8 & $>$22.2 & $>$21.3 & $>$21.0 & $>$18.5 &$>$17.7  &$>$16.7  \\ 
 111109A & 1109-04:46 & ~~1.80 & 8/8 & $>$21.9 & $>$22.6 & $>$22.2 & $>$22.1 & $>$19.7 &$>$19.0  &$>$17.9   \\ 
 111121A\footnotemark[1] & 1122-07:24 &14.96 & 35/32 & $>$15.0 & $>$13.9 & $>$13.5 & $>$13.0 & $>$12.5 &$>$12.5  &$>$12.0  \\
 111207A\footnotemark[1] & 1209-06:06 & 39.82 &  74/60 & $>$24.8 & $>$25.2 & $>$24.7 & $>$24.7 & $>$22.5 &$>$21.9  &$>$21.0   \\
 120118A\footnotemark[1] & 0118-06:11 & ~~0.10 & 3/4 & $>$23.1 & $>$23.3 & $>$22.6 & $>$22.4 & $>$19.6 &$>$18.6  &$>$18.1    \\ 
 120118B\footnotemark[1] & 0119-01:30 & ~~8.49 & 23/27 & $>$24.8 & $>$24.7 & $>$23.6 & $>$22.8 & $>$20.3 &$>$19.8  &$>$18.5   \\ 
 120212A\footnotemark[1] & 0214-01:05 & 39.89 & 7/14 & 22.27$\pm$0.08 & 22.20$\pm$0.19 & $>$22.0 & $>$21.8 & $>$20.4 &$>$20.0  &$>$18.7   \\ 
 {\bf 120215A}\footnotemark[1] & 0217-00:34 & 47.88 & 11/16 & $>$23.2 & 23.01$\pm$0.35 & 22.33$\pm$0.34 & 21.74$\pm$0.28 & $>$19.9 &$>$19.6  &$>$17.5   \\
 120229A & 0301-00:12 & ~~9.61& 16/20 & $>$22.9 & $>$23.4 & $>$22.9 & $>$22.9 & $>$20.3 &$>$19.9  &$>$18.9  \\
 {\bf 120302A}\footnotemark[1] & 0304-01:47 & 47.82 & 53/60 & $>$22.2 & 22.27$\pm$0.29 & 21.80$\pm$0.25 & 21.33$\pm$0.14 & 21.16$\pm$0.41 & $>$20.7 & $>$19.8 \\
 120311B\footnotemark[4] & 0312-07:34 & 16.43 & 0/80 & -- & -- & -- & -- & $>$22.3 &$>$21.8  &$>$20.8       \\ 
 120312A & 0313-09:31 & 17.41 & 19/20 & $>$23.7 & $>$23.8 & $>$23.3 & $>$23.0 & $>$20.6 &$>$20.0  &$>$19.1   \\ 
 120320A & 0321-07:43 & 19.78 & 92/76 & $>$25.3 & $>$25.4 & $>$24.7 & $>$24.2 & $>$21.9 &$>$21.4  &$>$20.8   \\
 120328A\footnotemark[1] & 0328-08:24 & ~~5.29& 75/60 & $>$26.0 & 25.75$\pm$0.32 & 24.60$\pm$0.40 & 23.79$\pm$0.35 & $>$21.6 &$>$21.2  &$>$20.4   \\ 
 120403B & 0405-00:47 & 28.22 & 50/40 & $>$22.7 & $>$23.5 & $>$22.9 & $>$22.5 & $>$20.3 &$>$19.8  &$>$18.9    \\ 
 120419A\footnotemark[1] & 0419-23:38 & 10.69& 11/15 & $>$21.4 & $>$22.2 & $>$21.2 & $>$21.0 & $>$18.8 &$>$16.9  &$>$16.1  \\
 120612A & 0613-23:39 & 45.56&  46/49 &  $>$24.7 & $>$24.3 & $>$23.5 & $>$22.8 & $>$20.4 &$>$19.8  &$>$19.2   \\ 
 120703A & 0705-09:25 & 39.99 & 77/78 & 23.80$\pm$0.28 & 22.99$\pm$0.21 & 22.65$\pm$0.10 & 22.49$\pm$0.10 & $>$22.2 &$>$21.7  &$>$21.1   \\
 120724A\footnotemark[1] & 0725-03:32 & 20.88 & 50/40 & $>$23.0 & $>$23.0 & $>$23.0 & $>$23.0 & $>$20.0 & $>$19.0  & $>$17.5   \\ 
 120805A\footnotemark[1] & 0805-23:29 & ~~2.01& 9/15 & $>$20.4 & $>$20.7 & $>$20.3 & $>$20.1 & $>$17.8 &$>$17.4  &$>$16.7   \\ 
 120807A & 0807-23:59 & 16.82 & 76/61 & $>$24.6 & $>$24.7 & $>$23.8 & $>$23.8 & $>$20.1 &$>$19.8  &$>$19.1   \\ 
 120817A\footnotemark[1] & 0818-04:28 & 21.64 & 9/13 & $>$21.2 & $>$21.1 & $>$20.6 & $>$20.3 & $>$18.2 &$>$17.9  &$>$17.3   \\ 
 120907A & 0908-07:23 &30.98 & 25/20 & $>$23.0 & $>$23.1 & $>$22.8 & $>$22.1 & $>$19.7 &$>$19.3  &$>$18.3    \\
 121014A\footnotemark[1] & 1015-09:20 & 13.13&  4/24 &  $>$21.3 & $>$21.7 & $>$21.3 & $>$21.0 & $>$17.6 &$>$17.5  &$>$17.2    \\
 121028A\footnotemark[4] & 1029-00:06 & 19.02&  0/6  &  -- & -- & -- & -- & $>$19.0 & $>$18.5 & $>$17.7 \\ 
 121102A\footnotemark[1] & 1103-00:20 & 21.88 &  12/17 &  $>$22.9 & $>$22.8 & $>$22.3 & $>$21.8 & $>$18.3 &$>$17.7  &$>$17.2    \\
 121226A\footnotemark[1] & 1227-07:05 & 11.92& 150/120 & -- & -- & -- & -- & -- & -- & --  \\
 {\bf 130211A}\footnotemark[1]$^,$\footnotemark[5]$\!\!$ & 0211-06:57 &~~3.34 & 94/85 & 24.73$\pm$0.21 & 23.82$\pm$0.08 & 23.22$\pm$0.09 & 22.95$\pm$0.11 & 21.34$\pm$0.22 &$>$20.3  &$>$19.4     \\
 130216A & 0217-00:27 & ~~2.19& 3/3 & $>$22.5 & $>$22.6 & $>$22.0 & $>$21.9 & $>$19.4 &$>$18.9  &$>$18.1  \\ 
 130216B & 0218-01:07 & 30.15& 44/30 & $>$24.0 & $>$24.0 & $>$23.6 & $>$23.4 & $>$20.7 &$>$20.2  &$>$19.0  \\ 
 130306A & 0307-08:55 & ~~9.07& 58/60 & $>$22.1 & $>$22.1 & $>$21.6 & $>$21.3 & $>$18.6 &$>$18.1  &$>$17.8  \\ 
 130313A & 0314-06:38 & 14.50 & 31/39 & $>$25.2 & $>$25.1 & $>$24.3 & $>$24.0 & $>$21.2 &$>$20.5  &$>$19.4   \\ 
 130603B\footnotemark[1] & 0604-23:01 & 31.20& 7/8 &   -- & -- & -- & -- & -- & --  & --  \\ 
 130606B\footnotemark[6] & 0607-04:50 & 16.91& 12/15 & $>$24.4 & $>$23.8 & -- & -- & $>$21.4 &$>$20.8  &$>$19.2  \\ 
 130610A & 0611-02:27 & 23.25 & 25/20  & 23.33$\pm$0.14 & 23.21$\pm$0.17 & 22.88$\pm$0.22 & 22.54$\pm$0.29 & $>$20.5 &$>$20.3  &$>$18.6   \\ 
 130612A\footnotemark[1] & 0613-03:47 & 24.41 & 75/60 & 24.96$\pm$0.15 & 24.54$\pm$0.14 & 24.31$\pm$0.28 & 24.14$\pm$0.29 & $>$21.2 & $>$20.6 & $>$19.8   \\ 
 {\bf 130725A}\footnotemark[1] & 0727-01:40 & 38.05 & 37/30 & 23.53$\pm$0.16 & 23.17$\pm$0.16 & -- & -- & $>$20.3 &$>$19.8  &$>$18.8      \\
 130807A\footnotemark[1] & 0808-06:07 & 19.69 & 12/16 & $>$22.1 & $>$21.2 & $>$20.8 & $>$19.9 & $>$17.8 &$>$17.1  &$>$16.6     \\
 131018A & 1019-07:42 & 18.90 & 75/65 & $>$24.0 & $>$24.5 & $>$24.2 & $>$23.9 & $>$21.3 &$>$20.8  &$>$19.8    \\ 
 131209A\footnotemark[1] & 1211-08:59 & 43.85 &  0/12 &  -- & -- & -- & -- & $>$20.1 &$>$19.8 &$>$19.2  \\
 131231A & 0101-01:13 & 20.46 & 8/20 & 18.81$\pm$0.03 & 18.47$\pm$0.03 & 18.51$\pm$0.04 & 18.38$\pm$0.03 & 18.28$\pm$0.07 & 18.12$\pm$0.10 & 18.06$\pm$0.11  \\
 140114A & 0116-08:15 & 44.29& 34/36 & $>$22.5 & $>$23.2 & $>$22.9 & $>$22.6 & $>$20.5 &$>$20.0  &$>$18.1     \\ 
 140129A\footnotemark[1] & 0130-01:18 & 21.90& 17/18 & 22.62$\pm$0.08 & 22.41$\pm$0.08 & 22.10$\pm$0.11 & 21.16$\pm$0.09 & $>$19.9 &$>$19.4  &$>$16.4  \\ 
 140209A\footnotemark[1] & 0210-01:24 & 17.88 & 67/54 & $>$22.6 & $>$22.3 & 21.69$\pm$0.18 & 21.62$\pm$0.14 & 20.37$\pm$0.16 & 19.79$\pm$0.30 & 19.35$\pm$0.26    \\ 
 140331A\footnotemark[1] & 0401-04:28 & 22.64 & 25/20 &   $>$24.4 & $>$24.1 & $>$23.7 & $>$23.5 & $>$21.2 &$>$20.7  &$>$19.5 \\ 
 {\bf 140412A}\footnotemark[1] & 0412-23:45 & ~~1.40 & 26/25 & $>$23.5 & $>$24.3 & $>$23.7 & 23.32$\pm$0.32 & $>$20.4 &$>$19.9  &$>$19.3  \\
 140614B & 0614-07:57 & ~~1.31 & 16/20  & $>$22.1 & $>$21.8 & $>$21.4 & $>$21.2 & $>$19.0 &$>$18.5  &$>$17.5     \\ 
 {\bf 140619A}\footnotemark[1] & 0620-09:25 & 21.77& 62/51 & 24.25$\pm$0.21 & 23.82$\pm$0.19 & 23.53$\pm$0.30 & 22.82$\pm$0.16 & $>$20.5 & $>$20.2  &$>$19.5  \\
 {\bf 140710B}\footnotemark[1] & 0710-23:07 & ~~1.49 &  9/10 & $>$21.6 & 21.79$\pm$0.24 & 20.39$\pm$0.15 & 19.51$\pm$0.18 & 18.48$\pm$0.19 & 17.64$\pm$0.15 & 16.76$\pm$0.14 \\
 140719A\footnotemark[1] & 0719-23:47 & 17.89& 62/65& $>$24.4 & 23.91$\pm$0.10 & 23.51$\pm$0.17 & $>$23.0 & $>$20.3 &$>$19.8  &$>$18.6 \\
 140719B & 0720-09:31 & 12.69& 8/7 &  $>$23.6 & $>$23.8 & $>$23.2 & $>$22.6 & $>$20.1 &$>$19.4  &$>$18.6    \\
 140927A\footnotemark[7] & 0927-05:30 & ~~0.25 & 4/4 &  -- & -- & -- & -- & -- & --  & --     \\
 141207A\footnotemark[8] & 1209-07:00 & 35.81 & 35/40 & -- & -- & -- & -- & -- & --  & --     \\
 141212A                 & 1213-03:00 & 14.77 & 137/110 & $>$25.5 & $>$25.1 & $>$24.5 & $>$24.5 & $>$21.3 & $>$20.7 & $>$19.9 \\
 150101B\footnotemark[9] & 0106-07:45 & 112.36~ & 24/20 & $>$18.0 & $>$18.5 & $>$18.5 & $>$18.2 & $>$16.5 &$>$16.0  &$>$15.0   \\ 
 150123A\footnotemark[10] & 0124-04:04 & 13.04 & 24/20 & $>$24.2 & $>$23.9 & $>$23.5 & $>$23.5 & $>$20.4 &$>$20.0  &$>$18.8  \\
 150428B\footnotemark[11] & 0428-05:51 & ~~2.65 & 3/3 & $>$21.5 & $>$21.2 & $>$20.9 & $>$20.8 & $>$18.7 &$>$18.2  &$>$17.2  \\ 
 {\bf 150518A}\footnotemark[1] & 0520-04:16 & 30.56 & 50/40 & 21.89$\pm$0.04 & 21.33$\pm$0.03 & 21.09$\pm$0.04 & 20.67$\pm$0.04 & 20.35$\pm$0.11 & 20.06$\pm$0.12 & 20.00$\pm$0.24  \\ 
 150831B\footnotemark[12] & 0901-02:17 & ~~3.96 & 28/30  & $>$21.3 & $>$20.6 & $>$20.1 & $>$19.7 & $>$16.9 &$>$16.6  &$>$15.9  \\ 
 150902A\footnotemark[12] & 0904-00:35 & 30.99 & 89/16 & $>$25.0 & $>$25.0 & $>$24.3 & $>$23.8 & $>$20.6 &$>$19.7  &$>$19.3   \\
 151004A\footnotemark[13] & 1004-23:55 & ~~5.77 & 15/11 & $>$22.1 & $>$22.2 & $>$21.1 & $>$21.1 & $>$19.1 &$>$18.5  &$>$17.7 \\
 151228A & 1228-08:09 & ~~2.95& 4/4 & $>$22.0 & $>$22.4 & $>$22.1 & $>$21.6 & $>$20.2 &$>$19.8 &$>$18.6 \\ 
 160104A & 0107-01:49 & 62.41 & 72/60 & 24.78$\pm$0.18 & 23.81$\pm$0.11 & 23.64$\pm$0.18 & 23.08$\pm$0.19 & $>$21.7 &$>$21.1  &$>$20.5      \\
 160127A & 0127-08:50 & ~~0.12 & 3.3/3.0 & 17.63$\pm$0.02 & 17.49$\pm$0.01 & 17.31$\pm$0.03 & 17.09$\pm$0.02 & 17.06$\pm$0.05 & 16.90$\pm$0.10 & 16.55$\pm$0.07\\
 160216A & 0217-09:30 & 14.32 & 30/36 & $>$23.7 & $>$23.8 & $>$23.0 & $>$22.7 & $>$20.2 &$>$19.7 &$>$18.8 \\ 
 160223B & 0224-01:00 & 15.01 & 39/32 & $>$22.8 & $>$23.5 & $>$23.2 & $>$23.3 & $>$20.7 &$>$20.2  &$>$19.0     \\
 160625B & 0627-08:59 & 34.31 & 90/100 & 19.51$\pm$0.03 & 19.38$\pm$0.03 & 19.36$\pm$0.03 & 19.23$\pm$0.03 & 19.23$\pm$0.06 & 18.98$\pm$0.07 & 18.96$\pm$0.14 \\
 160726A\footnotemark[14] & 0726-11:01 & ~~9.45 & 0/9 & -- & -- & -- & -- & $>$18.3 &$>$17.9 &$>$17.0  \\ 
 160801A & 0802-00:30 & 15.03 & 54/45 & $>$25.2 & $>$24.8 & $>$23.8 & $>$23.2 & $>$20.8 &$>$20.1  &$>$18.9 \\
   \noalign{\smallskip}
   \hline
   \noalign{\smallskip}
   \noalign{\noindent $^{1}$ See Notes on individual sources.
          ~~$^{2}$ \gp\rp\ip\zp\ images only cover 70\% of the error circle.
          ~~$^{3}$  Electronic noise in \zp-band: no detection.
          ~~$^{4}$ Afterglow position is not covered in  \gp\rp\ip\zp.
          ~~$^{5}$  This is for the newly identified afterglow, and not
                    the one retracted by \cite{SudilovskyGreiner2013}.
          ~~$^{6}$ Temporally not operating \ip\zp\ detectors.
          ~~$^{7}$ The PSF of a known \rp=14.1/$J$=13.0 mag star fills the full 1\farcs7 XRT error circle, so the photometry is purely determined by the noise of that known star.
          ~~$^{8}$ The GROND pointing does cover the first reported Swift/XRT source, but not the second (believed to be the afterglow).
          ~~$^{9}$ These limits are estimates against the bright galaxy.
          ~~$^{10}$ The GROND pointings towards Swift/XRT
    sources \#1 and \#2 \citep{Melandri+2015} cover only 15\% of the MAXI
    error circle \citep{Fukushima+2015}.
          ~~$^{11}$ The PSF of a known star fills half of the 2\farcs3 XRT error circle.
          ~~$^{12}$ Two non-fading objects within the Swift/XRT error circle.
          ~~$^{13}$ Three objects within the 7\asec\ Swift/XRT error circle,
             all seen on archival SkyMapper images.
          ~~$^{14}$ Only observed in twilight with $JHK$.
   }
   \noalign{\vspace{-0.35cm}}
 \label{unpub}
 \end{longtable}

\twocolumn
\normalsize

\section{Notes on individual sources I}

This section provides previously unpublished GROND information on
  either new afterglow discoveries, new host detections or new redshifts
  as summarised in Tab. \ref{summary}. 
  GRBs with GROND upper limits in cases of no previous
optical/NIR afterglow reports or if GROND upper limits are
deeper than those reported by other groups are listed separately in
Appendix A.
We refrain from reviewing the full afterglow observation history for
each source, but just mention what is important for the context of
the GROND observation and the afterglow characterisation.

All GROND magnitudes (including $JHK$) are in the AB system, and not
corrected for foreground extinction unless specifically mentioned.
Upper limits are at 3$\sigma$ confidence, if not noted otherwise.

\begin{table}[th]
  \caption{Summary of new findings. \label{summary}}
   \vspace{-0.2cm}
      \begin{tabular}{ll}
      \hline
      \noalign{\smallskip}
      new afterglows (AG) & 081127, 101017A, 120215A, \\
                     & 120302A, 130211A, 130725A,  \\
                     & 140412A, 140619A, 140710B, \\
                     & 150518A \\
      new candidate AG & 120328A \\
      new host detections & 070917, 071117, 080523, 101017A, \\
                          & 110818A, 140412A, 140619A \\
      new/updated redshifts & 080516,  130211A, 140209A, 140619A\\
      \noalign{\smallskip}
      \hline
      \end{tabular}
\end{table}

\subsection{Newly discovered afterglows or candidates}

\subsubsection{GRB 081127}

\begin{figure}[ht]
  \includegraphics[width=8.8cm]{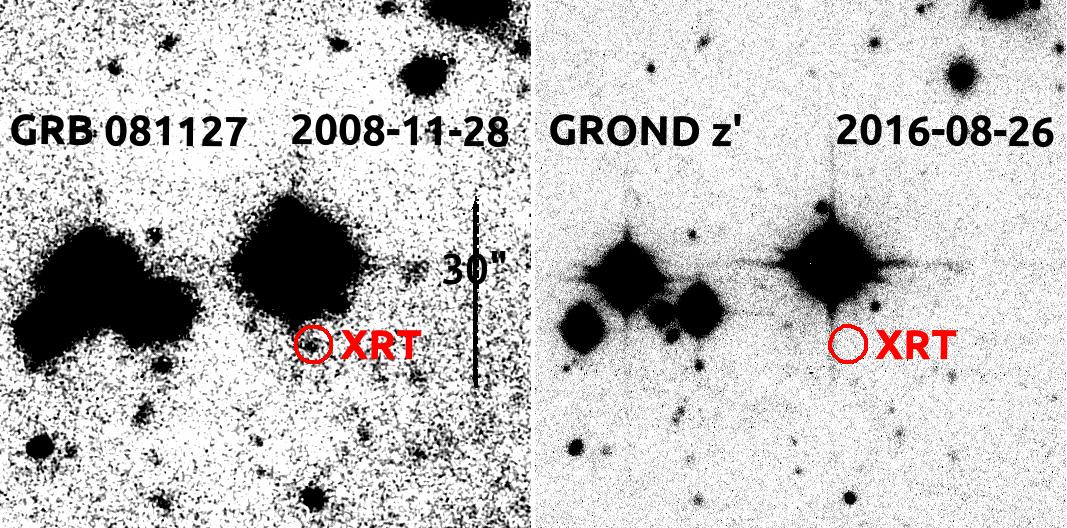}
  \caption[081127]{GROND \zp-band image of GRB 081127 from the 55 min co-add 
    of the first night (left), showing the optical afterglow within the
    1\farcs8 XRT  error circle (drawn here with a 3\asec\ radius
    for better visibility),
    and the second epoch from 26 Aug. 2016 (right).
  \label{081127}}
\end{figure}

For this Swift/BAT-detected GRB a bright, fading X-ray afterglow
was immediately found \citep{Mao+2008}, but no UVOT afterglow was detected
\citep{HollandMao2008}. There is no GCN Circular reporting any
optical or radio observations. GROND started observing
about 18 hrs after the GRB, and a source is found in the
\gp\rp\ip\zp\ bands within the 1\farcs8 XRT error circle .
Photometry in the bluer bands (Tab. \ref{unpub}) might be affected
by a bright star about 10\asec\ north of this source (Fig. \ref{081127}).
A second epoch on Aug. 26, 2016 at better seeing conditions shows no
emission anymore, implying fading by at least 2.5 mag in the two bluest filters
bands (\gp $>$ 25.9 mag, \rp $>$ 25.7 mag, \ip $>$ 24.8 mag, \zp $>$ 24.3 mag).
We thus propose this source, at
RA (J2000) = 22:08:15.43,  
Decl.(J2000) = +06:51:02 (error $\pm$0\farcs3)
as the optical afterglow of GRB 081127.

\subsubsection{GRB 101017A}\label{sect:101017A}

\begin{figure}[ht]
  \includegraphics[width=8.8cm]{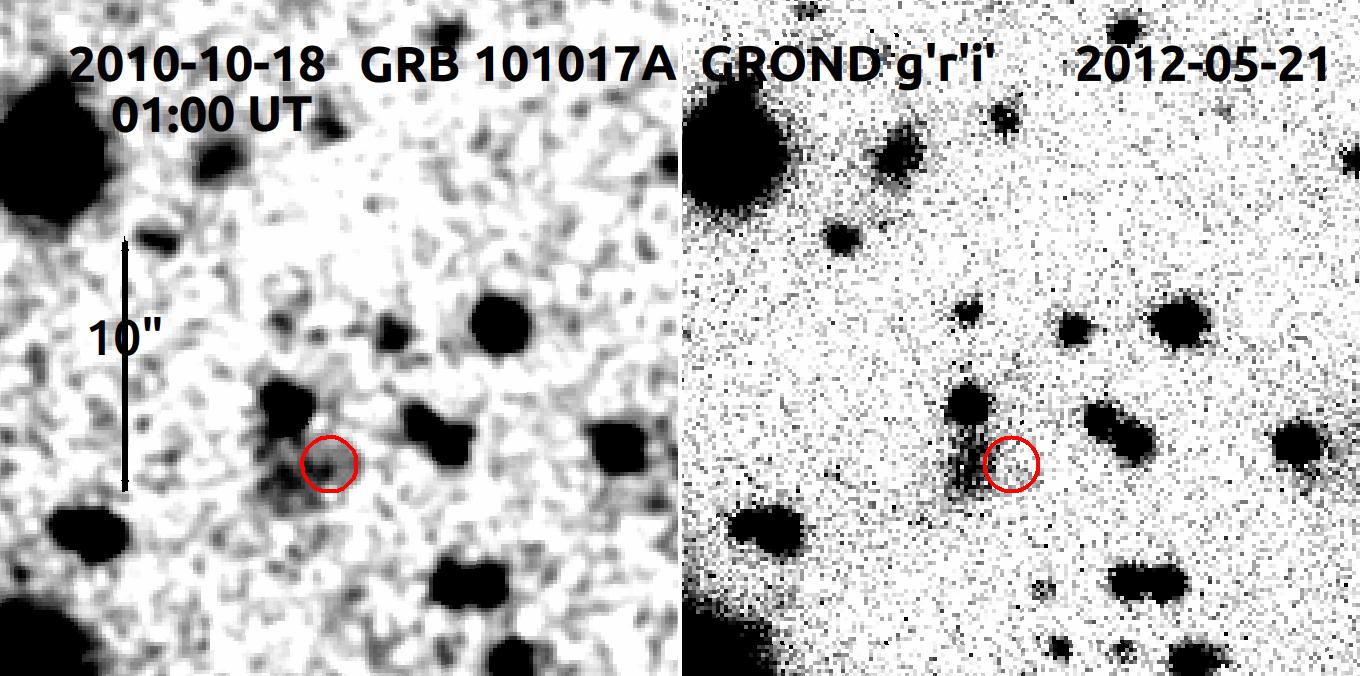}
  \caption[101017A]{GROND images from the co-add of \gp\rp\ip\ stacks
    of the afterglow (left, red circle) and  the host galaxy (right)
    of GRB 101017A.
    The circle denotes the GROND position of the afterglow, slightly North
    but consistent with the UVOT position.
  \label{101017A}}
\end{figure}

XRT and UVOT observations of this long and bright Swift-detected GRB
started already 81 s after the trigger, providing an X-ray position and
a UVOT counterpart at 20th mag \citep{Siegel+2010a}. GROND observations
started immediately after the end of twilight, on 2010-10-18 00:13 UT
and lasted for 2 hrs. A second observation was obtained on 2012-05-21,
at better seeing conditions. This clearly reveals extended emission
very close to the UVOT position, which we suggest to be the host
galaxy of GRB 101017A (right panel of Fig. \ref{101017A}).
In stacked images of the first epoch we also identify emission
in excess of the host emission (left panel of Fig. \ref{101017A}),
about 3 mag fainter than the UVOT detection. We measure a position of
RA(J2000) =  19\h 25\m 32\fss52 and Decl.(J2000) = -35\degr 08\amin 40\fss9
with $\pm$0\farcs3, slightly north but consistent with the UVOT position. 
For the host galaxy, we measure
\gp = 23.6$\pm$0.1 mag,
\rp = 22.9$\pm$0.1 mag,
\ip = 22.3$\pm$0.2 mag,
\zp = 22.0$\pm$0.2 mag.
We note that the $U$-band detection with Swift/UVOT \citep{Siegel2010}
suggests a redshift of $<$2.5.

\subsubsection{GRB 120215A}
\begin{figure}[ht]
  \includegraphics[width=8.9cm]{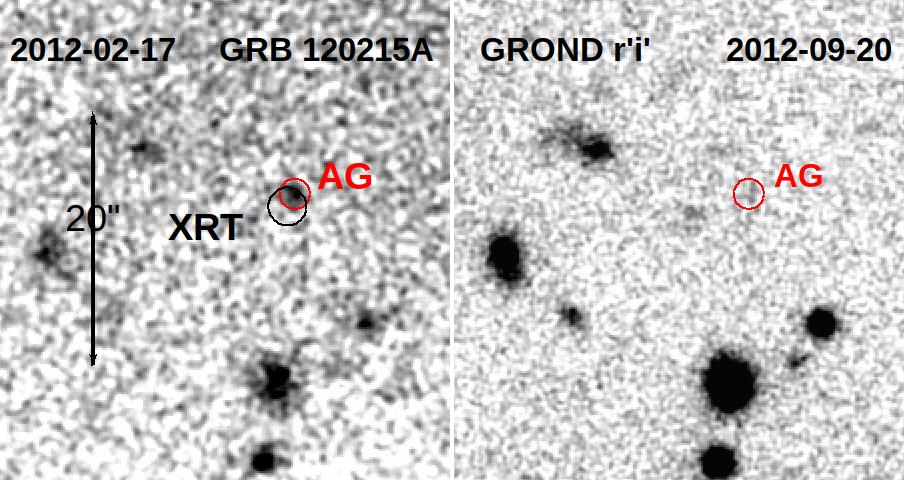}
  \caption[120215A]{GROND \rp\ip-band images from the 2 days after 
    GRB 120215A (left), and seven months later (right)
    The black circle denotes the Swift/XRT position of the afterglow,
    the red circle the afterglow.
  \label{120215A}}
\end{figure}

Based on Skynet observations starting 69 sec after the
Swift-detected GRB,
\cite{LaCluyze+2012} reported the lack of an immediate counterpart
detection, but a later (20 min post-burst) faint source at
$B$=20.32$^{+0.36}_{-0.27}$ mag and
$I$=19.10$^{+0.29}_{-0.23}$ mag.
No significant variability was found.
GROND observations only commenced 2 days later, and reveal a faint source at
the edge of the Swift/XRT error circle \citep{Evans+2012}, at
RA (J2000) = 02:00:11.42,
Decl. (J2000) =08:48:08.6,
with an error of 0\farcs2 (Fig. \ref{120215A}).
The results of the forced photometry
are given in Tab. \ref{unpub}. Assuming that this is the same source
as reported by \cite{LaCluyze+2012} (though their report does not
provide a position), this suggests fading by about 3 mag.
A (longer) GROND observation at better seeing was obtained on 2012-09-20,
providing upper limits about 2 mag deeper, thus establishing substantial
fading of at least 5 mag, and thus the very likely afterglow nature:
\gp $>$ 25.4 mag,
\rp $>$ 25.0 mag,
\ip $>$ 24.6 mag,
\zp $>$ 24.0 mag, J$>$21.2 mag, H$>$20.7 mag, K$>$20.0 mag.

\subsubsection{GRB 120302A}

This Swift-BAT detected GRB was found in automated ground analysis
\citep{Sakamoto+2012}, and the first GROND observation covered
the later-derived X-ray afterglow position only in the $JHK$ channels.
The second GROND observation revealed the newly found NIR source at the
X-ray position at a similar brightness level \citep{Elliott+2012b},
see Tab. \ref{unpub}, but given that
(i) the source was not visible in the SDSS, though brighter than
the SDSS depth, and
(ii) the SED slope was a GRB afterglow-typical powerlaw with slope 1.4,
we had argued for this source as the optical counterpart candidate
\citep{Elliott+2012b}.
A third GROND epoch was taken three weeks later (starting March 26, 00:52 UT)
which provided clear evidence for fading, thus securely identifying
the optical afterglow:
\gp = 25.00$\pm$0.22 mag,
\rp = 24.79$\pm$0.25 mag,
\ip $>$ 24.1 mag,
\zp $>$ 23.8 mag,
J $>$ 21.5 mag,
H $>$ 20.9 mag,
K $>$ 20.2 mag.

\begin{figure}[ht]
  \includegraphics[width=8.9cm]{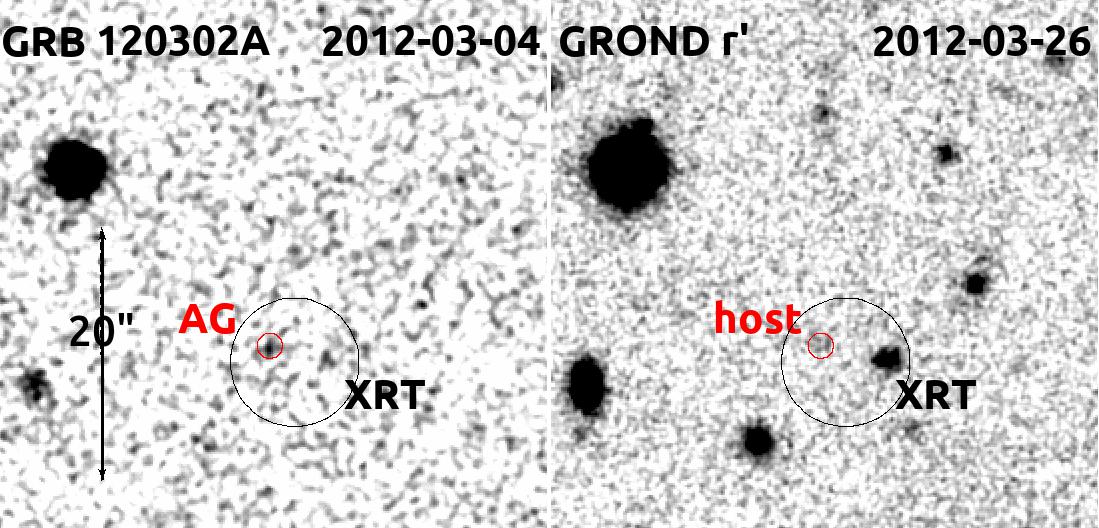}
  \caption[120302A]{GROND \rp-band images from the 2 days after 
    GRB 120302A (left), and three weeks later (right)
    The black circle denotes the Swift/XRT position of the afterglow,
    the red circle the afterglow/host.
  \label{120302A}}
\end{figure}

\subsubsection{GRB 120328A}
\begin{figure}[ht]
  \includegraphics[width=8.8cm]{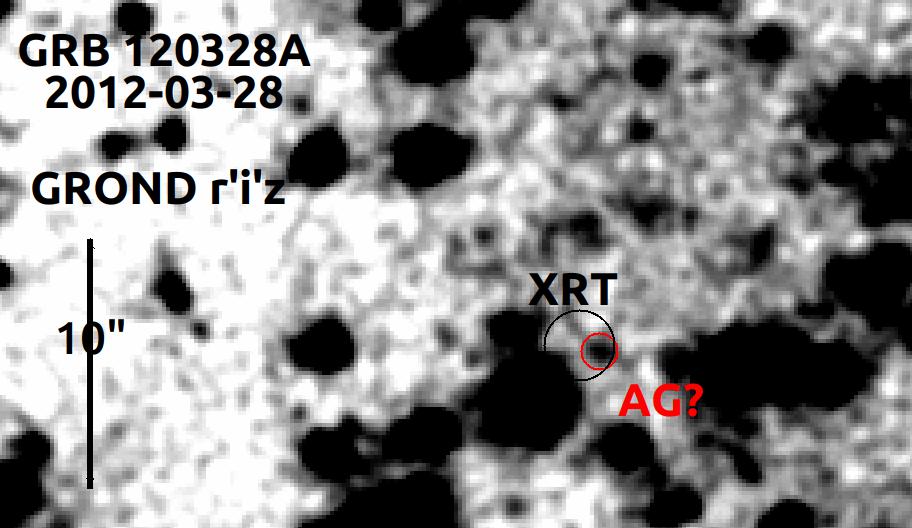}
  \caption[120328A]{GROND image from the co-add of \rp\ip\zp\ stacks
    of the afterglow candidate (red circle) of GRB 120328A.
    The black circle denotes the Swift/XRT position of the X-ray afterglow.
  \label{120328A}}
\end{figure}

The bright X-ray afterglow of this
Swift-detected long-duration GRB was rapidly
found with Swift/XRT observation \citep{Pagani+2012}.
Skynet observations at 4 min after the GRB did not reveal
an optical afterglow, with 3-sigma limiting magnitudes of
V=17.92 mag, R=18.73 mag, I=18.15 mag \citep{Haislip+2012}.
GROND observations started about 2.5 hrs after the GRB. In stacked
observations we detect a faint source in \rp\ (Tab. \ref{unpub})
within the 1\farcs4 Swift/XRT error box \citep{Beardmore+2012}.
This source is marginally visible in \rp\ip\zp\ (Fig. \ref{120328A}),
for which forced photometry \rp =  25.75$\pm$0.32 mag,
gives \ip = 24.60$\pm$0.40 mag, and \zp = 23.79$\pm$0.35 mag.
No second epoch
has been obtained, so no statement about fading can be made.
While the SED looks consistent with a red afterglow, we caution about
the substantial foreground galactic A$_{\rm V} = 2.3$ mag.

\subsubsection{GRB 130211A} \label{sect:130211A}
\begin{figure}[ht]
  \includegraphics[width=8.8cm]{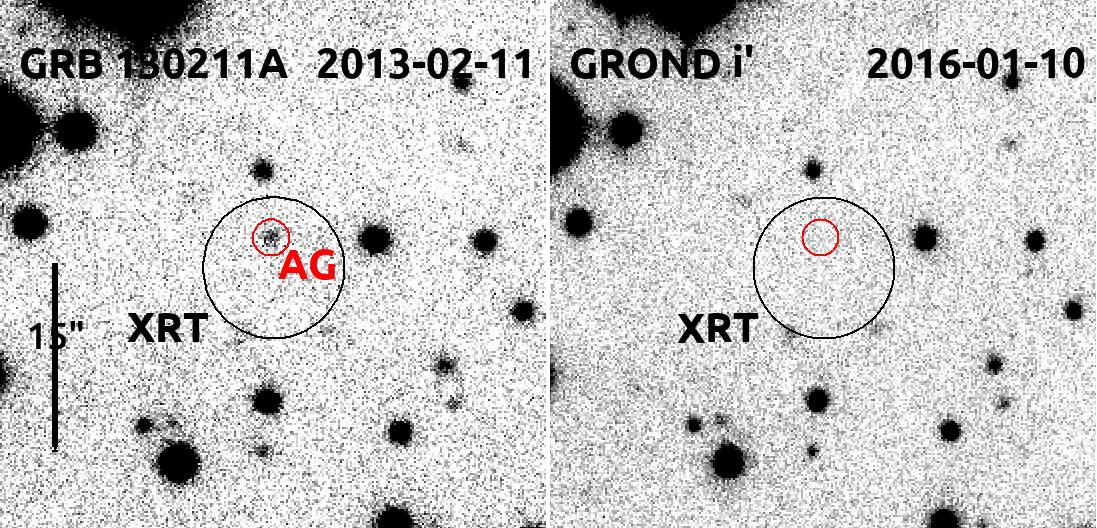}
  \caption[130211A]{GROND image from the 94 min \ip\ stack (left),
    and a later epoch (right), demonstrating the fading of the afterglow.
    The large circle denotes the Swift/XRT afterglow error circle,
    the small/red one encircles the afterglow position.
  \label{130211A}}
\end{figure}

\begin{figure}[ht]
  \includegraphics[width=8.8cm]{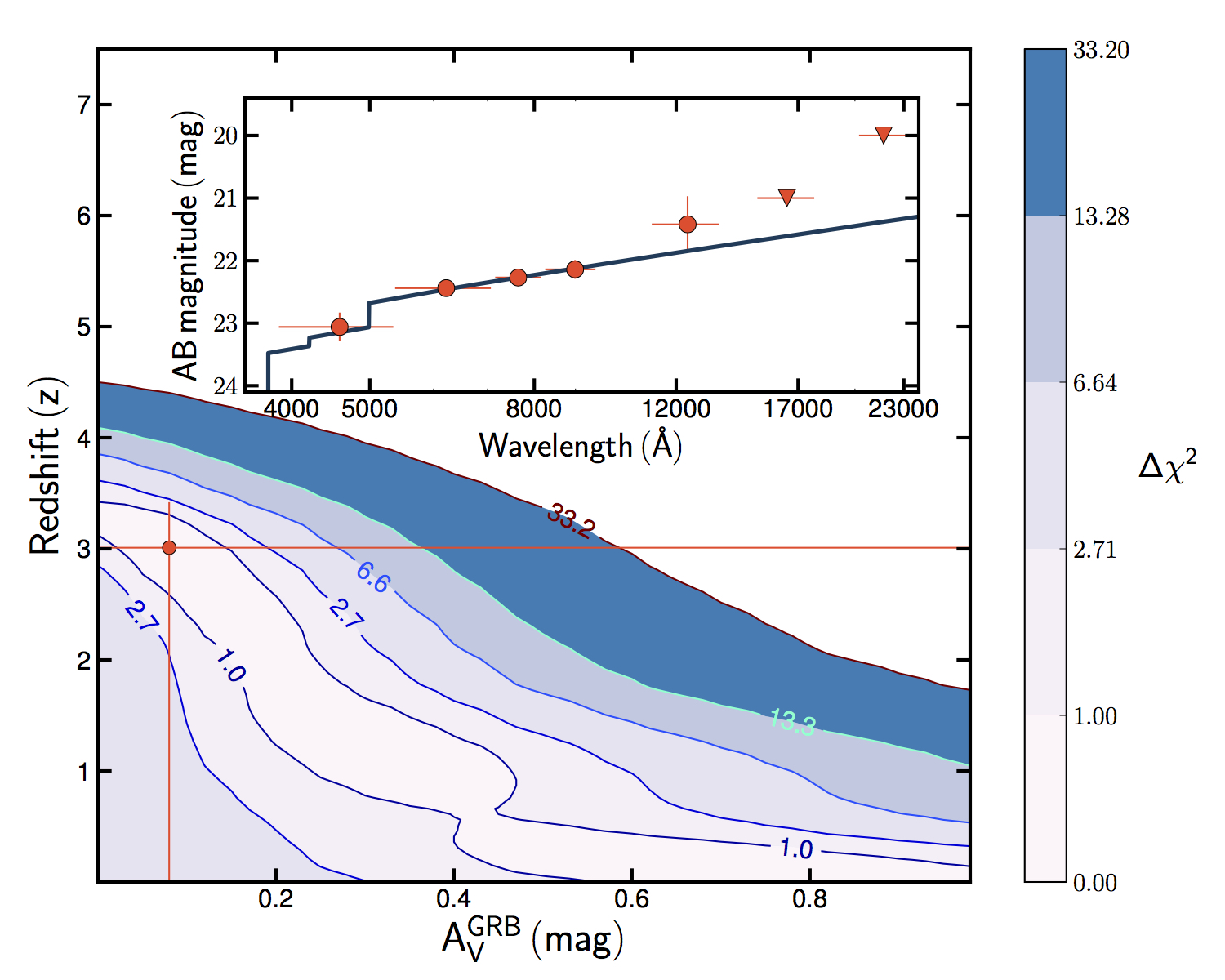}
  \caption[130211A_AV]{Fit to the GRB 130211A afterglow SED, using
    the foreground extinction-corrected GROND data with SMC-type
    host-intrinsic extinction.
  \label{130211A_AV}}
\end{figure}

GROND observations of this Swift-detected GRB \citep{Oates+2013}
started within 30 min under poor weather conditions, but
the optical source detected in the Swift/XRT error circle
\citep{Knust+2013} turned out to be constant \citep{SudilovskyGreiner2013}.
Later re-analysis of the GROND data using the UK Swift/XRT
repository listing of the UVOT-enhanced X-ray position
(which had shifted by 8\asec) revealed a fading optical source within
the XRT localisation, at
RA(J2000) = 09\h 50\m 08\fss66,
Decl.(J2000) = -42\degr 20\amin 38\farcs0, $\pm$0\farcs3 (Fig. \ref{130211A}).
During the first 25 min observations (mid-time 4:10 UT),
this object is only seen in the \rp\ip\ bands, at
\rp = 23.79$\pm$0.28 mag,
\ip = 22.88$\pm$0.21 mag.
Stacking of several OBs at later times
(longer exposure and better conditions) leads to detections in five bands, with
the magnitudes as reported in Tab. \ref{unpub}.
A late epoch was taken on Jan 10th, 2016, resulting in no detection
in any band, with upper limits of \rp $>$25.8 mag, \ip $>$24.9 mag,
\zp $>$24.4 mag, providing evidence of $>$2 mag fading.
Despite the large foreground reddening of E$_{(B-V)}=0.53$ mag,
there is indication for
a Ly-$\alpha$ drop of the \gp-band, suggesting a redshift of
$z \approx 3$ (Fig. \ref{130211A_AV}). The above deep upper limits
for any host emission at and around the afterglow position are consistent
with such an interpretation.

\subsubsection{GRB 130725A}
Due to observing preference given to GRB 130725B,
the Swift-detected GRB 130725A was observed
with GROND only one night later. Weather conditions were mediocre, but
a source is detected within the 1\farcs8 Swift/XRT error circle
\citep{Zhang2013} in the \gp\rp\ bands (the \ip\zp\ bands suffered
from a temporary electronics problem, so no data are available).
This is about 3 mag fainter than the R = 20.8$\pm$0.2 mag candidate
reported by \cite{Kuroda+2013} for their observation at 2 hrs after
the GRB, thus establishing  this to be the afterglow. We measure
RA(J2000) = 15\h 20\m 07\fss71, 
Decl.(J2000) = +00\degr 37\amin 39\farcs8, with an error of $\pm$0\farcs2.

\begin{figure}[ht]
  \includegraphics[width=8.8cm]{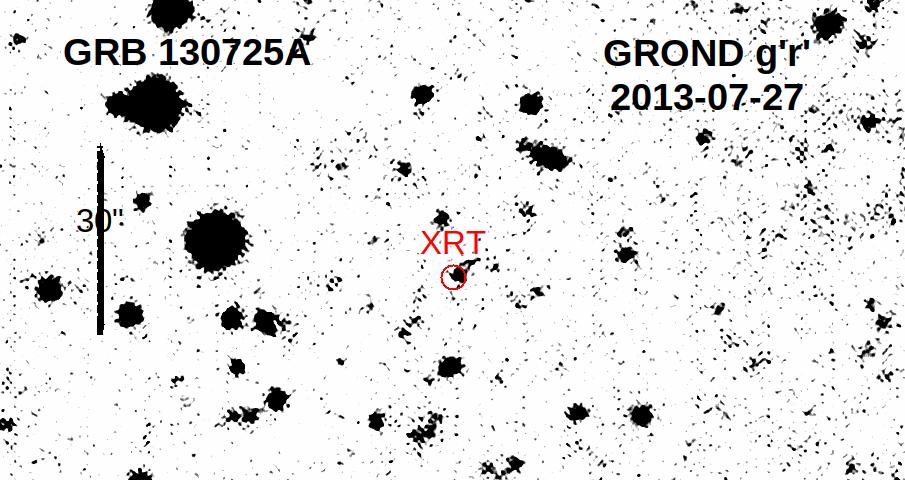}
  \caption[130725A]{GROND image from the co-add of \gp\rp stacks
    of the two OBs.
    The circle denotes the Swift/XRT afterglow position.
  \label{130725A}}
\end{figure}

\subsubsection{GRB 140412A}\label{sect:140412A}

\begin{figure}[ht]
  \includegraphics[width=8.8cm]{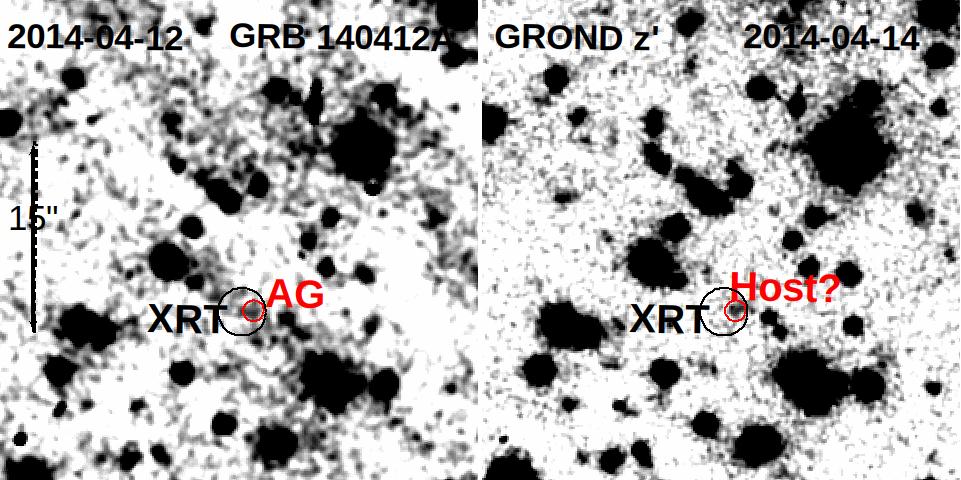}
  \caption[140412A]{GROND images from the co-add of 25 min from the
    first night (left, afterglow), and 100 min of the third night (right, host).
    The black circle denotes the Swift/XRT position of the X-ray afterglow of
    GRB 140412A,
    and the red circle marks the position of the afterglow/host.
  \label{140412A}}
\end{figure}

GROND observations started about 1 hr after this Swift-detected
GRB, at the end of
evening twilight, under very good conditions. No fading source is seen
within or close to the Swift/XRT error circle of the fading X-ray afterglow
\citep{DElia+2014, Goad+2014}. Observations continued for the whole night,
and further 2-hr long observations were done in the second and third night.
A stack of 194 min exposure towards the end of the first night reveals
a faint source in the \zp-band (\zp = 24.22$\pm$0.24 mag)
at the edge of the XRT error circle.
Stacking of the first four exposures (totalling $\approx$25 min)
uncovers this source at \zp = 23.32$\pm$0.32 mag (see Tab. \ref{unpub}),
and a stack of the third night (totalling $\approx$100 min) finds it
at \zp = 24.25$\pm$0.33 mag (Fig. \ref{140412A}). 
While formally consistent with each other within 3$\sigma$,
we propose the early emission to be the afterglow, and the faint
emission, being constant over 48 hrs, to stem from the host galaxy.
The unusual GROND non-detection in the somewhat more sensitive \rp-band
can be explained by the substantial Galactic foreground extinction
of $A_{\rm V}$ = 0.57 mag \citep{sfd98}.

\subsubsection{GRB 140619A}\label{sect:140619A}

\begin{figure}[ht]
  \includegraphics[width=8.8cm]{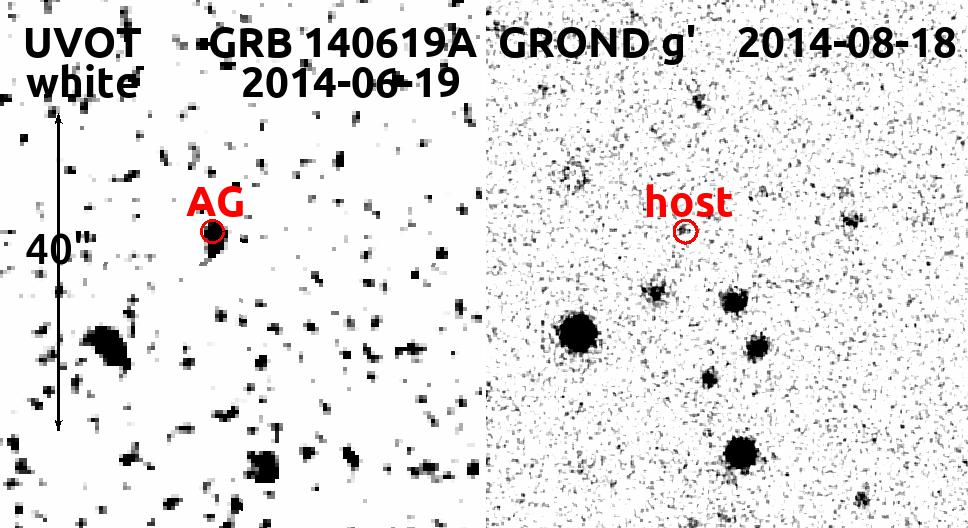}
  \caption[140619A]{Swift/UVOT white (left) and GROND \gp\ image (right)
    of GRB 140619A. The $\approx$3 mag fading confirms the afterglow
    nature of the UVOT detection, and the GROND detection two months
    later is likely the GRB host galaxy. The circle (radius 1\farcs4) is
    to guide
    the eye; the Swift/XRT error circle (1\farcs7) is omitted
    for better legibility.
  \label{140619A}}
\end{figure}

GROND observations started about 22 hrs after the Swift GRB trigger
\citep{DePasquale+2014}, and a source consistent with the Swift/UVOT
position \citep{Siegel+2014} is detected in the \gp\rp\ip\zp\ bands.
This fading of 3 mag since the UVOT epoch establishes the afterglow nature.
Another GROND observation was performed two months later, on Aug. 18, 2014,
at good seeing conditions. We still detect emission at this position
(Fig. \ref{140619A}),
about 0.6 mag fainter in \rp\ip\zp,
suggesting detection of the host, and that the emission seen during the
first GROND epoch is mostly afterglow (except in \gp).
We also obtain a marginal $J$-band detection, at J=21.31$\pm$0.32 mag.
If confirmed, the very red \zp-$J$ = 2.4$\pm$0.5 could be interpreted
as the Ca-HK break, suggesting a redshift of 1.9$\pm$0.5.

\subsubsection{GRB 140710B}

\begin{figure}[ht]
  \includegraphics[width=8.8cm]{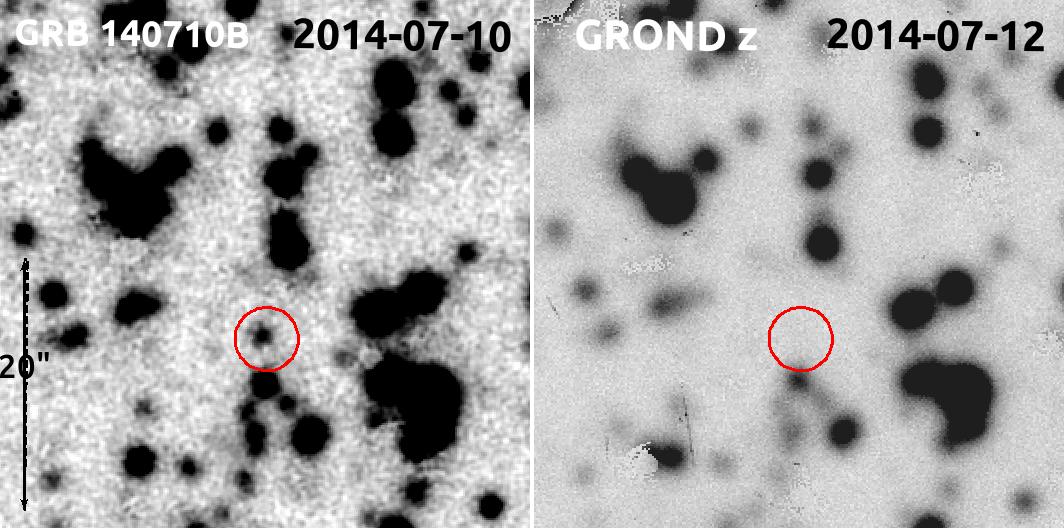}
  \includegraphics[width=8.8cm]{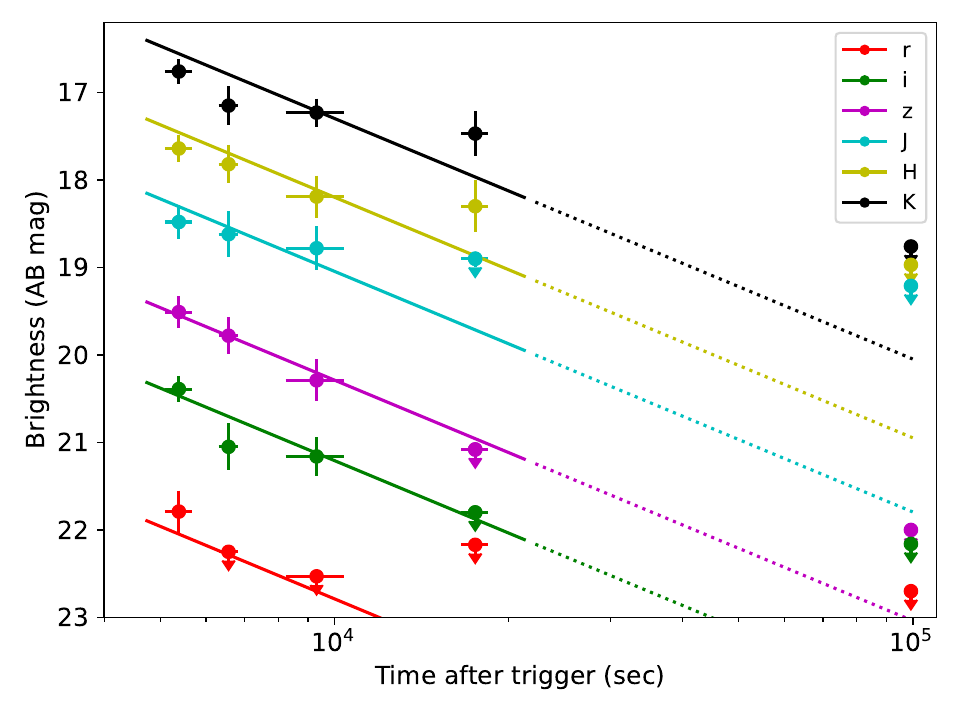}
  \caption[140710B]{GROND \zp-band image from the
    first 10 min OB from the
    first night (top left; mid-time 2014-07-10T23:09), the stack of two 30 min
    OBs from the next night (top right' mid-time  2014-07-12T01:13),
    and the GROND light curve with a best-fit slope
    of 1.2$\pm$0.1 (bottom).
    The 1\farcm7 INTEGRAL error circle is outside this figure,
    but the combined evidence of clear fading and typical powerlaw SED 
    identifies this source without doubt as the afterglow of GRB 140710B.
  \label{grb140710B}}
\end{figure}

GROND observations of this INTEGRAL-detected GRB \citep{Goetz+2014}
started about 80 min after the GRB trigger and continued for 4 hrs.
A second epoch was taken in the following night. Despite the crowded
low galactic latitude field, we identify (in all bands except \gp)
a fading source within the 1\farcm7 INTEGRAL/IBIS error box
(Fig. \ref{grb140710B}), at coordinates 
RA(J2000) = 13\h 38\m 31\fss68, 
Decl.(J2000) = $-$58\degr 35\amin 01\farcs3, $\pm$0\farcs3.
After correcting for the substantial Galactic foreground extinction
of A$_{\rm V}$ = 2.7 mag \citep{sfd98}, the SED is a straight powerlaw
with a slope of 1.4$\pm$0.1.
The combined evidence of fading nature and the afterglow-typical SED
suggests this to be the afterglow of GRB 140710B.

\subsubsection{GRB 150518A}

\begin{figure}[ht]
  \includegraphics[width=8.8cm]{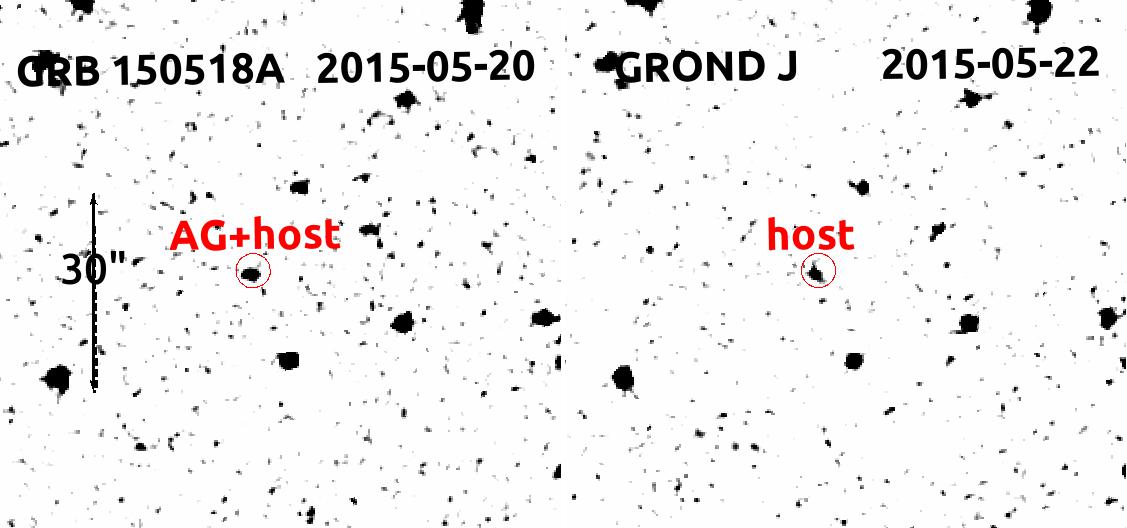}
  \caption[150518A]{GROND J-band images from the co-add of two 20 min OBs
      each from the May 20th (left), and May 22nd (right), respectively, of
      GRB 150518A. The red circle is the 2\farcs6 XRT error circle. Note also
      the slight shift of the centroid of the emission.
  \label{grb150518A}}
\end{figure}

Due to missing override permission, GROND observations of the
Swift/XRT-detected X-ray afterglow \citep{Sbarufatti+2015} of this
MAXI-detected GRB \citep{Kawamuro+2015} could only start 
30 hrs after the GRB trigger. We clearly detect the unresolved
afterglow plus host as first reported by \cite{Xu+2015}.
Our \gp-band detection is still 0.3 mag brighter than the
corresponding SDSS \gp-magnitude, while in the other visual
filters, we see an excess emission of only 0.1-0.2 mag.
A second GROND epoch was done on May 22nd, during which
we only see marginal \gp-excess relative to SDSS. However, we see
significant fading in the $JHK$-bands (down to $J$ = 20.66$\pm$0.12 mag)
between our two GROND epochs (Fig. \ref{grb150518A}),
providing additional evidence for the afterglow
being situated within the SDSS J153648.25+161946.9 galaxy.

\subsection{Newly detected host galaxies}

\subsubsection{GRB 070917}

\begin{figure}[ht]
  \includegraphics[width=8.8cm]{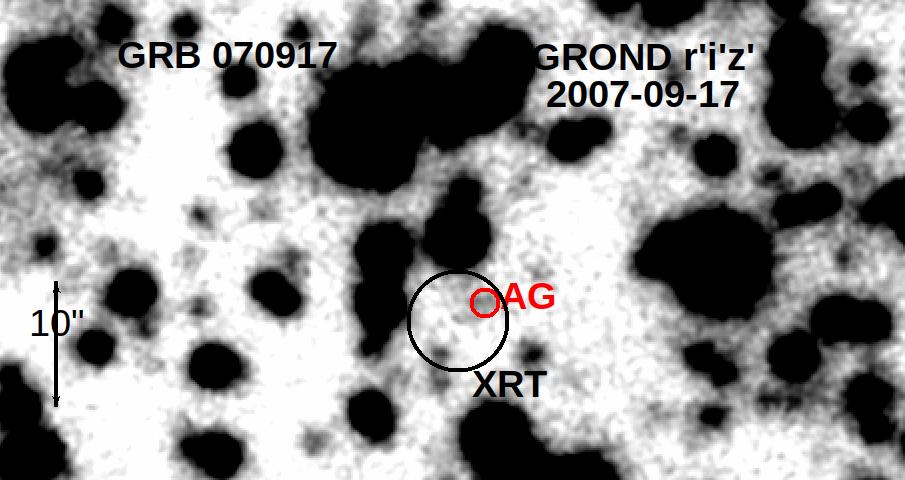}
  \caption[070917]{GROND image from the co-add of \rp\ip\zp stacks of
    GRB 070917 of the first night, showing the afterglow and the
    improved 4\asec\ X-ray error circle. 
  \label{070917}}
\end{figure}

This GRB was detected with Swift/BAT at 07:33:56 UT, with automatic slewing
of Swift disabled \citep{Cummings+2007}. 
Swift/XRT observations started 35 ks after the trigger, and localised
the X-ray afterglow to 8\asec\ accuracy \citep{Evans+2007}.
Optical observations with Gemini South revealed a faint source
inside the XRT error circle at R $\sim$ 22.0 mag, \ip $\sim$ 21.6 mag
\citep{Cenko2007}, which was found to fade with NOT observations,
finding R = 23.1 mag \citep{Fynbo+2007}.
GROND observations started at 23:48 UT, more than 16 hrs after the burst,
lasting about 4 hrs (see Fig. \ref{070917} for a finding chart).
A second epoch was done in the following night.
We confirm fading in \ip\zp\ between our first two epochs by about 0.5 mag,
amounting to about 1.3 mag fading relative to the early observations of
\cite{Cenko2007}. We have no detection in \gp\ (note the large Galactic
foreground reddening of E(B-V)=0.45 mag), and do not see fading in \rp,
suggesting contribution from the host galaxy. Indeed, our 2009 epoch,
performed at 0\farcs8 seeing, still shows faint emission consistent
with the second epoch brightness, at
\rp = 23.6 mag,
\ip = 23.3 mag,
\zp = 23.7 mag, and thus supporting a host interpretation.

\subsubsection{GRB 071117}

\begin{figure}[ht]
  \includegraphics[width=8.8cm]{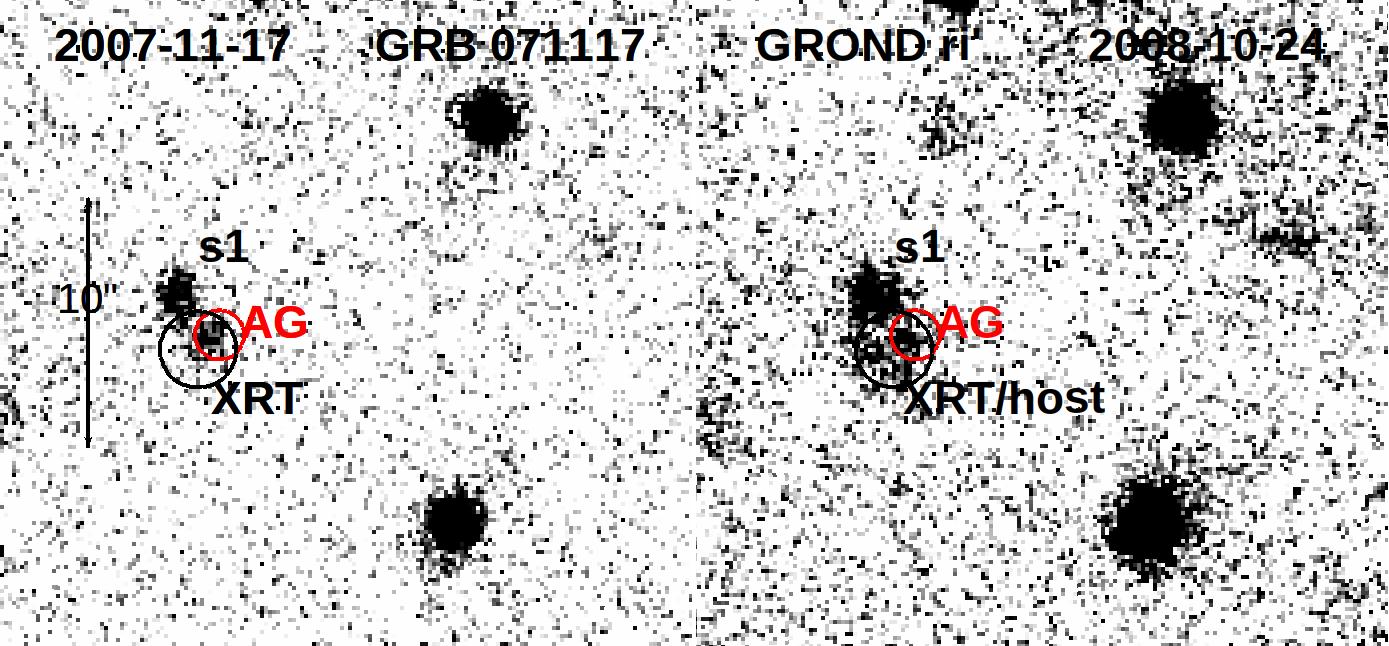}
  \caption[071117]{GROND images from the co-add of \rp\ip\ stacks
    of the GRB 071117 afterglow (left, red circle) and  the host galaxy (right).
    The latest (retrieved 2021) UVOT-enhanced Swift/XRT position is
    overplotted, which is nearly exactly centred on the host galaxy.
  \label{071117}}
\end{figure}

This 5-sec duration GRB was detected with Swift/BAT at 14:50:06 UT,
and the X-ray localisation was delayed by $\sim$45 min due to an Earth
limb constraint \citep{Ukwatta+2007}.
GROND observations started $\sim$9.5 hrs after the GRB, and continued
for 5 hrs. Further epochs were done in the following two nights
(Nov. 19 and 20), as well as on Nov. 22, and 24. A final epoch
was done on October 24, 2008. Similar to the finding of
\cite{Bloom2007a, Bloom2007b} with Gemini South, we find 2 optical sources
in the initial XRT error circle (at that time not yet UVOT-corrected),
with s1 being constant, and s2 clearly fading between the first two epochs.
Our late-time epoch from 2008 shows extended emission underlying
source s2, which we interprete as the host. A spectrum with the ESO VLT/FORS1
instrument revealed a weak emission line at 8688 \AA, interpreted as
[OII], with an inferred redshift of z=1.331 \citep{Jakobsson+2007}. 
Due to the relatively bright host, we detect the afterglow only
in \rp\ip\zp; already in the first epoch the host dominates in the \gp-band.
Fig. \ref{071117} shows \rp\ip\ co-adds of a 20 min observation from
the first night (left panel), and a 80 min stack of the last epoch.
The host galaxy is detected at
\gp = 23.5$\pm$0.2 mag,
\rp = 23.8$\pm$0.1 mag,
\ip = 23.3$\pm$0.1 mag,
\zp $>$23.8 mag,
J$>$21.1 mag,
H$>$20.7 mag,
K$>$19.5 mag.

\subsubsection{GRB 080523}

This $\sim$15 sec duration Swift-detected GRB \citep{Stroh+2008}
had an immediate X-ray afterglow detection. \cite{Fynbo+2008}
reported the ESO/VLT detection of a source inside the X-ray error circle,
and \cite{Malesani+2008} reported fading by at least 1.5 mag based on
a second VLT observation. Since no magnitudes were reported, we
list in Tab. \ref{unpub} the \griz\ magnitudes of the GROND observations
during the first night (nearly simultaneous to that of the VLT),
and confirm the fading based on a GROND observation on Nov 19, 2014.
In the second epoch, no source is detected longward of the \ip-band,
with upper limits of
\ip $>$ 23.8 mag, \zp $>$ 23.4 mag, $J>$ 20.5 mag, $H>$ 19.9 mag, $K>$ 18.8 mag.
However, we clearly detect a source in the optical, with
\gp = 24.02$\pm$0.29 mag and \rp = 24.69$\pm$0.30 mag.
The spatial coincidence and the substantially bluer colour wrt. the afterglow
suggest that this might be the host galaxy, providing
an opportunity to determine the redshift which was not possible to deduce
from the afterglow spectrum apart from a $z<3$ limit \citep{Malesani+2008}.

\begin{figure}[ht]
  \includegraphics[width=8.8cm]{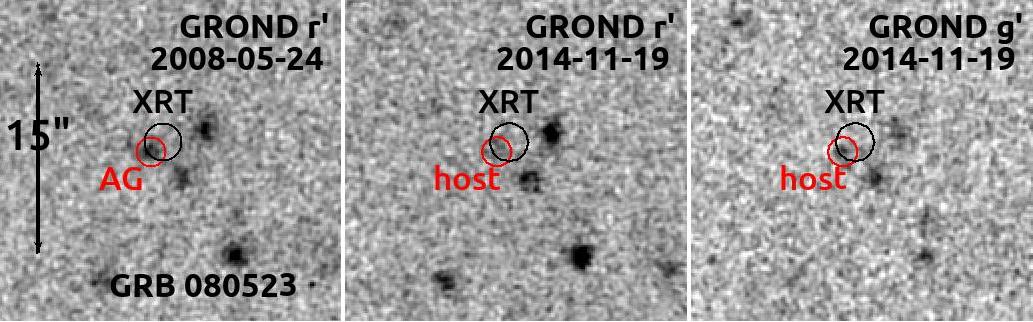}
  \caption[080523]{GROND images of the GRB 080523 X-ray afterglow position
    (black circle), showing the optical afterglow (left, red circle) from
    the first night's \rp-band observation (as given in Tab. \ref{unpub}),
    the \rp-band observation 6 yrs later demonstrating the fading of
    the afterglow (middle), and the \gp-band observation of the same epoch
    showing the strong blue colour of the emission directly underneath
    the afterglow position, consistent with a putative host galaxy.
  \label{080523}}
\end{figure}

\subsubsection{GRB 101017A}

See above sect. \ref{sect:101017A}

\subsubsection{GRB 110818A}

GROND observations of this Swift-detected GRB
were only possible in the second night.
We do detect a source at the ESO/VLT afterglow candidate position
\citep{DAvanzo+2011},
at about 2 mag fainter brightness (see Tab. \ref{unpub}),
thus establishing fading, and confirming the afterglow nature (which so
far was only implicit from the redshift measurement via absorption lines
\citep{DAvanzo+2011, Kruehler+2015}.
In the third night, we still detect the source, at similar magnitudes,
suggesting that we see the host galaxy at z=3.3609 \citep{DAvanzo+2011, Kruehler+2015}.

\subsubsection{GRB 140412A}

See above sect. \ref{sect:140412A}

\subsubsection{GRB 140619A}

See above sect. \ref{sect:140619A}

\subsection{New or updated photometric redshifts}

\subsubsection{GRB 080516}

\begin{figure}[ht]
  \includegraphics[width=8.8cm]{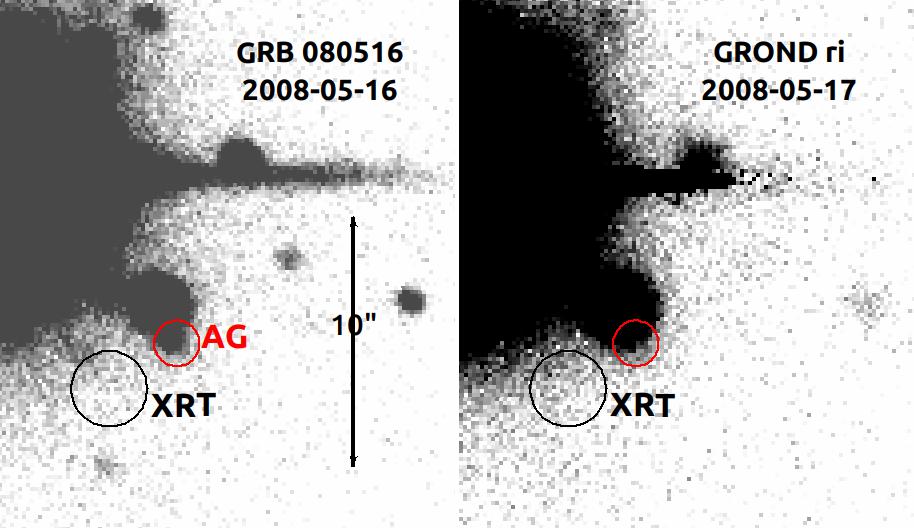}
  \caption[080516]{GROND images from the co-add of \rp\ip\ stacks
    of the GRB 080516 afterglow (left, red circle) at 28 min post-trigger,
    and  the following night (23.5 hrs post-trigger).
  \label{080516}}
\end{figure}

This multi-peaked Swift-detected GRB \citep{Holland+2008} is located
in (behind) the Galactic plane, yet an  X-ray afterglow was readily
discovered which in the beginning faded only slowly, before changing
to a nominal decay after 4800 s \citep{PageHolland2008}.
GROND observations started 8 min after the GRB trigger and revealed
an afterglow candidate \citep{fkjg08} at the corner of the X-ray error
circle (Fig. \ref{080516}), which had faded below the sensitivity threshold
22 hrs later \citep{fkg08}. The SED of stacked observations of the
first night shows a clear Ly drop-out, leading us to report a
photometric redshift of 3.2$\pm$0.3 \citep{fkg08}. A re-analysis
of the data with PS1 (DR2) photometry in the field (also provided
in Tab. \ref{unpub}) suggests an even
higher redshift. Using a foreground Galactic E(B-V)=0.35 mag
\citep{Schlafly+11} and the SMC dust extinction law returns
a photometric redshift of 4.1$\pm$0.1 with a spectral slope of
$\beta = 0.36 \pm 0.14$ and no additional intrinsic extinction.
LMC or MW dust lead to somewhat lower redshift solutions (3.8 and 4.0,
respectively), but at an unusually flat $\beta \approx 0$.

\begin{figure}[ht]
  \includegraphics[width=8.8cm]{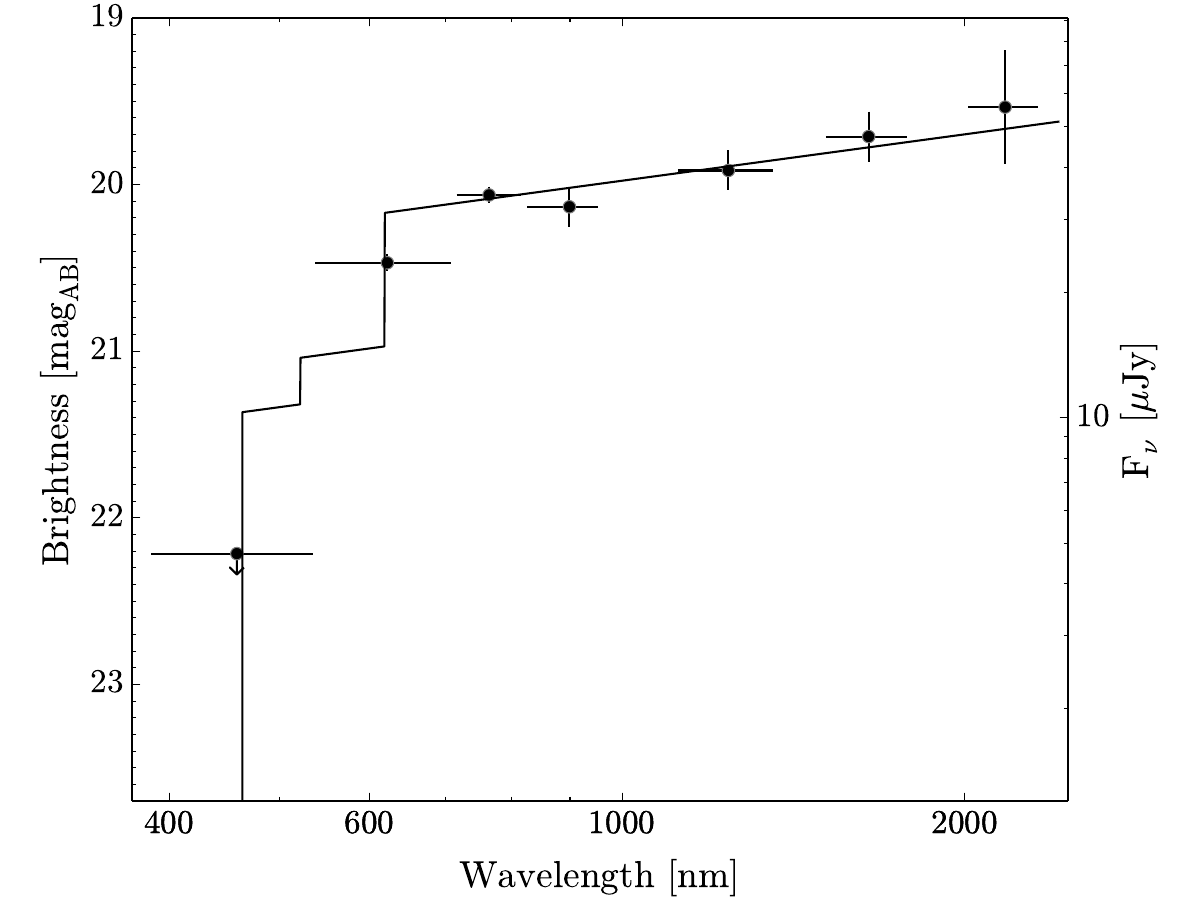}
  \caption[sed080516]{Spectral energy distribution of the GRB 080516 afterglow
   at 28 min after the Swift trigger, corrected for the foreground
   Galactic extinction, providing a best-fit photometric redshift
   of 4.1$\pm$0.1.
  \label{sed080516}}
\end{figure}

\subsubsection{GRB 130211A}

See above sect. \ref{sect:130211A}

\subsubsection{GRB 140209A}
\begin{figure}[ht]
  \includegraphics[width=8.8cm]{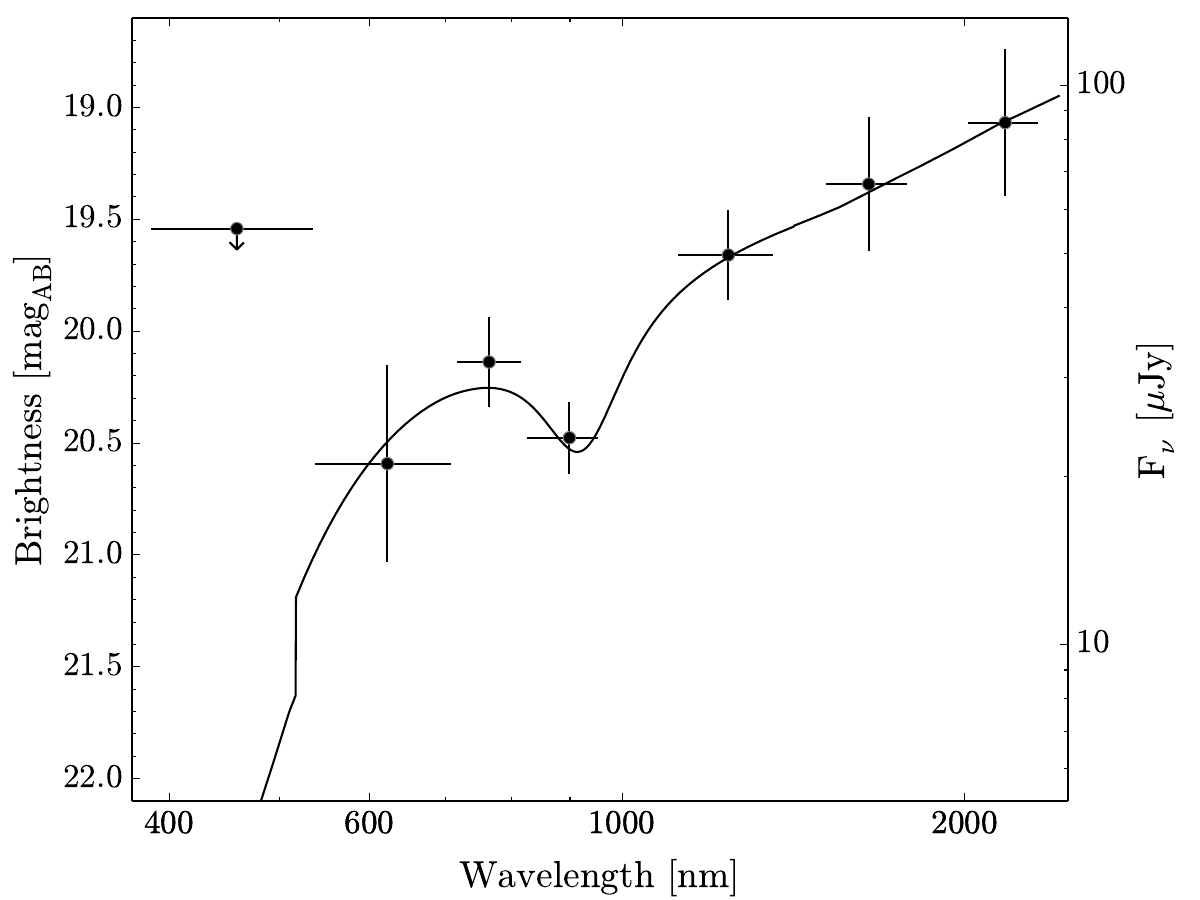}
  \caption[140209A]{Spectral energy distribution of a stack of 
    GROND images of the first epoch (2014 Feb. 10, 00:40--02:14 UT)
    of GRB 140209A.
    The data are corrected for  the foreground $A_{\rm V}$=2.4 mag.
  \label{140209A}}
\end{figure}

Swift could not slew in response to its BAT trigger
due to a Moon-constraint,
but despite the arcmin-scale error box and substantial foreground
extinction ($A_{\rm V}$ = 2.4 mag) an afterglow was quickly identified
which faded from R=17.6 to R=19.4 mag within the first 40 minutes
\citep{Perley2014}. GROND observations were performed after 17 hrs
(see Tab. \ref{unpub}), and 41 hrs, demonstrating further decline
over \rp = 22.7 mag (forced detection) to \rp $>$ 23.4 mag.
We derive a position of
RA(J2000) = 05\h 25\m 19\fss06, 
Decl.(J2000) = +32\degr 29\amin 53\farcs1, $\pm$0\farcs3,
consistent with \cite{Perley2014}.
The SED is clearly red, and the foreground extinction prevents
a \gp-detection. The noteworthy feature of the SED is a clear dip
in the \zp-band (Fig. \ref{140209A}), interpreted as the 2175 \AA\ feature
similarly to GRB 070802 \citep{kkg08}. It is 
only the sixth GRB known to show this feature, after
 GRB 070802 \citep{kkg08, Eliasdottir+2009},
 GRB 080607 \citep{Prochaska+2009, Perley+2011},
 GRBs 080605 and 080805  \citep{Zafar+2012}, and
 GRB 180325A \citep{Zafar+2018}.
The inferred redshift is $z = 3.2 \pm 0.3$.
We note that this is independent of the foreground extinction, but the
intrinsic host extinction has substantial uncertainty, both statistically
($A_{\rm V}^{host}$ = 0.8$\pm$0.2 mag) as well as depending on the extinction
law (not included in the error estimate).

\subsubsection{GRB 140619A}

See above sect. \ref{sect:140619A}

\section{Statistics}

\subsection{Delay time and detection statistics}

With a dedicated instrument, an automated telescope response on GRB triggers
and (for most of the time) a generous override permission,
it is interesting to look at the distribution of response times.
For a total of 69 (out of 514) GRBs we managed to start observations 
immediately, within 30 min after the GRB trigger. This 13\% fraction
is to be compared against the expectation of about 30\% (for a mean
8 hr duration of dark time per day) minus the losses due to
weather, technical issues, and satellite constraints. In particular
the latter is a large factor for observations from South America:
For about 40\% of the time (6 out of 15 orbits per day), GRB detection
is largely impossible from satellites in low-Earth orbit (like Swift or Fermi)
while flying over South America, due to the passage
of the South-Atlantic anomaly.
Thus, our success rate for immediate follow-up is close to
what is possible from La Silla (Chile).

\begin{figure}[t]
\vspace{-0.2cm}
\hspace{-0.3cm}\includegraphics[angle=270, width=9.7cm]{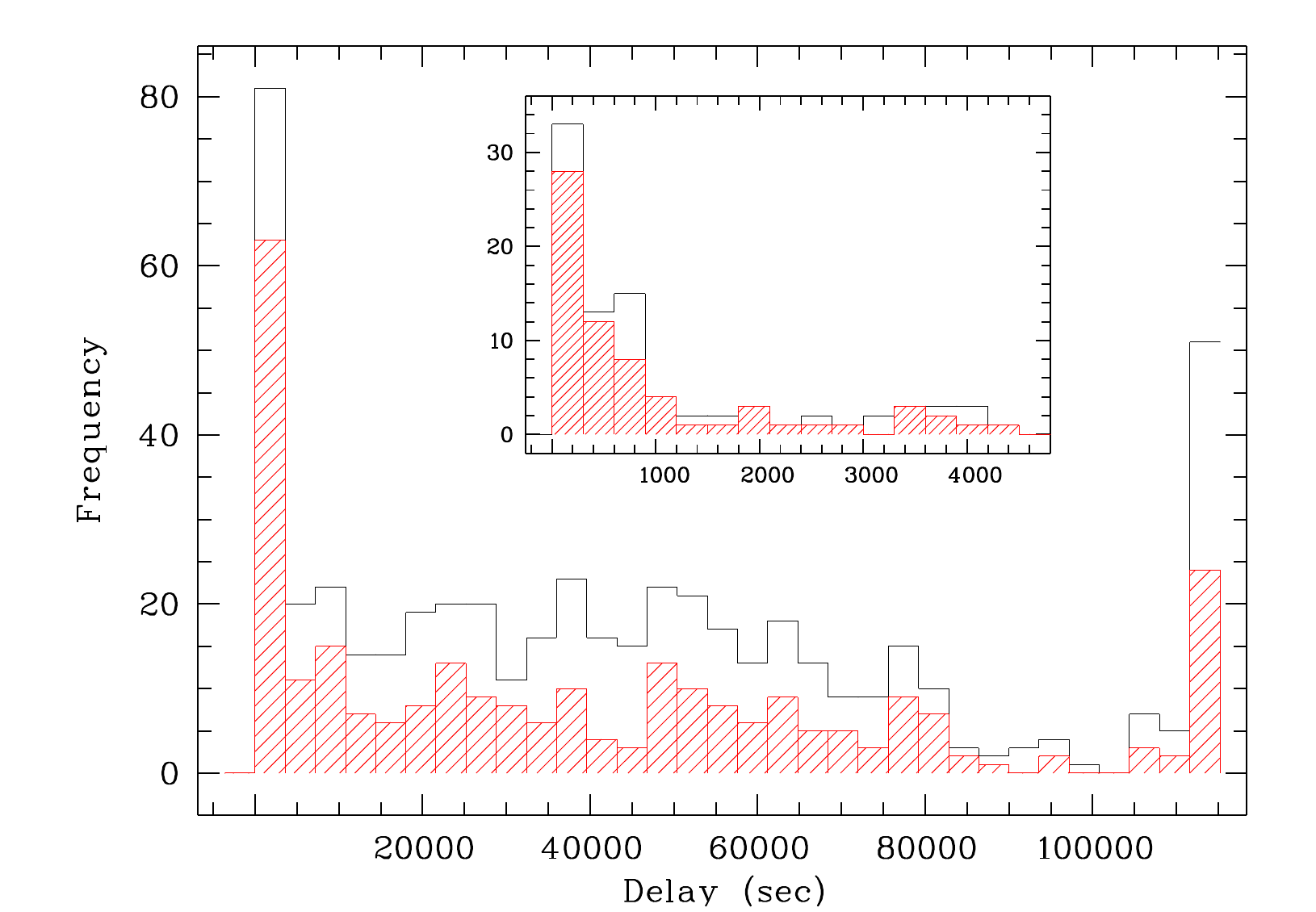}
\caption{Histogram of the time between GRB alert and start of the GROND
  observation in 1 hr bins, showing all 514 GRBs (black), and those with optical
  afterglow detection (red). The inset shows a zoom of the distribution
  during the first hour, in 5 min bins.
  The fastest reaction time was 78 s (for GRB 130514A).}
\label{grondGRBs_delay}
\end{figure}

The other important quantity is the detection rate of the optical/NIR
afterglows of GRBs, which
is detailed in Fig. \ref{grondGRBs_delay} and Tab. \ref{frac}.
Not surprisingly, the discovery fraction is significantly higher when
observations started during the first 30 min after a GRB.
However, compared to our first counting after 3 years of operation
\citep{gkk11}, this is down from 91\% to 81\%. This can partially
be explained with our increasingly more aggressive follow-up even at bad
observing conditions, which increased the likelihood of non-detections
due to lower sensitivity. However, the dominant factor for the overall
decreased afterglow detection fraction is likely just small number
statistics in the beginning \citep{gkk11}. Also not surprisingly, the
discovery fraction is substantially larger for long- than for
short-duration GRBs. This is readily explained by the generally brighter
afterglows of long GRBs.
Finally, we note that due to our use of NIR filter bands,
there is no detection bias wrt. extinction towards the Galactic plane,
i.e. the ratio of detected/non-detected afterglows is independent
of Galactic latitude (Fig. \ref{grondGRBs}).

\begin{table}[bh]
  \caption{GROND afterglow detection fraction,
    separately for the 465 long- and 49 short-duration GRBs,
     as a function of time delay of the start of the observation after the 
     GRB trigger, based on a total of 514 bursts.
     The relatively high detection rate at $>$24 hrs is
     biased by the Fermi/LAT bursts which are, on average, more energetic
     with correspondingly brighter afterglows. }
   \vspace{-0.2cm}
      \begin{tabular}{ccccc}
      \hline
      \noalign{\smallskip}
      $\!\!$Delay  & \multicolumn{2}{c}{detected vs. observed (\%)} & \multicolumn{2}{c}{fraction of total observed} \\
        (hrs)     &  long & short & long & short \\
      \noalign{\smallskip}
      \hline
      \noalign{\smallskip}
      $<$0.5       & 50/~62 (81\%)& 4/~7 (57\%)   & 13\% & 14\%\\
     $\!\!$0.5--4  & 38/~60 (63\%)& 4/~8 (50\%)   & 13\% & 16\%\\
       4--8        & 33/~61 (54\%)& 3/12 (25\%)   & 13\% & 24\%\\
       8--12       & 28/~61 (46\%)& 0/~5 ~\,(0\%) & 13\% & 10\%\\
     $\!\!$12--24  & 76/151 (50\%)& 4/14 (29\%)   & 32\% & 29\%\\
      $>$24        & 32/~70 (46\%)& 0/~3 ~\,(0\%) & 15\% & ~\,6\% \\
      \noalign{\smallskip}
      \hline
   \end{tabular}
   \label{frac}
\end{table}

Somewhat unexpected, however, is the fact that our
discovery fraction for long GRBs declines only slowly
over the next hours (Tab. \ref{frac}). Since our sensitivity (limiting
magnitude) does not change with  the delay time to the GRB,
this suggests that other effects play a role.
Partly, this is an observational bias:
initially bright afterglows can be detected for longer times than
faint ones, independent of their decay slope. But plateaus
(080129 \citep{gkm09}, 091127 \citep{Filgas+2011b}, 121217A \citep{Elliott14},
191016A \citep{Pereyra+2022}, 191221B \citep{Zhu+2024})
or even re-brightening episodes
(081029 \citep{ngk11}, 100621A \citep{gkn13}, 100814A \citep{Nardini+2014})
seem to play a larger than expected role at least during the first 24 hrs
of a GRB.
This is also suggested by our sample of the 60 best-covered
(with GROND) afterglow light curves, where about one third
have only decayed by about 2-3 magnitudes over the first 12 hours
(see future part IV of this publication series).

For later times, i.e. beyond 24 hrs after a GRB, we caution that
there is a technical bias: Fermi/LAT positions have
typically been reported after one day, delaying early observations.
Since on average those GRBs are more energetic
\citep{mkr10, Cenko+2011} with correspondingly brighter afterglows,
their detection rate certainly enhances our late-time detection fraction.

\begin{figure}[th]
\hspace{-0.4cm}\includegraphics[angle=270, width=9.7cm]{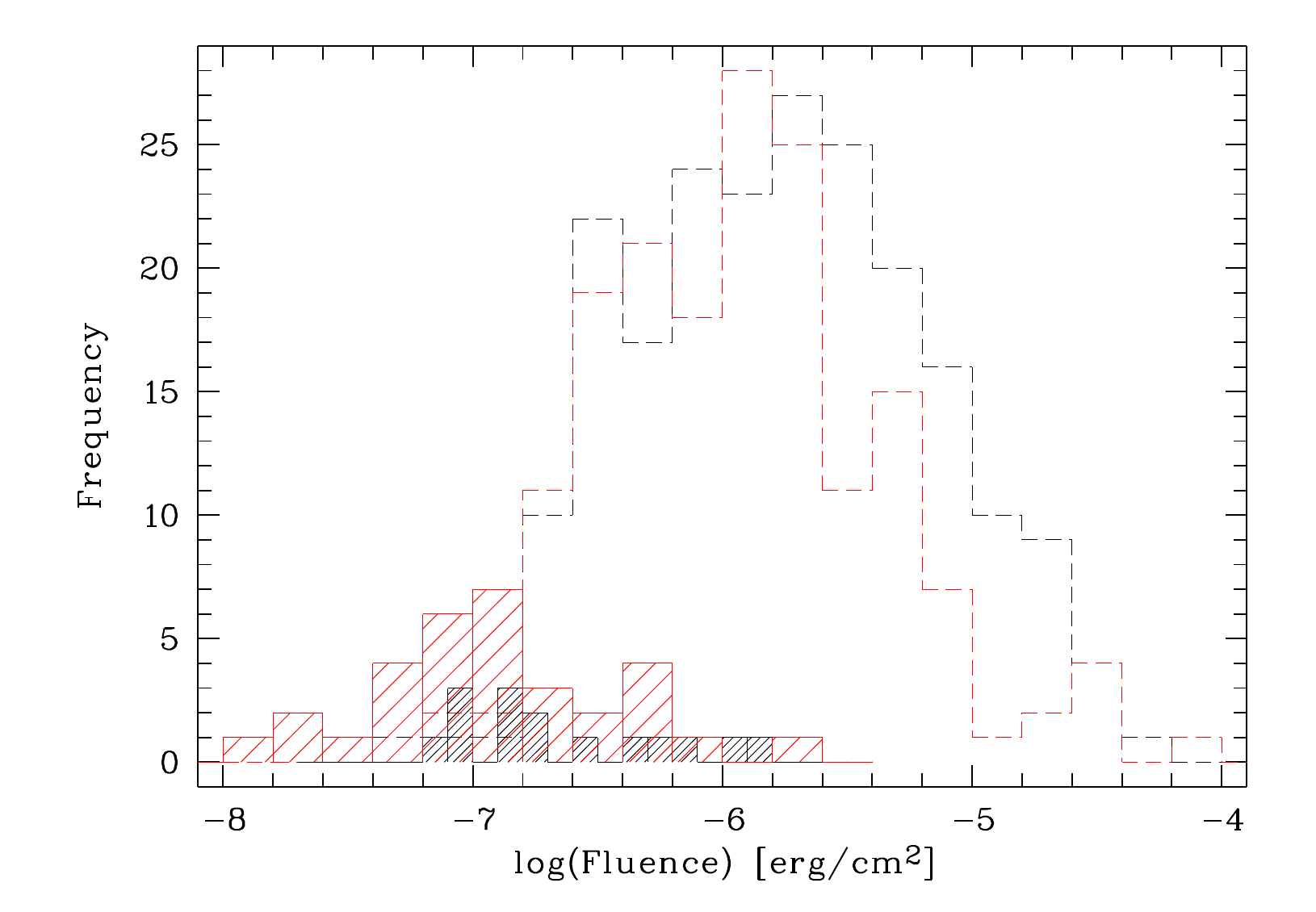}
\caption{Histogram of the 15--150 keV GRB fluences as measured with Swift/BAT
  for the detected (black) and non-detected (red) afterglows.
  For each category, we show 
  the long- (dashed lines) and short-duration (solid lines, hashed) 
  GRBs separately. There is only a weak tendency for bright GRBs
  to be better detectable than faint GRBs.
\label{BATfluence}}
\end{figure}

It might also be interesting to compare against the nominal UVOT
detection rate of 25\% \citep{Roming+2009}. For our sample of
62 GROND-observed (within 30 min) long-duration GRBs, we exclude
3 GRBs without and 6 GRBs with late Swift slews. From the remaining
53 GRBs, UVOT detected 22 afterglows, corresponding to a 41\% fraction,
while GROND detected 45 (85\%).
Out of the 23 GROND- but not UVOT-detected afterglows, 
9 have a redshift $>$3,
9 are below the typical UVOT sensitivity of $B \sim 22.0$ mag (incl. 4 with a redshift $<$3),
3 suffer from large (A$_V > 1$ mag) foreground absorption,
one was reported with a 2.5$\sigma$ detection at the GROND afterglow position,
and one should have been detectable with a different filter exposure
distribution. Thus, from this limited sample,
we derive the following fractions for the three reasons of
UVOT-non-detection (each with at least a 5\% error):
39\% due to high redshift, 13\% due to (foreground)
dust extinction, and 48\% due to simply not enough sensitivity.

A comparison with the gamma-ray fluence is made in Fig. \ref{BATfluence},
showing  histograms of the Swift/BAT 15-150 keV gamma-ray
energy fluence distribution of the GRB prompt emission \citep{Lien+2016},
computed for simple powerlaw fits, for detected and non-detected
afterglows, separately for long- and short-duration GRBs. The nearly
identical distributions demonstrate the well-known lack of correlation
between prompt emission and afterglow detection rate.

\begin{figure}[ht]
\centering
\includegraphics[width=7.2cm, viewport=128 60 350 280, clip]{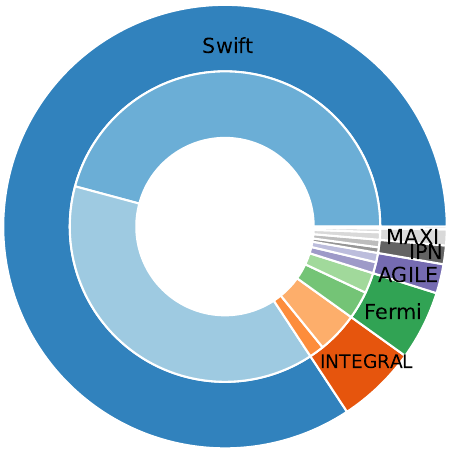}
\vspace{-0.3cm}
\caption{Pie diagram of the satellite-origin of the 514 GROND-observed GRBs:
  the outer circle represents the overall ratio of the missions providing
  the trigger, the inner circle depicts the ratio of  detected (darker colour)
  vs. non-detected (lighter colours) optical/NIR afterglows.
  The thin slice above
  MAXI is the optically identified iPTF14yb transient, which later was
  related to GRB 140226A based on a Mars/Odyssey, Konus-Wind and INTEGRAL data.
  The vast majority of
  GROND-observed are from Swift, and thus have XRT-positions and spectra.
  \label{grondpie}}
\vspace{-0.3cm}
\end{figure}

\begin{figure}[ht]
\centering
\includegraphics[width=7.9cm, viewport=26 5 270 210, clip]{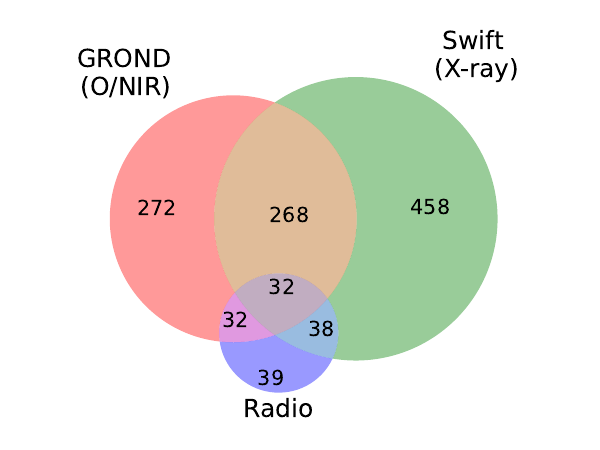}
\vspace{-0.3cm}
\caption{Venn diagram of the detected (!) afterglows
  out of the sample of 514 GROND-observed GRBs visible
  from Chile between 21 May 2007 and 1 Oct 2016, i.e. 458 of the 514
  GROND-observed GRBs have an immediate Swift/XRT detection, 272 a GROND
  detection, 268 a combined GROND and Swift/XRT detection, and so on.
  Radio detections are taken from \url{https://www.mpe.mpg.de/~jcg/grbgen.html}.
  \label{grondVenn}}
\vspace{-0.3cm}
\end{figure}

Figs. \ref{grondpie} and \ref{grondVenn} finally provide a
distribution of observed GRBs over satellite origin, and a visual comparison
of the detection rates in X-rays, optical/NIR and radio based on our
sample of 514 GRBs.

\begin{figure}[ht]
\vspace{-0.3cm}
\hspace{-0.7cm}\includegraphics[angle=270,width=10.5cm]{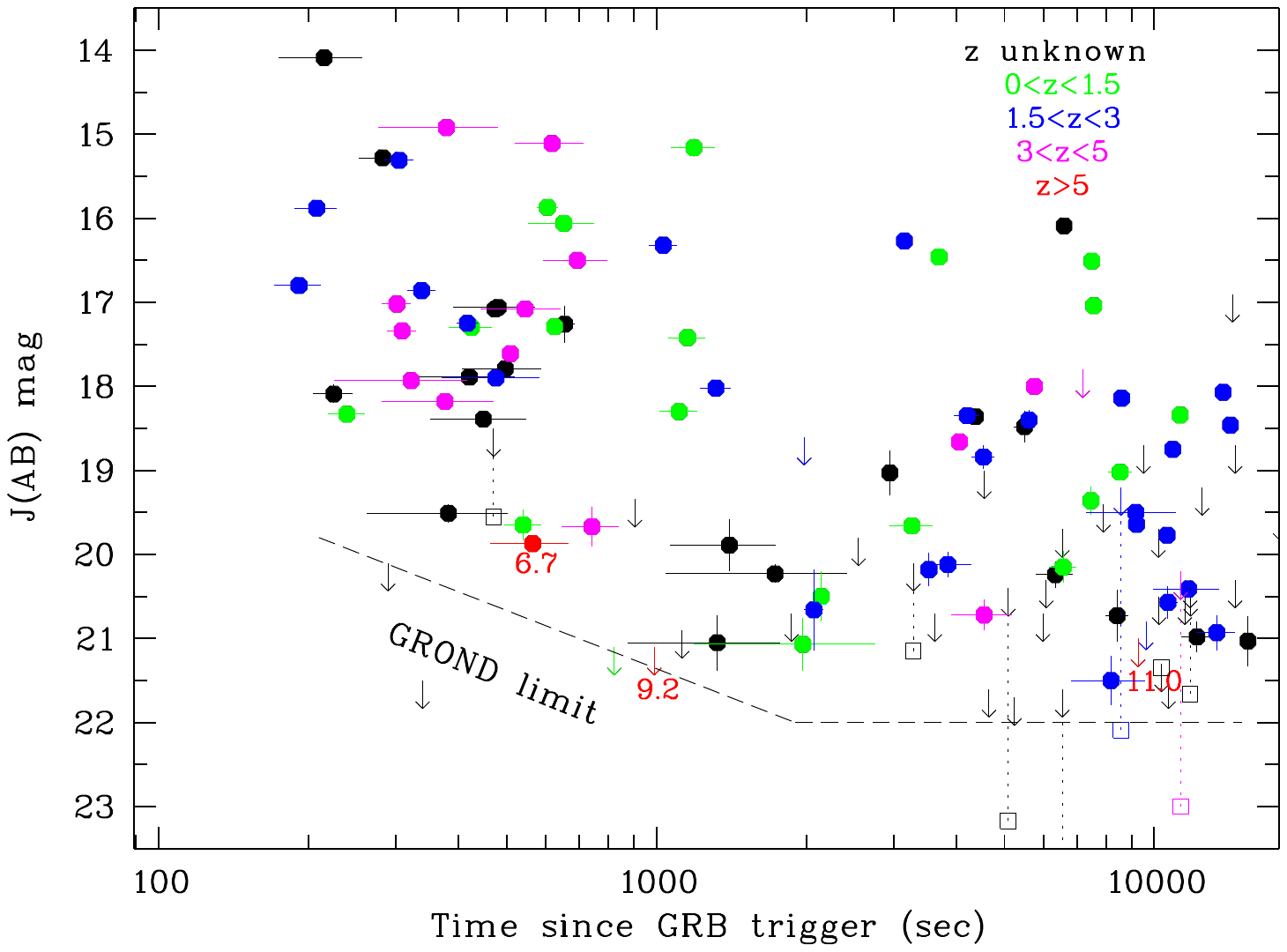}
\caption{Sample of 122 long GRB afterglows observed with GROND
within the first 4 hrs, coloured according to redshift.
For bright afterglows, these are the first 4-min exposures;
for fainter afterglows,
stacking of multiple exposures has been applied (as indicated by horizontal
'error' bars) and a compromise between
amount of stacking and time after the GRB has been adopted.
The sensitivity limit of GROND@2.2m is given as dashed line, assuming a start
of the exposure
at 150 s and an exposure of 60 s, and thereafter improving as time progresses:
at early times this is close to the typical $t^{-1}$ afterglow fading rate.
For those afterglows with $J$-band upper limits but \zp-detections, we plot
an open square at the expected $J$-band value
(using a mean colour of $z$-$J$ = 0.15 mag), and connecting this with the
downwards arrow with a dotted line (GRB 120328A at 6534 s has \zp = 23.79$\pm$0.35, outside of the plot).
The sky background limits our sensitivity to $J$(AB) \lax 22 mag, irrespective
of exposure duration.
\label{GRBsobserved}}
\end{figure}

\subsection{Peak brightness distribution and dominance of X-ray over optical flaring}

The brightness of the GRB afterglow at the first detection with GROND is
shown in Fig. \ref{GRBsobserved} for all observations starting within
4 hrs of the GRB trigger. We chose to show the $J({\rm AB})$ magnitudes
as compromise between minimizing the effect of dust extinction (A$_{\rm V}$)
wrt. the optical bands, and optimizing GROND's sensitivity.

A noteworthy aspect of this distribution is the difference in the dynamic
range of the
optical/NIR  vs. X-ray afterglows. Figure \ref{GRBsobserved} shows
the former to be of the order of a factor 600 (7$^{\rm m}$) during the
first 900 s. 
In contrast, for the sub-sample with Swift-XRT afterglows,
the difference between brightest and faintest X-ray emission is
a factor 10$^4$ (for this estimate, we have omitted the first 100 s
interval in order to not be biased by the prompt-to-afterglow
transition and the so-called tail emission).
This large diversity in the X-ray emission is consistent with earlier
  estimates \citep[e.g.][]{Berger+2003}.
Late-peaking optical forward
shock emission is rare beyond a few hundred seconds, and thus cannot
be the cause of this difference.
The missed afterglows are obviously fainter, but would have to
be at least 3 mag below our sensitivity line to make up the difference,
corresponding to $J$(AB) \gax\ 24--25, or \rp(AB) \gax\ 26--27 mag. 
Similarly unlikely are unaccounted for host-intrinsic absorption/extinction
effects in the detected optical/NIR afterglows:
while not being corrected for,
very substantial amounts of dust are needed, given our use of the
$J$-band fluxes with $A_{\rm J} = 0.29 A_{\rm V}$.
One possible explanation could be the prevalence of X-ray flaring
which is not accompanied by proportionally increased optical emission
\citep[e.g.][]{kgm09, Uehara+2010}.

\subsection{Low incidence of bright reverse shocks}

The brightness distribution of the early afterglow emission
(Fig. \ref{GRBsobserved}) is remarkable for another reason, namely
that very bright afterglows with $<$13 mag (AB) are very rare.
GRB 080319B, the first naked eye afterglow, is well-known in the community,
partly because it still is only one out of a handful afterglows brighter
than the above limit after the detection of about 900 optical
afterglows\footnote{\url{https://www.mpe.mpg.de/~jcg/grbgen.html}}.
This is consistent
with wide-field searches for afterglows in the 1990s
\citep{Greiner+1996, Hudec+1996}, and corroborated by the low discovery rate
of orphan afterglows in the wealth of present-day all-sky monitoring
observations.
Allowing a reverse shock  decline with $t^{-2}...t^{-3}$
\citep{KumarPanaitescu2000, NakarPiran2004}
during the first 1000\,s, and the canonical $t^{-1}$ at later times,
back-extrapolating the first readings of our 122 GRB afterglows
results in only 6 to rise above 13 mag (AB) at time $<$300 s after the GRB.
Such a 5\% rate of observable reverse shocks is already higher than
the about 20 suggested RS interpretations (out of 900 afterglows) in the
literature \citep[e.g.][]{Oganesyan+2021}, but consistent within the
statistics, and in line with theoretical considerations \citep{McMahon+2006}.

\subsection{Correlated optical and radio detection probability}

For a check of the relative detection probabilities in the optical
and radio bands we use the Swift/XRT sample. Among the 272 optically
detected afterglows, 32 (12\%) have a reported radio detection.
In contrast, only 6 (2.5\%) radio detections are reported for the 242
optically non-detected afterglows.
In their review of radio afterglow detections, \cite{ChandraFrail2012}
have shown (their Fig. 18) that for optically bright afterglows the
radio detection rate is substantially higher than for optically faint
afterglows. While our statistics is lower than that of \cite{ChandraFrail2012},
our finding is consistent with this correlation.

\subsection{The dark and high-z fraction}

The distribution of Fig. \ref{GRBsobserved} for the 2007--2010 GROND operation
phase had been used to estimate the fraction of dark bursts to 25--40\%
depending on definition, \citep{gkk11}. Since that sample had a afterglow
detection rate of 90\% and a redshift (z) completeness of 92\%, the fraction of
high-redshift GRBs could be separated off, revealing 5.5$\pm$2.8\%
of GRBs at $z>5$ \citep{gkk11}. The present sample, more than 3x larger,
unfortunately does not reach these completeness levels: the
detection rate is 71\% (87/122), and the redshift completeness is 56\%
(69/122). Thus, the dark and the high-z fraction cannot be improved.

\section{Conclusions}

The dedicated and systematic GRB follow-up program with GROND
at the 2.2\,m telescope has substantially increased the
detection rate of GRB afterglows,
by e.g. (i)  being first in reporting
the afterglow detection in 75 cases (only surpassed by Swift/UVOT with
92 afterglows, and counting), or (ii) by providing afterglow identifications
for 32 GRBs (corresponding to 12\% of all GROND-detected afterglows)
which no other group has detected.
  This is particularly true for fainter afterglows, beyond the reach
  of robotic telescopes (typically $<$1\,m), or dust-extinguished GRBs.  
The GROND program also helped
in identifying high-redshift GRBs.
While GROND's record-breaking (photometric) redshift GRB 080913
  \citep{gkf09}
  lasted for only a few months (when the z=8.2 GRB 090423 took over),
  the number of identified $3<z<6$ redshifts, easily seen with GROND as
\gp- or \rp-drop-outs, has substantially decreased since 2017.
GROND's 7-channels were very useful for photometric redshift estimates,
allowing to select particularly interesting GRBs for spectroscopic
follow-up.
The lack of a filter bluer than 400 nm has limited
the frequent  application to redshifts \gax3,
though with contemporaneous Swift/UVOT observations \citep{ksg11}
this method could be extended down to $z \approx 1$,
thus covering the peak of the GRB redshift distribution.

The multi-channel coverage of GROND results in a SED from 400-2400 nm
  which is providing additional benefits for various applications,
  e.g. in identifying the afterglow among many candidates due to the powerlaw
  shape of the SED, measuring of dust extinction (local as well as
  host-intrinsic), increased confidence in low-significance objects
  by requiring detections in more than one spectral band. More details
  of these benefits, examples and non-GRB applications are summarised
  in \cite{jcg19}.
  
In terms of detection fraction, a 2\,m class
telescope on the ground is still not sufficient to identify
all afterglows, even in the NIR bands, as we still miss the
$\approx$20\% faintest ones.
For a higher rate of follow-up of GRBs detected from low-Earth orbit,
an observing site away from the longitudes of the South-Atlantic Anomaly
would be beneficial. Alternatively, a NIR multi-channel camera on a
small satellite can substantially increase the detection rate: with
a 50\,cm diametre telescope, a detection fraction close to 100\%
in the J or H band is feasible \citep{Thomas+2022, GreinerLaux2022}.

\begin{acknowledgement}

SK acknowledges support by DFG grants Kl 766/11-3, 13-1 and 13-2.
Part of the funding for GROND (both hardware as well as personnel)
was generously granted from the Leibniz-Prize to Prof. G. Hasinger
(DFG grant HA 1850/28-1).
This work made use of data supplied by the UK {\it Swift} Science Data Centre 
at the University of Leicester.
\end{acknowledgement}

\bigskip

\noindent {\small {\it Facilities:} Max Planck:2.2m (GROND), 
                  Swift}

\clearpage

\xentrystretch{0.001}
\begingroup\small
\topcaption{All 517 (514 plus the 3 with large error boxes; see sect. 3)
  GROND-observed GRBs. The ``L/S'' in the 2nd column denotes
  the long/short duration classification, and ``y/n'' in the 5th column refers to the afterglow detection with GROND.
  \vspace{-0.2cm}
\label{allgrbs}}
\tablefirsthead{%
\toprule
    GRB    & $\!\!$L/S$\!\!$ & GROND-Start & $\Delta$t~~~ & $\!\!$Det.$\!\!$ & References \\
\midrule}
\tablehead{%
\multicolumn{4}{@{}l}{Table \thetable, continued}\\
\midrule[\heavyrulewidth]
    GRB    & $\!\!$L/S$\!\!$ & GROND-Start & $\Delta$t~~~ & $\!\!$Det.$\!\!$ & References \\
\midrule}
\tabletail{\bottomrule}
\tablelasttail{\bottomrule}
\begin{xtabular}{@{}lccrcl@{}}
 070521  & L & 0521-07:01:45 &     634 & n &  6449     \\
 070621  & L & 0622-04:44:08 &   19588 & n &           \\ 
 070628  & L & 0628-22:21:57 &   27654 & y & 6590     \\ 
 070729  & S & 0729-06:37:27 &   22294 & n & (1) \\ 
 070802  & L & 0802-07:23:20 &     955 & y & 6694, (2) (3)     \\ 
 070917  & L & 0917-23:49:24 &   58528 & y &           \\ 
 071010  & L & 1010-03:57:36 &     984 & y & (3, 4)  \\ 
 071021  & L & 1022-00:03:34 &   51721 & n &           \\ 
 071025  & L & 1026-01:50:14 &   78080 & y & (5, 41)  \\ 
 071028B & L & 1030-00:23:51 &  164405 & y & (6) \\ 
 071031  & L & 1031-01:10:21 &     225 & y & 7021, (3, 7)  \\ 
 071112B & S & 1113-00:11:25 &   20874 & n & (1)     \\ 
 071112C & L & 1113-02:13:26 &   27629 & y & (8) \\ 
 071117  & L & 1118-00:15:09 &   33903 & y &           \\ 
 071227  & S & 1228-00:20:04 &   14777 & n & (1, 9)   \\ 
 080120  & L & 0121-07:04:02 &   48932 & y &  7197     \\ 
 080129  & L & 0130-06:10:18 &     195 & y &  7231, (3, 10) \\ 
 080130  & L & 0131-08:11:30 &   75451 & n &           \\ 
 080207  & L & 0208-06:35:44 &   32723 & n &  7279, (11)  \\ 
 080210  & L & 0210-09:02:10 &    4325 & y &  7283, (3)     \\ 
 080212  & L & 0213-05:32:39 &   43086 & y &  7303     \\ 
 080218B & L & 0219-00:38:14 &    2427 & n &  7319, (3, 11)  \\ 
 080319B & L & 0325-06:30:41 &  519472 & n &           \\ 
 080319D & L & 0320-00:15:17 &   25808 & y &  7503     \\ 
 080325  & L & 0325-09:50:02 &   20445 & n &  7520     \\ 
 080330  & L & 0330-03:44:23 &     187 & y &  7545, (3, 12)   \\ 
 080405  & L & 0406-23:27:43 &  137327 & n &           \\ 
 080408  & L & 0408-23:25:13 &   18745 & y &  7581     \\ 
 080411  & L & 0411-23:18:13 &    7361 & y &  7586, (3)     \\ 
 080413A & L & 0413-05:14:19 &    8400 & y &  (3)    \\ 
 080413B & L & 0413-08:56:36 &     324 & y &  7599, (3, 13) \\ 
 080414  & L & 0415-04:36:28 &   21773 & n &           \\ 
 080514B & L & 0515-07:04:36 &   76120 & y &  7722, (14) \\ 
 080516  & L & 0516-00:25:32 &     505 & y &  7740, 7741, 7747, (3)     \\ 
 080520  & L & 0521-01:21:12 &   10848 & y &  7756, (3)     \\ 
 080523  & L & 0524-07:24:58 &   36187 & y &           \\ 
 080604  & L & 0605-00:29:23 &   61342 & y &           \\ 
 080605  & L & 0606-01:10:56 &    4979 & y &  7851, 7834 (15)  \\ 
 080613  & L & 0614-00:26:44 &   53483 & n &  7880     \\ 
 080613B & L & 0613-22:49:01 &   41784 & n &  7879     \\ 
 080623  & L & 0623-22:50:20 &   44692 & n &  7902     \\ 
 080702B & L & 0703-08:25:41 &  112500 & n &  7943     \\ 
 080703  & L & 0703-23:04:28 &   14655 & y &  7944     \\ 
 080707  & L & 0707-09:59:54 &    5521 & y &  7948, (3)     \\ 
 080710  & L & 0710-07:19:29 &     379 & y & (3, 16)     \\ 
 080714  & L & 0714-22:49:54 &   17818 & y &  7984     \\ 
 080723  & L & 0724-02:46:25 &   80821 & n &         \\ 
 080723B & L & 0724-23:47:35 &  123916 & n &  8079     \\ 
 080727B & L & 0727-23:08:12 &   53688 & y &  8045     \\ 
 080802  & L & 0802-23:08:00 &   28539 & n &           \\ 
 080804  & L & 0804-23:38:59 &    1125 & y &  8075, (3)     \\ 
 080805  & L & 0805-07:45:35 &     241 & y &  8060, (3, 15) \\ 
 080810  & L & 0811-03:07:42 &   50250 & n &           \\ 
 080822B & L & 0823-08:05:53 &   39781 & n &           \\ 
 080825B & L & 0826-00:24:55 &   23895 & y & (17, 42)    \\ 
 080828  & L & 0828-23:47:47 &   73982 & n &           \\ 
 080905  & S & 0906-05:22:14 &   62600 & n & (1)     \\ 
 080905B & L & 0906-07:45:06 &   53361 & n &           \\ 
 080906  & L & 0907-00:14:06 &   38450 & y &  8194, (17)  \\ 
 080913  & L & 0913-06:51:11 &     257 & y &  8223, 8218, (3, 18, 42) \\
 080915  & L & 0915-00:07:38 &     289 & n &  8268, (3, 11)    \\
 080915B & L & 0915-23:31:17 &   27462 & n &           \\
 080916  & L & 0917-00:01:14 &   51354 & y &  8266  \\
 080916C & L & 0917-07:57:08 &  114263 & y &  8257, 8272, (19, 42)\\
 080919  & S & 0919-00:08:30 &     197 & n &  (1, 3)     \\
 080922  & L & 0922-23:33:53 &   44996 & n &           \\
 080928  & L & 0929-04:23:45 &   48133 & y &  8296, (20) \\
 081003B & L & 1004-00:32:31 &   13455 & n &           \\
 081007  & L & 1007-05:38:21 &     869 & y &  (3, 21, 22) \\
 081008  & L & 1008-23:44:01 &   13552 & y &  (3, 23)   \\
 081012  & L & 1013-07:43:17 &   66774 & n &  8373, (10)   \\
 081016  & L & 1016-23:49:24 &   61042 & n &           \\
 081016B & L & 1017-00:46:52 &   17978 & n &           \\
 081028  & L & 1028-06:19:21 &   21261 & y &  8424     \\
 081029  & L & 1029-01:50:06 &     370 & y &  8437, (3, 24)  \\
 081104  & L & 1105-07:42:50 &   79688 & n &  8482     \\
 081105  & L & 1106-02:01:27 &   45315 & n &  8492, (11)  \\
 081109  & L & 1110-00:07:57 &   61551 & y &  8510, (25)  \\
 081118  & L & 1119-01:07:09 &   36633 & y &  8529     \\ 
 081121  & L & 1122-00:29:23 &   14031 & y &  8540, (3)     \\ 
 081127  & L & 1128-00:25:57 &   62449 & y &           \\ 
 081204  & L & 1205-00:32:09 &   28034 & n &  8627, (11)  \\ 
 081221  & L & 1222-01:07:42 &   31591 & y &  8719     \\ 
 081222  & L & 1222-05:08:56 &     896 & y &  8693, (3)     \\ 
 081226  & S & 1226-01:14:50 &     673 & y &  8731, (1, 3) \\ 
 081226B & S & 1227-01:30:13 &   47822 & n &  (1)    \\ 
 081228  & L & 1228-01:24:51 &     431 & y &  8745, (3, 17) \\ 
 081230  & L & 1231-00:48:16 &   15124 & y &  8760, (17) \\ 
 090102  & L & 0102-05:25:55 &    9010 & y &  8771, (3, 26) \\ 
 090113  & L & 0114-00:37:14 &   21395 & n &  8812     \\ 
 090117  & L & 0118-01:40:23 &   37109 & n &  8820     \\ 
 090118  & L & 0119-00:55:27 &   39685 & n &  8826     \\ 
 090123  & L & 0124-00:56:22 &   61466 & y &  8849     \\ 
 090129  & L & 0130-08:51:52 &   42277 & n &           \\ 
 090201  & L & 0202-00:22:17 &   23715 & n &           \\ 
 090205  & L & 0206-05:26:40 &   23006 & y &  8888, (42)     \\ 
 090305  & S & 0305-05:47:50 &    1679 & y &  (1, 3)     \\ 
 090306B & L & 0308-04:34:30 &  106048 & n &  8958     \\ 
 090307  & L & 0307-04:54:15 &    4058 & n &  8959     \\ 
 090308  & L & 0309-00:01:20 &   21597 & n &  8966     \\ 
 090309  & L & 0310-07:22:43 &   28410 & n &  8975     \\ 
 090313  & L & 0313-09:13:25 &     418 & y &  8983, (3)     \\ 
 090323  & L & 0324-02:50:31 &   96469 & y &  9026, (27, 28)\\
 090324  & L & 0326-09:39:24 &  197474 & n &           \\ 
 090328  & L & 0328-23:37:16 &   50430 & y &  9054, (28)  \\ 
 090401B & L & 0402-00:32:27 &   57422 & y &           \\ 
 090404  & L & 0405-06:18:48 &   51738 & n &  9096     \\ 
 090407  & L & 0407-23:33:59 &   47134 & n &  9109     \\ 
 090410  & L & 0411-10:17:20 &   62368 & n &  9128     \\ 
 090418B & L & 0419-03:02:39 &   65017 & n &  9178     \\ 
 090418  & L & 0419-09:17:40 &   79800 & y &           \\ 
 090419  & L & 0419-23:13:58 &   34227 & y &  9169     \\ 
 090423  & L & 0423-23:09:03 &   54824 & y &  9215, (29, 42) \\ 
 090424  & L & 0425-02:52:33 &   45624 & y &  9245     \\ 
 090426  & S & 0427-01:08:11 &   44364 & y &  9268, (30)  \\ 
 090427  & L & 0429-06:22:09 &  111342 & n &  9319     \\ 
 090429  & L & 0430-01:06:15 &   72756 & n &  9303,  \\ 
 090429B & L & 0429-05:41:32 &     689 & n &  9283, (3, 31, 42) \\ 
 090509  & L & 0509-05:14:19 &     256 & y &  9326     \\ 
 090510  & S & 0510-06:33:49 &   22249 & y &  9352, (28, 32) \\ 
 090516  & L & 0516-23:02:14 &   52464 & y &  9382, (42) \\ 
 090518  & L & 0518-23:00:21 &   75937 & n &  9395     \\ 
 090519  & L & 0519-22:53:06 &    6250 & y &  9408, (3)     \\ 
 090520  & L & 0521-09:35:44 &  115108 & n &  9420     \\ 
 090529  & L & 0531-03:32:34 &  134399 & y &           \\ 
 090530  & L & 0531-00:35:56 &   76658 & y &  9458, (17)   \\ 
 090531  & L & 0531-01:47:22 &     125 & n &  9456     \\ 
 090531B & S & 0601-00:03:21 &   19645 & n &  9480     \\
 090625B & L & 0626-03:58:48 &   52346 & n &  9578     \\ 
 090628  & L & 0629-02:10:30 &   17418 & n &  9591     \\
 090809  & L & 0810-01:55:25 &   30251 & y &           \\
 090812  & L & 0812-08:13:22 &    7856 & y &  9773, (3)     \\
 090814  & L & 0814-00:56:53 &     274 & y &  9794, (3)     \\ 
 090827  & L & 0829-01:37:11 &  109845 & n &           \\ 
 090831C & L & 0901-08:10:15 &   38390 & n &  9862     \\ 
 090902B & L & 0903-00:22:00 &   47812 & y &  9874, (28)  \\ 
 090904B & L & 0904-01:28:20 &     242 & y &  9901, (3)     \\ 
 090915  & L & 0917-00:19:47 &  117851 & n &  9923     \\ 
 090926  & L & 0926-23:39:11 &   69524 & y &        (33)  \\ 
 090926B & L & 0927-01:54:24 &   14316 & y &           \\ 
 090927  & L & 0928-03:01:00 &   60824 & y &  (1)     \\ 
 090929  & L & 0930-03:27:32 &   82464 & n &           \\ 
 090929B & L & 0930-07:40:53 &   77506 & y &           \\ 
 091010  & L & 1010-23:45:12 &   75723 & y & 10008     \\ 
 091015  & L & 1017-01:12:57 &   94360 & n &           \\ 
 091018  & L & 1018-23:49:55 &   10896 & y & 10039, (3, 34) \\ 
 091026  & L & 1026-23:59:45 &   38895 & n & 10091     \\ 
 091029  & L & 1029-03:57:42 &     320 & y & 10098, (3, 35) \\ 
 091102  & L & 1103-00:02:28 &   34070 & n & 10123     \\ 
 091109  & L & 1110-00:08:26 &   69043 & y & 10158     \\ 
 091109B & S & 1110-03:31:26 &   20543 & n & (1)     \\ 
 091117  & S & 1119-00:46:29 &  111720 & n & (1)      \\ 
 091127  & L & 1128-00:21:59 &    3374 & y & 10195, (3, 36)  \\ 
 091208B & L & 1209-00:51:58 &   54121 & y & 10271     \\ 
 091221  & L & 1222-00:49:32 &   14200 & y & 10286, (3)     \\ 
 091230  & L & 1230-06:30:27 &     177 & y & 10299     \\ 
 100103A & L & 0104-00:48:20 &   25548 & n & 10314     \\ 
 100115A & L & 0116-00:47:36 &   48737 & y &           \\ 
 100117A & S & 0118-00:46:23 &   13204 & y & (1, 3)    \\ 
 100203A & L & 0205-02:20:46 &  114579 & n &           \\ 
 100205A & L & 0205-06:37:58 &    8355 & n & 10383, (3)     \\ 
 100206A & S & 0207-00:30:34 &   39629 & n & 10396, (1)    \\ 
 100219A & L & 0220-00:30:34 &   33288 & y & 10439, (37, 42) \\ 
 100225A$^1$$\!$ & L & 0226-08:40:35 &  107704 & n &           \\ 
 100316B & L & 0316-08:16:06 &     870 & y & 10493, (3)     \\ 
 100316C & L & 0317-00:44:45 &   56806 & n & 10516     \\ 
 100316D & L & 0317-00:27:33 &   42163 & y & 10514, (38) \\ 
 100324A & L & 0324-23:41:20 &   83993 & n & 10545     \\ 
 100331B & L & 0401-09:29:10 &   44432 & n & 10572     \\ 
 100401A & L & 0403-09:21:00 &  180808 & n & 10577     \\ 
 100413A & L & 0414-05:43:07 &   43779 & n & 10592     \\ 
 100414A & L & 0417-03:12:09 &  262307 & y & 10607     \\ 
 100418A & L & 0419-04:50:48 &   27640 & y & 10617, 10616  \\ 
 100423A & L & 0423-00:37:24 &     145 & y & 10652     \\ 
 100424A & L & 0425-00:16:12 &   27810 & n & 10676     \\ 
 100425A & L & 0425-05:17:44 &    8819 & y & 10683     \\ 
 100504A & L & 0505-01:30:03 &   22144 & n & 10717     \\ 
 100508A & L & 0508-22:48:17 &   48455 & y & 10734     \\ 
 100518A & L & 0519-04:38:27 &   61492 & y & 10782, (42)  \\ 
 100522A & L & 0523-09:18:02 &  106330 & n & 10793     \\ 
 100526A & L & 0527-06:00:07 &   48837 & n &           \\ 
 100526B & L & 0527-09:37:55 &   52637 & n & 10812     \\ 
 100528A & L & 0530-07:50:34 &  194549 & y & 10815     \\ 
 100606A & L & 0607-04:30:04 &   33443 & n & 10835     \\ 
 100615A & L & 0615-02:05:29 &     386 & n & 10844     \\ 
 100621A & L & 0621-03:06:52 &     200 & y & 10874, (39)  \\ 
 100625A & S & 0626-06:08:58 &   41790 & n & 10906, (1)    \\ 
 100628A & S & 0629-00:46:14 &   59374 & n & 10910, (40) \\ 
 100702A & S & 0702-04:28:18 &   12271 & n &        (1)     \\ 
 100704A & L & 0704-23:24:05 &   71317 & n &           \\ 
 100707A & L & 0708-07:57:54 &  112276 & n &           \\ 
 100713A & L & 0713-23:49:18 &   33192 & n & 10956     \\ 
 100724A & L & 0724-00:45:18 &     179 & y & 10969     \\ 
 100725A & L & 0725-23:04:46 &   57114 & n & 10991     \\ 
 100728A & L & 0728-09:26:54 &   25710 & y & 11020     \\ 
 100728B & L & 0728-10:34:26 &     155 & y & 11013     \\ 
 100814A & L & 0814-03:57:25 &     434 & y & 11091, (41)\\ 
 100816A & L & 0818-06:33:46 &  194155 & n &           \\ 
 100823A & L & 0824-03:56:37 &   37862 & n &           \\ 
 100901A & L & 0904-06:25:58 &  233508 & y &           \\ 
 100902A & L & 0904-08:10:24 &  131910 & y & 11209     \\ 
 100905A & L & 0906-04:25:54 &   47860 & y & (42)  \\ 
 100915B & L & 0915-08:06:07 &    8188 & n &           \\ 
 100917A & L & 0917-23:32:02 &   66517 & n &           \\ 
 100924A & L & 0925-00:31:42 &   74014 & y & 11297     \\ 
 100928A & L & 0928-23:35:52 &   76560 & n &           \\ 
 101008A & L & 1009-00:27:47 &   27872 & y & 11326     \\ 
 101011A & L & 1012-00:14:08 &   26133 & n & 11337     \\ 
 101017A & L & 1018-00:13:08 &   49221 & y &           \\ 
 101023A & L & 1023-23:52:56 &    3764 & y & 11369     \\ 
 101024A & L & 1026-07:41:56 &  158547 & y & 11383     \\ 
 101030A & L & 1031-08:19:12 &   62563 & y & 11391     \\ 
 101114A & L & 1116-00:36:55 &  173045 & n & 11409     \\ 
 101117B & L & 1118-05:15:08 &   36105 & n & 11421     \\ 
 101129A & S & 1130-06:20:01 &   52829 & n &      (1)     \\ 
 101201A & L & 1202-00:29:09 &   52041 & y & 11430     \\ 
 101204A & L & 1207-05:58:04 &  244576 & y & 11445     \\ 
 101219A & S & 1219-03:53:20 &    4911 & n & 11471, (1)    \\ 
 101219B & L & 1220-00:30:32 &   28959 & y & 11478     \\ 
 110128A & L & 0129-07:05:49 &  105676 & y &           \\ 
 110206A & L & 0207-00:32:52 &   23087 & y & 11665     \\ 
 110207A & L & 0208-00:33:10 &   47750 & n &           \\ 
 110213B & L & 0215-00:26:09 &  122061 & y & 11743     \\ 
 110223B & L & 0224-00:03:26 &    9458 & y & 11756     \\ 
 110305A & L & 0305-06:40:43 &     162 & y & 11774     \\ 
 110312A & L & 0312-23:56:55 &   21678 & n & 11785     \\ 
 110318B & L & 0319-07:40:51 &   58349 & n & 11809     \\ 
 110319B & L & 0320-08:24:33 &   46231 & n & 11815     \\ 
 110407A & L & 0408-00:48:43 &   38522 & y & 11906     \\ 
 110414A & L & 0414-23:17:42 &   56127 & n & 11939     \\ 
 110420A & L & 0421-09:33:36 &   81072 & y & 11954     \\ 
 110420B & S & 0421-06:36:08 &   28437 & n & 11949     \\ 
 110426A$^1$$\!$ & L & 0427-00:54:26 &   35150 & n &     \\ 
 110519A & L & 0520-04:09:27 &   93431 & n &           \\ 
 110625A & L & 0626-02:25:48 &   19040 & n & 12096     \\ 
 110709B & L & 0709-22:33:16 &    3637 & n & 12129, (43) \\ 
 110715A & L & 0718-00:35:11 &  213589 & y & 12169, (44) \\ 
 110719A & L & 0720-09:01:07 &   96716 & n &           \\ 
 110721A & L & 0722-06:33:35 &   92752 & n & 12192     \\ 
 110731A & L & 0803-04:07:10 &  233860 & y &       (45)   \\ 
 110818A & L & 0820-03:39:52 &  113763 & n &           \\ 
 110825A & L & 0829-07:50:03 &  364992 & n & 12403     \\ 
 110915A & L & 0915-23:27:20 &   36396 & n & 12353     \\ 
 110915B & L & 0916-06:07:59 &   42220 & n & 12356     \\ 
 110918A & L & 0920-02:34:45 &  104868 & y & 12366, (46) \\ 
 110921A & L & 0921-23:41:57 &   35437 & n &           \\ 
 111005A & L & 1005-23:24:57 &   55183 & n & 12417, (47) \\ 
 111008A & L & 1009-04:35:54 &   22976 & y & 12428, (42) \\ 
 111018A & L & 1018-23:49:34 &   22990 & n & 12454     \\ 
 111020A & S & 1020-23:50:15 &   62186 & n &           \\ 
 111022A & L & 1023-00:00:56 &   28432 & n &           \\ 
 111026A & L & 1026-23:56:43 &   61754 & n &           \\ 
 111107A & L & 1107-03:35:02 &    9878 & y & 12536     \\ 
 111109A & L & 1109-04:03:37 &    3951 & n &           \\ 
 111117A & S & 1118-02:04:43 &   49862 & y & 12568, (48) \\ 
 111121A & S & 1122-06:53:52 &   52648 & n &           \\ 
 111123A & L & 1124-06:34:34 &   44473 & n & 12595     \\ 
 111129A & L & 1130-00:27:14 &   29340 & y & 12605     \\ 
 111204A & L & 1205-00:37:37 &   39609 & n & 12622     \\ 
 111205A & L & 1209-07:19:05 &  324495 & n & 12636     \\ 
 111207A & L & 1209-05:17:26 &  140427 & n &           \\ 
 111209A & L & 1210-00:54:57 &   63769 & y & 12647, (49, 50) \\ 
 111210A & L & 1211-08:06:34 &   62971 & n & 12662     \\ 
 111211A & L & 1212-07:45:54 &   34101 & y & 12668     \\ 
 111212A & L & 1213-00:36:17 &   54790 & n & 12679     \\ 
 111228A & L & 1229-04:55:21 &   47438 & y & 12757, (8)  \\ 
 111229A & L & 1230-00:47:05 &    7753 & y & 12775     \\ 
 120118A & L & 0118-06:07:26 &     162 & n &           \\ 
 120118B & L & 0119-01:02:42 &   28941 & n &           \\ 
 120119A & L & 0119-04:07:58 &     208 & y & 12863     \\ 
 120202A & L & 0203-06:32:12 &   31915 & y & 12913     \\ 
 120211A & L & 0212-00:29:06 &   45038 & n & 12931     \\ 
 120212A & L & 0214-00:39:04 &  142062 & y &           \\ 
 120215A & L & 0217-00:23:24 &  171744 & y &           \\ 
 120224A & L & 0225-00:14:08 &   70452 & n & 12988     \\ 
 120229A & S & 0229-23:59:15 &   33844 & n &           \\ 
 120302A & L & 0303-00:33:42 &   81492 & y & 13003     \\ 
 120311A & L & 0311-08:04:51 &    9073 & y & 13048     \\ 
 120311B & L & 0312-05:51:27 &   52997 & n &           \\ 
 120312A & L & 0313-09:08:43 &   61335 & n &           \\ 
 120320A & L & 0321-06:40:43 &   67468 & n &           \\ 
 120324A & L & 0324-09:06:40 &   11249 & n & 13098     \\ 
 120327A & L & 0327-03:34:22 &    2346 & y & 13129, (51)  \\ 
 120328A & L & 0328-05:40:42 &    8486 & y? &           \\ 
 120401A & L & 0401-23:32:07 &   65272 & y & 13219     \\ 
 120403B & L & 0405-00:15:38 &   99702 & n &           \\ 
 120404A & L & 0405-07:01:16 &   65414 & y & 13229, (52) \\ 
 120419A & L & 0419-23:22:41 &   37576 & n &           \\ 
 120422A & L & 0422-23:37:30 &   59127 & y & 13256, (53) \\ 
 120514A & L & 0514-04:02:50 &   10201 & n & 13298     \\ 
 120521A & S & 0521-22:55:02 &   60920 & n & 13335     \\ 
 120521B & L & 0522-00:36:36 &   55728 & n & 13328     \\ 
 120522A & L & 0523-23:12:26 &  158479 & n &           \\
 120612A & L & 0613-22:51:25 &  161166 & n &           \\ 
 120624B & L & 0625-23:39:21 &   90926 & n & 13393     \\ 
 120701A & L & 0701-09:11:30 &    4848 & y & 13409     \\ 
 120703A & L & 0704-09:10:31 &   56709 & y &           \\ 
 120711A & L & 0711-08:34:09 &   20961 & y & 13438, (54) \\ 
 120712A & L & 0712-22:57:14 &   33287 & y & 13457, (42) \\ 
 120714A & L & 0714-23:04:47 &   55081 & n & 13479     \\ 
 120714B & L & 0715-03:05:38 &   20812 & y & 13478 (8) \\ 
 120716A & L & 0719-05:07:07 &  216123 & y & 13492     \\ 
 120722A & L & 0722-23:20:33 &   37627 & y & 13506     \\ 
 120724A & L & 0725-03:01:49 &   73367 & n &           \\ 
 120728A & L & 0728-23:05:38 &    2427 & y & 13526, 13542     \\ 
 120804A & S & 0804-02:22:41 &    5307 & n & 13579     \\ 
 120805A & L & 0805-23:09:11 &    6062 & n &           \\ 
 120807A & L & 0807-23:09:49 &   57612 & n &           \\ 
 120815A & L & 0815-02:16:11 &     133 & y & 13648, (55) \\ 
 120817A & L & 0818-04:13:26 &   77024 & n &           \\ 
 120819A & L & 0819-23:46:02 &   38148 & y & 13688     \\ 
 120821A & L & 0821-23:47:25 &   37420 & n & 13697     \\ 
 120907A & L & 0908-07:01:09 &  110206 & n &           \\ 
 120909A & L & 0909-01:44:39 &     156 & y & 13729     \\ 
 120919A & L & 0921-23:35:14 &  231032 & n &          \\ 
 120922A & L & 0922-23:30:00 &    3572 & y & 13795     \\ 
 120923A & L & 0924-00:10:00 &   67894 & n & 13811, (42, 56) \\ 
 120927A & L & 0928-09:04:59 &   37453 & n & 13827     \\ 
 121001A & L & 1001-23:37:05 &   18843 & y & 13840     \\ 
 121014A & L & 1015-09:00:13 &   46097 & n &           \\ 
 121017A & L & 1017-23:49:53 &   15985 & n & 13881     \\ 
 121024A & L & 1024-05:53:40 &   10648 & y & 13891     \\ 
 121027A & L & 1029-08:09:44 &  178635 & y & 13926     \\
 121028A & L & 1028-23:57:39 &   67988 & n &           \\
 121102A & L & 1103-00:01:54 &   77692 & n &           \\ 
 121117A & L & 1118-00:15:43 &   55487 & y & 13976     \\ 
 121123A & L & 1124-00:23:13 &   51632 & y & 13992     \\ 
 121201A & L & 1202-00:28:27 &   43365 & y & 14031     \\ 
 121209A & L & 1210-00:19:47 &    8436 & n & 14049     \\ 
 121217A & L & 1217-07:21:11 &     204 & y & 14091, (57)  \\ 
 121226A & S & 1227-05:20:45 &   36662 & n &           \\ 
 121229A & L & 1229-05:09:12 &     531 & y & 14117     \\ 
 130131B & L & 0201-04:06:41 &   32193 & n & 14173     \\ 
 130206A & L & 0207-01:11:50 &   20122 & n & 14191     \\ 
 130211A & L & 0211-03:51:23 &     891 & y & 14196, 14358 \\ 
 130215A & L & 0216-00:25:16 &   82426 & y & 14221, (58)  \\ 
 130216A & L & 0217-00:23:31 &    7687 & n &           \\ 
 130216B & L & 0218-00:37:17 &  106745 & n &           \\ 
 130306A & L & 0307-08:07:24 &   29783 & n &           \\
 130310A$^1$ & L & 0312-06:57:18 &  125256 & n &           \\
 130313A & S & 0314-05:59:41 &   49890 & n &           \\ 
 130315A & L & 0315-23:52:44 &   40032 & y & 14325     \\ 
 130408A & L & 0408-23:23:47 &    5529 & y & 14364     \\ 
 130418A & L & 0419-01:20:33 &   22780 & y & 14386     \\ 
 130427A & L & 0427-23:18:47 &   55850 & y &        (59)   \\
 130513A & L & 0513-22:54:29 &   54989 & n & 14639     \\ 
 130514A & L & 0514-07:14:59 &      78 & y & 14634     \\ 
 130514B & L & 0514-23:11:03 &   35071 & n & 14657     \\ 
 130515A & S & 0515-02:12:08 &    3051 & n & 14655     \\ 
 130529A & L & 0530-07:17:19 &   72036 & n & 14730     \\ 
 130603B & S & 0604-22:54:28 &  111914 & n &           \\ 
 130605A & L & 0605-23:43:57 &     135 & y & 14774     \\ 
 130606A & L & 0607-03:29:55 &   23116 & y & 14807, (42)  \\ 
 130606B & L & 0607-03:19:05 &   55412 & n &           \\ 
 130609B & L & 0611-09:30:10 &  129090 & y & 14866     \\ 
 130610A & L & 0611-02:08:50 &   82597 & y &           \\ 
 130612A & L & 0613-03:01:39 &   85157 & y &           \\ 
 130615A & L & 0615-10:17:21 &    1956 & y & 14898     \\ 
 130623A & L & 0625-09:14:46 &  163922 & y & 15080     \\ 
 130725A & L & 0727-01:10:48 &  135217 & y &           \\ 
 130725B & L & 0726-00:47:24 &   25676 & y & 15034     \\
 130727A & L & 0727-23:43:06 &   25066 & y & 15052     \\
 130803A & L & 0803-23:13:20 &   47428 & n & 15066     \\ 
 130807A & L & 0808-06:07:38 &   70915 & n &           \\ 
 130812A & L & 0813-10:13:36 &   42644 & n & 15091     \\ 
 130816A & L & 0816-01:49:42 &     167 & y & 15101     \\ 
 130816B & L & 0816-23:14:17 &   66039 & y & 15107     \\ 
 130831A & L & 0901-03:35:03 &   52247 & y & (8, 60, 61) \\ 
 130831B & L & 0831-23:22:13 &   34433 & n & 15158     \\ 
 130903A & L & 0903-07:07:05 &   22773 & y & 15170     \\ 
 130912A & S & 0912-08:50:08 &     910 & y & 15214     \\ 
 130925A & L & 0925-04:18:31 &    1328 & y & 15247, (62, 63) \\ 
 131002B & L & 1003-03:15:17 &   58865 & n & 15297     \\ 
 131004A & S & 1004-23:41:07 &    7204 & y & 15309     \\ 
 131011A & L & 1013-03:30:56 &  121402 & y & 15328     \\ 
 131014A & L & 1016-07:06:58 &  155485 & n & 15347     \\ 
 131018A & L & 1019-06:41:25 &   64417 & n &           \\ 
 131024A & L & 1025-00:10:34 &   42254 & n & 15373     \\ 
 131117A & L & 1117-00:36:34 &     150 & y & 15493, (42)  \\ 
 131128A & L & 1129-00:27:15 &   33651 & y & 15547     \\ 
 131202A & L & 1203-01:01:10 &   35341 & n & 15586     \\ 
 131205A & L & 1206-05:07:26 &   71328 & y & 15585     \\ 
 131209A & L & 1211-08:48:22 &  157225 & n &           \\ 
 131218A & L & 1219-01:18:21 &   15169 & n & 15606     \\ 
 131227A & L & 1228-02:07:52 &   76981 & n & 15632     \\ 
 131229A & L & 1229-06:51:53 &     749 & n & 15636     \\ 
 131231A & L & 0101-01:09:39 &   73463 & y &           \\ 
 140102A & L & 0103-07:51:17 &   38020 & y & 15665     \\ 
 140114A & L & 0116-07:37:08 &  157168 & n &           \\ 
 140129A & L & 0130-00:56:58 &   77579 & y &          \\ 
 140209A & L & 0210-00:32:22 &   61285 & y &           \\ 
 140213A & L & 0214-00:26:06 &   18250 & y & 15829     \\ 
 140226A & L & 0227-08:10:08 &   79631 & y & 15887, (64) \\ 
 140301A & L & 0302-00:09:03 &   31454 & y & 15904     \\ 
 140302A & L & 0302-08:21:12 &     494 & y & 15903     \\ 
 140311A & L & 0312-03:08:02 &   21766 & y &       (42) \\ 
 140331A & L & 0401-03:16:51 &   77223 & y &           \\ 
 140408A & L & 0409-06:16:31 &   61237 & n & 16089, 16341 \\ 
 140412A & L & 0412-23:19:45 &    3536 & y &           \\ 
 140413A & L & 0413-00:13:31 &     231 & y & 16099     \\ 
 140428A & L & 0429-00:43:14 &    7344 & y & 16191, (42) \\ 
 140506A & L & 0509-07:09:56 &  208940 & y &        (65)  \\ 
 140509A & L & 0509-10:10:06 &   28073 & y & 16992     \\ 
 140512A & L & 0513-03:59:17 &   30448 & y & 16249, 16257    \\ 
 140515A & L & 0515-22:57:45 &   49509 & y & 16280, (42) \\ 
 140614A & L & 0614-01:08:35 &     216 & y & 16392, (42) \\ 
 140614B & L & 0614-08:22:07 &    6236 & n &           \\ 
 140619A & L & 0620-07:25:54 &   71239 & y &           \\ 
 140622A & S & 0622-10:00:44 &    1480 & n & 16435,  (66)   \\ 
 140626A & L & 0626-00:44:56 &     835 & y & 16458     \\ 
 140628A & L & 0629-08:40:15 &   68678 & y & 16476     \\ 
 140706A & L & 0707-09:01:43 &   48490 & y & 16534     \\ 
 140710A & L & 0710-10:49:44 &    1984 & y & 16565     \\
 140710B & L & 0712-00:28:02 &   96628 & y &           \\
 140716A & L & 0717-10:16:14 &   85697 & y & 16601     \\ 
 140719A & L & 0719-23:04:27 &   61832 & n & 16611     \\ 
 140719B & L & 0720-09:22:39 &   45184 & n &           \\ 
 140730A & L & 0802-08:36:18 &  219147 & n & 16664     \\ 
 140801A & L & 0802-09:43:54 &   53041 & y & 16666, (67) \\ 
 140815A & L & 0816-08:17:25 &   37340 & n & 16693     \\ 
 140818A & L & 0818-23:15:06 &   52990 & y & 16726     \\ 
 140818B & L & 0818-23:43:50 &   17974 & y & 16713     \\ 
 140919A & L & 0919-23:30:53 &   29738 & y & 16831, 16836    \\ 
 140927A & L & 0927-05:26:13 &     662 & n &           \\ 
 140928A & L & 0929-08:43:34 &   80020 & y & 16849     \\ 
 140930B & S & 1001-02:26:24 &   24282 & y & 16872,  (66)   \\ 
 141004A & L & 1006-05:31:55 &  108661 & y & 16899     \\ 
 141005A & L & 1005-23:49:41 &   66995 & n & 16898     \\ 
 141017A & L & 1018-02:56:24 &   66656 & y & 16926     \\ 
 141022A & L & 1022-01:39:07 &     685 & n & 16939    \\ 
 141026A & L & 1026-02:51:28 &     877 & y & 16953     \\ 
 141028A & L & 1028-23:56:37 &   46910 & y & 16977     \\ 
 141031A & L & 1031-07:45:46 &    1640 & n & 16996     \\ 
 141109A & L & 1109-06:51:31 &    3696 & y & 17041     \\ 
 141121A & L & 1121-05:51:02 &    7219 & y & 17078     \\ 
 141207A & L & 1209-06:30:26 &  127145 & n &           \\
 141212A & S & 1213-01:22:07 &   47286 & n &           \\
 141221A & L & 1221-08:09:16 &     126 & y & 17212     \\ 
 141225A & L & 1226-05:57:17 &   24970 & y & 17248     \\ 
 150101B & S & 0106-07:29:00 &  403526 & n &           \\ 
 150120B & L & 0121-01:24:50 &   64549 & y & 17336     \\ 
 150123A & L & 0124-03:19:17 &   44247 & n &           \\ 
 150201A & L & 0202-00:48:45 &   39706 & n & 17372     \\ 
 150202A & L & 0203-00:43:46 &    5624 & n & 17381     \\ 
 150203A & L & 0203-04:20:38 &     691 & n & 17386     \\ 
 150204A & L & 0205-01:33:31 &   68543 & n & 17430     \\ 
 150206A & L & 0207-00:31:52 &   36110 & y & 17419     \\ 
 150222A & L & 0223-03:39:52 &   38630 & n & 17495, 17506 \\ 
 150301A & S & 0301-06:02:49 &   17901 & n & 17513     \\ 
 150301B & L & 0302-00:08:13 &   16209 & y & 17522     \\ 
 150318A & L & 0318-08:27:54 &    4981 & n & 17602     \\ 
 150423A & S & 0423-06:32:00 &     236 & y & 17729, 17732  \\ 
 150424A & S & 0424-23:09:08 &   55571 & y & 17757, (68)  \\ 
 150428A & L & 0428-02:24:07 &    3207 & n & 17767     \\ 
 150428B & L & 0428-05:47:13 &    9202 & n &           \\
 150430A & L & 0430-07:06:25 &   24297 & n & 17796     \\
 150514A & L & 0515-23:36:04 &  104459 & y & 17821     \\ 
 150518A & L & 0520-03:46:30 &  108256 & y &           \\ 
 150821A & L & 0822-01:54:00 &   58199 & y & 18195     \\ 
 150831A & S & 0831-23:21:37 &   46045 & n & 18219     \\ 
 150831B & L & 0901-01:49:43 &   12616 & n &           \\ 
 150902A & L & 0903-23:33:04 &  107845 & n &           \\ 
 150907B & L & 0908-02:03:03 &    9396 & n & 18252     \\ 
 150910A & L & 0911-05:07:48 &   72180 & y & 18277    \\ 
 150911A & L & 0912-05:59:25 &   40744 & n & 18313     \\ 
 150915A & L & 0915-23:26:16 &    7672 & y & 18317     \\ 
 151004A & L & 1004-23:39:30 &   19826 & n &           \\
 151023A & L & 1023-23:52:47 &   36582 & n & 18461      \\
 151027B & L & 1028-06:23:00 &   27740 & y & 18507, (42, 69) \\ 
 151029A & L & 1029-08:35:12 &    2718 & y & 18522, 18523    \\ 
 151031A & L & 1101-06:42:20 &   89510 & y & 18538, 18550    \\ 
 151111A & L & 1112-00:09:10 &   56147 & y & 18578, 18585    \\ 
 151112A & L & 1113-00:10:26 &   37538 & y & 18588, 18595, (42) \\ 
 151114A & L & 1115-04:58:40 &   68346 & y & 18599, 18607    \\ 
 151120A & L & 1120-08:34:02 &     672 & y & 18624     \\ 
 151127A & S & 1127-09:12:30 &     221 & n & 18645     \\ 
 151205B & L & 1206-00:31:40 &   10106 & n & 18674     \\ 
 151210A & L & 1210-03:17:35 &     279 & y & 18680     \\ 
 151212A & L & 1213-03:27:48 &   48441 & y & 18688     \\ 
 151215A & L & 1215-04:08:29 &    4021 & y & 18694     \\ 
 151228A & S & 1228-08:00:44 &   17732 & n &      \\ 
 160104A & L & 0107-01:10:09 &  222359 & y &           \\ 
 160117B & L & 0118-00:53:54 &   39267 & y & 18887     \\ 
 160119A & L & 0119-07:53:00 &   17212 & y & 18902     \\ 
 160121A & L & 0122-04:09:00 &   51503 & y & 18921     \\ 
 160123A & L & 0123-09:12:24 &     844 & n & 18936     \\ 
 160127A & L & 0127-08:46:01 &     174 & y &           \\ 
 160131A & L & 0201-00:44:10 &   59019 & y & 18967     \\ 
 160203A & L & 0203-02:18:39 &     329 & y & 18980     \\ 
 160206A & L & 0207-00:40:00 &   78437 & n & 18997     \\ 
 160216A & L & 0217-08:33:29 &   48178 & n &           \\ 
 160220A & L & 0220-05:43:53 &   15508 & n & 19027     \\ 
 160220B & L & 0221-07:14:43 &   72230 & n & 19042     \\ 
 160221A & L & 0222-05:12:21 &   19359 & n & 19054     \\ 
 160223A & L & 0223-03:30:41 &    6376 & y & 19058     \\ 
 160223B & L & 0224-00:35:21 &   52583 & n &           \\ 
 160228A & L & 0229-00:02:36 &   23284 & y & 19114     \\ 
 160303A & S & 0304-03:27:00 &   59538 & y & 19144     \\ 
 160314A & L & 0315-23:53:26 &  130835 & y & 19200     \\ 
 160325A & L & 0325-23:31:47 &   59504 & n & 19233     \\ 
 160410A & S & 0410-05:41:30 &    1902 & y & 19272, (70) \\
 160411A & S & 0411-09:16:17 &   28046 & n & 19292     \\ 
 160412A & L & 0412-23:11:20 &   67646 & n & 19314     \\ 
 160422A & L & 0422-23:02:24 &   39803 & y & 19352     \\ 
 160425A & L & 0426-04:38:15 &   18724 & y & 19349     \\ 
 160501A & L & 0501-04:27:21 &   13610 & n & 19375     \\ 
 160506A & L & 0506-06:47:16 &   11881 & n & 19396     \\ 
 160607A & L & 0608-08:29:40 &   76539 & y & 19512     \\ 
 160625A & L & 0626-22:57:06 &   87395 & n & 19629     \\ 
 160625B & L & 0627-08:12:26 &  120730 & y &           \\ 
 160630A & L & 0630-04:10:58 &     409 & y & 19625     \\ 
 160712A & L & 0713-03:18:21 &   26685 & n & 19693     \\ 
 160726A & S & 0726-10:52:12 &   33485 & n &           \\ 
 160801A & L & 0801-23:35:50 &   50819 & n &           \\ 
 160804A & L & 0804-23:12:02 &   77955 & y & 19774     \\ 
 160927A & S & 0927-23:26:10 &   19281 & y & 19959     \\ 
 161001A & S & 1001-02:52:00 &    6404 & y & 19975     \\ 
\end{xtabular}
\tablebib{
  The 4- or 5-digit numbers denote the GCN (GRB Coordinate Network) number,
  and can be retrieved from https://gcn.nasa.gov/circulars/NNNNN.
  The numbers in brackets are references as follows:
  (1) \citet{NicuesaGuelbenzu+2012b};
  (2) \citet{kkg08b};
  (3) \citet{gkk11};
  (4) \citet{cdk08};
  (5) \citet{Perley+2010};
  (6) \citet{Clemens+2011};
  (7) \citet{kgm09};
  (8) \citet{Klose+2019};
  (9) \citet{NicuesaGuelbenzu+2014};
  (10) \citet{gkm09};
  (11) \citet{Rossi+2012};
  (12) \citet{Guidorzi+2009};
  (13) \citet{Filgas+2011a};
  (14) \citet{Rossi+2008b};
  (15) \citet{Zafar+2012};
  (16) \citet{kga10};
  (17) \citet{ksg11};
  (18) \citet{gkf09};
  (19) \citet{Greiner+2009};
  (20) \citet{Rossi+2011};
  (21) \citet{dbm08};
  (22) \citet{Olivares+2015};
  (23) \citet{Yuan+2010};
  (24) \citet{ngk11};
  (25) \citet{kgs11b};
  (26) \citet{gkp10};
  (27) \citet{Savaglio+2012};
  (28) \citet{mkr10}
  (29) \citet{Tanvir+2009};
  (30) \citet{NicuesaGuelbenzu+2011};
  (31) \citet{clf11};
  (32) \citet{NicuesaGuelbenzu+2012a};
  (33) \citet{Rau+2010};
  (34) \citet{Wiersema+2012};
  (35) \citet{Filgas+2012};
  (36) \citet{Filgas+2011b};
  (37) \citet{Thone+2013};
  (38) \citet{Olivares+2012};
  (39) \citet{gkn13};
  (40) \citet{NicuesaGuelbenzu+2015};
  (41) \citet{Nardini+2014};
  (42) \citet{Bolmer+2018};
  (43) \citet{Zauderer+2013};
  (44) \citet{Sanchez-Ramirez+2017};
  (45) \citet{Ackermann+2013};
  (46) \citet{Elliott13};
  (47) \citet{Tanga+2018};
  (48) \citet{Margutti+2012};
  (49) \citet{Greiner+2015};
  (50) \citet{Kann+2018};
  (51) \citet{Melandri+2017};
  (52) \citet{Guidorzi+2014};
  (53) \citet{Schulze+2014};
  (54) \citet{Martin-Carillo+2014};
  (55) \citet{kgs13};
  (56) \citet{Tanvir+2018};
  (57) \citet{Elliott14};
  (58) \citet{Cano+14};
  (59) \citet{Perley+2014};
  (60) \citet{Klose+2013a};
  (61) \citet{Cano+14};
  (62) \citet{gyk14};
  (63) \citet{Schady+2015};
  (64) \citet{Cenko+2015};
  (65) \citet{Fynbo+2014};
  (66) \citet{Pandey+2019};
  (67) \citet{Lipunov+2016};
  (68) \citet{Knust+2017};
  (69) \citet{Greiner+2018};
  (70) \citet{Fernandez+2022}.
  }
  $^1$ Very large error box, thus not listed in Tab. \ref{unpub}. 
\endgroup

\appendix

\section{Notes on individual sources II}

This section provides unpublished GROND information on
(i) known afterglows or
(ii) upper limits in cases of no previous optical/NIR afterglow reports or
(iii) if GROND upper limits are deeper than those reported by other groups or
(iv) host galaxy candidates in case of no afterglow detection. 
This includes GRBs with previous GROND GCN
if new information is available (such as GRB 110721A or 151023A).

All GROND magnitudes (including $JHK$) are in the AB system, and not
corrected for foreground extinction unless specifically mentioned.
Upper limits are at 3$\sigma$ confidence.

\subsection{GRB 071021}

This long-duration (T90 = 225 sec) GRB triggered Swift/BAT at
09:41:33 UT, providing an X-ray position shortly after \citep{Sakamoto+2007}.
Despite early optical observations, no afterglow was found. 
Deep NIR observations 11.25 hrs post-burst revealed a ``faint source''
(no magnitudes given) in the $HK$
bands \citep{CastroTirado+2007}, spurring speculations on a high-redshift GRB.
Observations with Subaru/IRCS about 20 hrs post-burst measured
  $K \approx 21$ mag \citep{Terada+2007}, confirmed to be in the Vega system
  (priv. comm. N. Kawai).
  Several years later, deep imaging detected the host galaxy at this
  position down to the $B$ band \citep{Perley+2013}, and VLT spectroscopy
  revealed a redshift of  $z = 2.452$ \citep{Kruehler+2012} via host
  emission lines. Interestingly, \cite{Perley+2013} measured $K = 20.25\pm0.23$
  which means that the Subaru/IRCS detection was already that of the host
  galaxy, not that of the afterglow emission. Re-analysis and photometric
  calibration of the TNG $H/K$ data from \cite{CastroTirado+2007} show
  that the brightness of the ``faint source'' is fully compatible with
  the host measurements of  \cite{Perley+2013}.
GROND observations started 14 hrs after the GRB (lasting for 2 hrs),
thus between the TNG and Subaru observing times.
A second epoch was done in the following night,
though it is less deep.
In the first epoch, we only detect the ``faint source''  in the $K$ band
which in light of the above details is compatible with the detection of
the host galaxy.

\subsection{GRB 080405}

This GRB was detected with Swift/BAT and Mars Odyssey, but localisation
was delayed by $\sim$22 hrs \citep{CummingsHurley2008}. A Swift/XRT
observation at $\sim$43 hrs after the GRB found one X-ray source
within the BAT error circle, but no fading could be established.
GROND observations started $\sim$38 hrs after the GRB. There is
one bright (\rp $\sim$ 19 mag), a fainter (\rp $\sim$ 23 mag)
and an even fainter extended object within the 7\asec\ XRT error
circle \citep{LaParola+2008}, but all three sources were found
to not have faded in a GROND observation on Feb. 25, 2009.

\subsection{GRB 080414}

This INTEGRAL burst with 10 sec duration \citep{Mereghetti+2008}
occurred at a galactic latitude of only 0\fdg5, behind a correspondingly
very large column density of E(B-V) = 7 mag. Follow-up observations
with Swift did not reveal an X-ray counterpart. GROND observations
started about 6 hrs after the GRB trigger, and did not reveal a
counterpart candidate.

\begin{figure}[ht]
  \includegraphics[width=8.8cm]{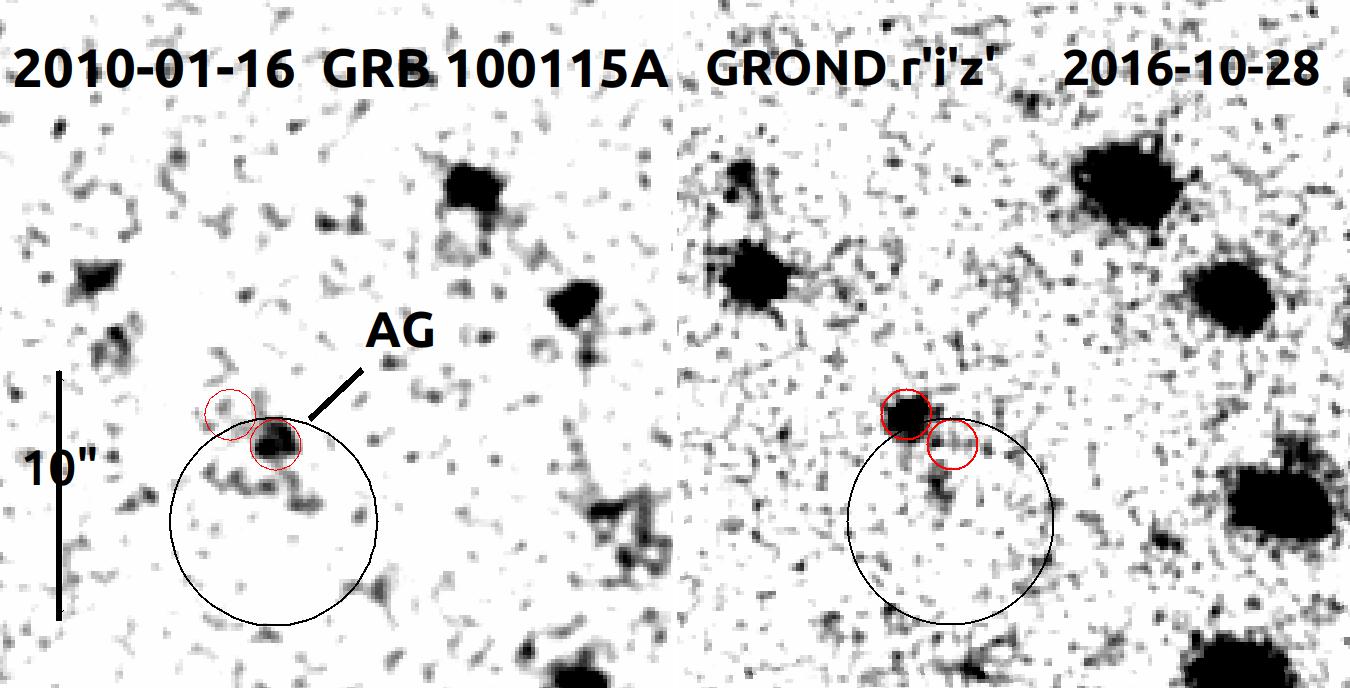}
  \caption[100115A]{GROND images from the co-add of \gp\rp\ip stacks
    of the afterglow (left) and  a late deep image (right).
    The large circle denotes the Swift/XRT afterglow position, the two
     small red ones the two components of object 'A' of \cite{Jelinek+2010}.
  \label{100115A}}
\end{figure}

\subsection{GRB 080905B}

For this Swift-detected GRB with XRT position, an afterglow was seen both
with UVOT \citep{Stroh+2009} as well as the ESO/VLT \citep{Vreeswijk+2009},
about 3\farcs8 off the centre of the galaxy 2MASSX J20065732-6233465.
The ESO/VLT observations, starting 8.3 hrs after the GRB, reveal a bright
source with R$\sim$20.2 mag, and a redshift (based on resonance absorption
features) of z=2.374 \citep{Vreeswijk+2009}.
GROND observations started 14.8 hrs after the burst, but due to very bad
seeing (2\farcs5) the afterglow is not resolved from the galaxy. The limits
in Tab. \ref{unpub} refer to the limiting magnitude of the
stack of the
first four individual 6 min integrations, but not to the brightness
of the well-detected unresolved host+afterglow emission!

\subsection{GRB 090827}

This GRB triggered Swift-BAT, but the onboard software could not identify
a position, thus the localisation was distributed after ground analysis
\citep{Cummings+2009}. Swift-XRT observations clearly identified an X-ray
counterpart \citep{Evans+2009}. GROND observations were taken at 4 epochs
over 24 days. There are two sources in the revised\footnote{\url{https://www.swift.ac.uk/xrt\_positions/}} (retrieved: 2020) 4\farcs2 error
circle, but both are constant in all our images. 
Due to one of the four read-out channels of the $K$-band detector temporarily
malfunctioning, there is no $K$-band coverage of the final XRT-position.

\subsection{GRB 100115A}
This GRB was found in ground analysis of a preplanned Swift slew maneuver
\citep{Cummings+2010}, but despite the late Swift/XRT follow-up,
an X-ray afterglow was identified \citep{Margutti+2010}.
No UVOT counterpart was detected, but \cite{Jelinek+2010} identified
an optical source with the NOT (called object 'A', which shows two components)
which \cite{Cucchiara+2010}
found to fade based on GMOS data from the first two nights.
GROND observations were taken at 3 epochs, one each in the first two nights,
the other in 2016.
During the first two epochs, we clearly see an object coincident with
the south-western component of object 'A', declining by about 1 mag.
The magnitudes of the first epoch are given in Tab. \ref{unpub}; the
others are: at mid-time 2010-01-17 01:08 UT
\gp$>$22.3 mag,
\rp=23.2$\pm$0.2 mag,
\ip$>$23.1 mag,
\zp$>$22.5 mag,
and at mid-time 2016-10-28 03:50 UT: 
\gp$>$25.3 mag,
\rp$>$25.2 mag,
\ip$>$24.3 mag,
\zp$>$24.1 mag,
J$>$21.0 mag,
H$>$20.4 mag,
K$>$18.7 mag.
Thus, the total amplitude in $r$ is $>$2.9 mag.
For the north-eastern component, we measure the following magnitudes:
\gp=24.10$\pm$0.12 mag,
\rp=23.43$\pm$0.07 mag,
\ip=23.34$\pm$0.14 mag,
\zp=23.42$\pm$0.20 mag.

\subsection{GRB 100316C}

For this Swift-detected GRB a rapid X-ray afterglow position was reported
\citep{Stamatikos+2010}, but neither Swift/UVOT (starting 81 s after the
trigger) nor BOOTES-3 observations (starting 4 s after the trigger)
revealed an optical afterglow \citep{Stamatikos+2010, Jelinek+2010b}.
GROND observations (15.8 hours after the GRB trigger) revealed a candidate
at 23rd mag \citep{Afonso+2010},
but a second epoch taken on August 6, 2010, showed this object at
the same magnitudes, therefore excluding this object as the afterglow.

\subsection{GRB 100816A}

The optical afterglow of this short GRB was detected with Swift/UVOT
at 17th magnitude
\citep{Oates+2010}, and then seen to rapidly fade to 20.5 mag after 2.8 hrs
\citep{Antonelli+2010}. Our GROND imaging, commencing on Aug. 18, 06:33 UT
(2.2 days after the GRB), did not reveal the counterpart anymore.
We clearly detect the nearby host galaxy
at z=0.80 \citep{Tanvir11123}
in all bands except $K$.

\subsection{GRB 100823A}

The NIR afterglow (source C) seen with LIRIS on the William Herschel telescope
by \cite{Levan+2010} at K(Vega)$\sim$19.5 mag
at 8 hrs after the GRB is about 2 mag below our limiting magnitudes, which
suffered from particularly bad observing conditions (seeing $\sim$3\asec).

\subsection{GRB 110721A}

Fermi/GBM triggered at 04:47:44 UT (trigger 332916465 / 110721200)
\citep{Tierney2011} on this FRED-like GRB with a duration of 24 s
and a 1-s peak photon flux of 31 ph/cm$^2$/s, among the brightest GRBs
seen with Fermi.
Combined analysis with Fermi/LAT data suggests a peak energy of
15$\pm$2 MeV during the early phase, making it the highest-E$_{peak}$
GRB \citep{Axelsson+2012}. This GRB got even more prominence as
being one for which significant polarisation of the prompt emission
was detected with IKAROS/GAP \citep{Yonetoku+2012}.

Swift/XRT observations of the Fermi/LAT
position did reveal one X-ray source, for which GROND observations
identified an optical counterpart candidate \citep{Greiner+2011}. Though the
7-channel GROND SED has a red powerlaw slope (1.2$\pm$0.2), consistent
with a GRB afterglow, the optical source did not show any fading
in the first 2 nights  \citep{Greiner+2011}. Gemini spectroscopy
suggested a redshift of z=0.382 based on  CaII H\&K \citep{Berger2011},
while an X-shooter spectrum did not provide any further clues
\citep{Selsing+2019}.

Here, we report a 3rd GROND epoch taken
on 2011 July 24 at 7:20--8:50 UT, which confirms the constant flux.
Also, an archival NTT/EFOSC $R$-band image taken on Aug 10, 2010,
a year prior to the GRB, shows this source at a similar magnitude.
We therefore conclude that the optical source is not the
afterglow of GRB 110721A. Instead, the powerlaw SED and the spectrum
suggest a blazar nature. The X-ray detection is consistent with
this interpretation, though we note that \cite{Poonam2011}
suggested PKS 2211-388 (which is at 9\asec\ distance) as the
source of the X-ray emission.

The lack of an optical counterpart is astonishing,
given that this is one of the brightest GRBs detected so far.

\subsection{GRB 111020A}

GRB 111020A triggered Swift/BAT at 06:33:50 UT, and a fading X-ray
afterglow was identified \citep{Mangano+2011} which was also detected
with XMM-Newton \citep{Campana2011}. While the total GRB duration is
only 0.5 sec, the substantial positive spectral lag argues
against a short-duration classification \citep{Sakamoto+2011}.
GROND observations started at 23:50, more than 17 hrs after the
GRB, and did not detect the optical afterglow. The GROND limits
are less constraining than the  Gemini observations
\citep{Fong+2012}, but we do detect their sources S2 and G3. 

\subsection{GRB 111026A}
This Swift/BAT detected GRB could not be followed up with XRT/UVOT
due to a pointing constraint.
We obtained only one epoch with GROND of the 3\amin\ BAT error box,
and the limits in Tab. \ref{unpub} are those
of SkyMapper (\gp\rp\ip\zp) and 2MASS ($JHK$). The GROND images
are about 1 mag (\gp\rp\ip\zp) and 2 mag  ($JHK$) deeper.

\subsection{GRB 111121A}

We do not detect the faint emission reported by \cite{Fong+2011}
for which neither confirmation nor evidence of fading exists.
The Swift/XRT afterglow position is within 4\asec\ of a 12th magnitude star,
so our upper limits are dominated by the noise of the image
subtraction of two stacks about 1 hr apart.

\subsection{GRB 111207A}

This 3-sec duration Swift-GRB was found in ground-analysis, and only reported
1.5 days post-burst \citep{Cummings+2011}. No X-ray counterpart was reported.
GROND observations in the following two nights
reach deep limits, but image subtraction does not reveal a convincing
counterpart candidate within the BAT error circle.

\subsection{GRB 120118A}
GROND observations started 140 sec after INTEGRAL trigger 6443,
but were offset in declination to cover the common error box with
the earlier sub-threshold trigger 6441. Unfortunately, the final
error box \citep{Goetz+2012}
is not covered in the \griz-channels, and only to about
30\% in $JHK$. The counterpart candidate of \cite{Trello+2012}
is not covered, even in $JHK$. 
The upper limits in Tab. \ref{unpub} refer to the
first 4 min of observation. \cite{GoetzBozzo2012} caution that these
two INTEGRAL triggers and the INTEGRAL trigger 6449 could be caused
by the high-mass
X-ray binary GX 301-4 which was active at this time. While GX 301-4
is covered by our images, it is saturated in all bands.

\subsection{GRB 120118B}
Observations with the TNG telescope 10 hrs after this Swift-BAT detected
GRB did not reveal a new source within the XRT error circle
\citep{Osborne+2012}, down to limiting magnitudes of
J(Vega)$>$20.5 mag and  H(Vega)$>$20.3 mag \cite{AvanzoPalazzi2012}.
However, they noted a faint NIR source close to the
Swift/XRT afterglow position.
Keck observations about one year later identified an extended source with
I(Vega)$\sim$23.8 mag at that position,
suggestive of the host galaxy, and obtained a redshift of
z = 2.943 \citep{Malesani+2013}.
GROND observations in the first night were taken about one hour earlier
than those with the TNG, but did not find a source, to the limits given in
Tab. \ref{unpub}.
GROND follow-up observations in February 2016 provide the following
host magnitudes (all in AB):
\gp = 25.57$\pm$0.18 mag,
\rp = 24.67$\pm$0.09 mag,
\ip = 24.48$\pm$0.15 mag,
\zp = 23.92$\pm$0.16 mag, J$>$21.6 mag, H$>$21.1 mag, K$>$19.9 mag.

\subsection{GRB 120212A}

The optical afterglow of this Swift-detected GRB \citep{Sonbas+2012}
was detected 14 min after trigger by the 2-m Faulkes Telescope South
\citep{Guidorzi120212}. GROND observations were only possible on the
following night because the telescope was closed due to too high humidity.
The afterglow is well detected, about 6 magnitudes fainter than
at discovery by \cite{Guidorzi120212}, but the photometry is
affected by the tail of the bright star to the North-East.

\subsection{GRB 120224A}

GROND observations of this Swift-detected GRB \citep{Saxton+2012} started
19.58 hrs after the GRB trigger, and revealed a source \citep{Elliott+2012}
at the edge of the revised error circle of the X-ray afterglow
\citep{Osborne+2012b}. GROND observations in the following night
(starting 2012-02-26T00:48) and 6 months later (starting 2012-08-26T07:10)
reveal the object at the same magnitude, namely
\gp = 25.75$\pm$0.28,
\rp = 24.00$\pm$0.09,
\ip = 23.00$\pm$0.08,
\zp = 22.45$\pm$0.06,
$J$ = 20.86$\pm$0.09,
$H$ = 20.95$\pm$0.19,
$K$ = 19.94$\pm$0.12.
This SED is well fit with a dusty, star-forming (3$\pm$0.2 \msun/yr)
galaxy at redshift 1, consistent with the Balmer-break redshift
of 1.1$\pm$0.2 as determined with  the VLT/X-shooter \citep{Kruehler+2015}.

\subsection{GRB 120324A}
For this Swift-detected GRB an X-ray afterglow was rapidly found
 \citep{Zhang+2012}, but due to the low galactic latitude the substantial
crowding and foreground absorption complicated the search for an optical/NIR
counterpart. The first reported uncatalogued object \citep{Guidorzi13092}
was found to be constant over a few hours and visible on old POSS2 plates
\citep{Im13097}, and a fainter source reported with no brightness information,
but positioned in the then revised X-ray error circle \citep{Im13097}
was also detected with GROND in several bands about 1.5 hrs earlier
\citep{Sudilovsky13098}. However, another GROND epoch taken 17 months later
(starting 2013-08-12T02:28) showed this source also at unchanged
brightness. Thus, no optical/NIR counterpart was identified for this GRB.

\subsection{GRB 120419A}
GROND observations of this INTEGRAL burst started about 11 hrs after
the GRB, and a second
epoch was obtained on 9th July, 2012. The GRB position is in a
very crowded field in the Galactic Plane, with large foreground
extinction (of order $A_{\rm V} \sim$ 40 mag, \citealt{sfd98}).
The 4\farcs5 Swift/XRT error circle of the X-ray afterglow \citep{Page+2012}
contains one bright (13 mag) and 5 fainter (18--19 mag) stars,
none of which are variable between the two GROND epochs.
Our limits refer to the first epoch on top of the diffuse
emission of unresolved sources.

\subsection{GRB 120724A}
Due to visibility constraints, GROND observations started only 21 hrs
after the GRB, under poor seeing conditions. Image subtraction with
a later epoch on March 23, 2013 did not reveal the afterglow
\citep{Guidorzi13511} anymore, but just the host galaxy at
redshift 1.48 \citep{Cucchiara+2012}: the residuals are completely
dominated by the noise of the image subtraction (corresponding to about
\rp\ip\zp $>$ 23 mag), while the formal 3$\sigma$ upper limits of the images
are about 1 mag deeper; thus we refrain from quoting individual limits
in the seven GROND filter bands.
The host magnitudes are
\gp = 22.19$\pm$0.04 mag,
\rp = 21.02$\pm$0.02 mag,
\ip = 20.59$\pm$0.02 mag,
\zp = 20.22$\pm$0.02 mag,
J = 19.98$\pm$0.13 mag,
H = 19.61$\pm$0.17 mag,
K = 19.50$\pm$0.30 mag.
We note that the brightness of the galaxy is nearly identical to the
\ip-magnitude given by \cite{Guidorzi13511} for the fading afterglow.

\subsection{GRB 120805A}
GROND observations started immediately after evening twilight,
but rolling-in clouds severely limited our image depth. The optical afterglow
\citep{Gorosabel+2012} is not detected. Another observation
on March 9, 2013 reached about 4 mags deeper, and we detect the
z=3.1 \citep{Kruehler+2015} host galaxy at the optical afterglow position,
at the following magnitudes:
\gp = 25.37$\pm$0.24 mag,
\rp = 24.07$\pm$0.10 mag,
\ip = 23.32$\pm$0.11 mag,
\zp = 23.25$\pm$0.16 mag,
and derive NIR upper limits of
$J >$ 21.1 mag,
$H >$ 20.4 mag,
$K >$ 19.7 mag.

\subsection{GRB 120817A}
The GROND observations in the first night (after 21.6 hrs) suffered
from particularly bad seeing (4\asec), thus the counterpart candidate
\citep{Guidorzi120817} is not resolved from the nearby brighter star.
In a 1-hr exposure three nights later we clearly detect this
object, at a magnitude consistent with \cite{Guidorzi120817},
and therefore exclude the afterglow interpretation.

\begin{figure}[ht]
  \includegraphics[width=8.8cm]{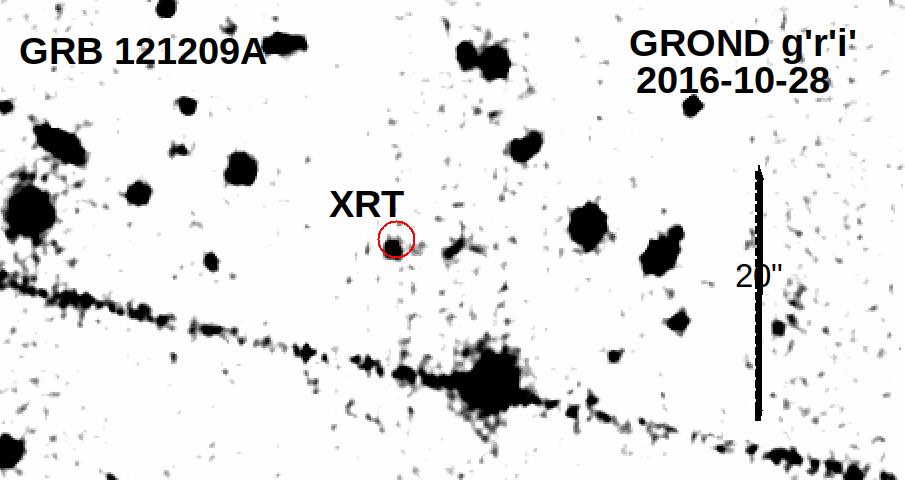}
  \caption[121209A]{GROND image from the co-add of \gp\rp\ip\ stacks
    of the host candidate of GRB 121209A within the Swift/XRT position
    of the afterglow (red circle).
  \label{121209A}}
\end{figure}

\subsection{GRB 121014A}
Due to the small Sun distance, no Swift/XRT follow-up could be done,
and GROND observations of the  2\amin\ Swift/BAT error circle
\citep{Barthelmy+2012} in all 7 channels could only be done for 4 minutes.
$JHK$ imaging continued for another 20 min during morning twilight.
No new source is found after comparison to PS1 or 2MASS, where the
$JHK$ upper limits in Tab. \ref{unpub} are those of 2MASS, while the
actually reached sensitivities are 
$J >$ 19.2 mag,
$H >$ 18.6 mag,
$K >$ 17.8 mag.

\subsection{GRB 121102A}
GROND observations did not reveal an afterglow. The 1\farcs5 Swift/XRT
error circle \citep{Osborne+2012c} does contain an object, but correcting
for a fraction of the foreground extinction $A_{\rm V}$ = 5.1 mag,
the colours suggest a B star. Follow-up observations in the subsequent
two nights show no fading of this object.

\subsection{GRB 121209A}
Based on observations in the first night, a detection of a faint source
in \rp\ip\zp\ within the Swift/XRT error circle was reported
\citep{Kruhler+2012GCN}. Another GROND observation on 2016 Oct. 28 under
good conditions returns similar magnitudes of this object:
\gp = 24.79$\pm$0.14 mag, 
\rp = 24.04$\pm$0.10 mag,
\ip = 24.10$\pm$0.24 mag,
\zp = 23.63$\pm$0.24 mag.
The PSF of this object is about 50\% larger
of that of stars of similar brightness. This makes it a possible
host galaxy candidate, rather than a chance coincidence with a
galactic stellar object.

\subsection{GRB 121226A}
GROND observations (under poor seeing conditions) started about
10 hrs after the GRB. In stacked
images of 150 min \gp\rp\ip\zp and 120 min $JHK$ exposure,
centred at Decl. 27 07:05 UT, we detect
the object first reported by \cite{CastroTirado+2012} within the
Swift/XRT error circle, at
\gp = 24.68$\pm$0.34 mag,
\rp = 24.05$\pm$0.13 mag,
\ip = 24.30$\pm$0.22 mag,
\zp = 24.00$\pm$0.23 mag,
J $>$22.6 mag,
H $>$22.0 mag,
K $>$21.0 mag.
In a later GROND observation on Feb. 15, 2013 we obtain very similar
magnitudes, supporting the conclusion of \cite{Malesani+2013a} of 
host emission, based on
the positional coincidence with the XRT position and of a radio
source \citep{Fong+2012a} (though also no radio variability has
been reported).
Since our early detection is close to the limit of the observations,
and thus the residuals are completely dominated by the noise of the
image subtraction,
we refrain from providing upper limits for the afterglow.

\subsection{GRB 130327B}
The X-ray source \citep{KrimmTroja2013} found with Swift/XRT within the
error circle of this Fermi/LAT-localised GRB \citep{Ohno+2013} remained constant
over about 10 days and is not considered to be the X-ray afterglow.
In GROND observations of this X-ray source, starting 3 days after the GRB (and because of this delay not included in Tab. \ref{unpub}),
we detect 2 objects within and another 2 at the border of the X-ray source,
all of which remained constant over 4 days of GROND observations.

\subsection{GRB 130603B}

GROND observations of this short Swift-detected GRB \citep{Melandri+2013}
were possible only at 31 hrs after the trigger. The
optical afterglow \citep{Levan+2013}
is not detected on top of the bright (about 21 mag in \rp\ip\zp)
host galaxy. The residuals are completely dominated by the
noise of the image subtraction, thus we refrain from giving individual
limits in the seven filter bands in Tab. \ref{unpub}.

\subsection{GRB 130612A}
Due to withdrawn override permission, GROND observations started
only 24 hr after the GRB, but still detected the rapidly fading
afterglow of this Swift-detected GRB with UVOT afterglow detection
\citep{Racusin+2013}. Another GROND observation 3 years later
(May 4, 2016) did not reveal any source down to \rp$>$25.5 mag,
implying another $>$1 mag fading.

\subsection{GRB 130807A}

Within the Swift/XRT error circle, we do not detect any new object
besides a source at 
RA(J2000) = 17\h 59\m 12\fss0, 
Decl.(J2000) = -27\degr 36\amin 59\asec,
which is known from the VVV survey (observations taken 2011).
Given the low galactic latitude (1\fdg9), the field is very dense,
and the upper limits in Tab. \ref{unpub} refer to rare not-populated
regions.

\subsection{GRB 131014A}

For this Fermi/LAT-detected GRB, a rapid Swift observation at 12 hrs
after the GRB trigger revealed an X-ray afterglow \citep{KenneaRogers2013}.
Three different faint objects have been found within or at the edge
of the X-ray error circle, and for one of these a constant brightness
over the first 2 days was established \citep{Kann+2013}. A further
GROND observation was executed starting 2013-10-18T07:21, which showed
that also sources \#2 and \#3 have not faded relative to the observations
on Oct. 16th, and therefore none of the three sources is the optical
afterglow of GRB 131014A.

\subsection{GRB 131209A}
The LAT localisation of this Fermi-detected GRB came with a 0\fdg9
error \citep{VianelloOmodei2013}, and was subsequently improved
by triangulation with Konus-MESSENGER to a 1\fdg5$\times$0\fdg2
annulus \citep{Hurley+2013}, in which Swift/XRT follow-up observations
found two X-ray sources, one of which was uncatalogued and showed
marginal fading \citep{GrupeBreeveld2013}. Further Swift/XRT
observations identified this source as variable, but unrelated
to GRB 131209A \citep{Grupe2013}, and due to the observed rapid
X-ray variability a Narrow Line Seyfert 1 galaxy or a blazar
nature was suggested. GROND observations targeted this
X-ray source on Dec. 11th in morning twilight (in NIR-only mode),
and again on Dec. 14th. At the border of the X-ray error circle
we identify an optical source
(at RA(J2000) = 09\h 07\m 18\fss8, 
Decl.(J2000) = -33\degr 51\amin 22\asec, $\pm$0\farcs5)
which has an optical/NIR SED consistent with a powerlaw:
\gp = 23.39$\pm$0.30 mag,
\rp = 21.42$\pm$0.05 mag,
\ip = 20.73$\pm$0.05 mag,
\zp = 20.36$\pm$0.06 mag,
J = 19.43$\pm$0.04 mag,
H = 19.15$\pm$0.04 mag,
K = 18.72$\pm$0.07 mag.
We derive a powerlaw slope of 1.33$\pm$0.07, and given the large
\gp-dropout, a Ly-absorption fit results in a redshift 3.84$\pm$0.17.
We propose this source J090718.8-335122 as the counterpart of the variable X-ray
source of \cite{Grupe2013}, and suggest a high-redshift blazar interpretation.
The upper limit for the GRB afterglow in $JHK$ is given in
Tab. \ref{unpub}, but we note that the GROND observation only
covered a tiny fraction ($\sim$10\%) of the combined Fermi/LAT circle
and triangulation annulus.

\begin{figure}[ht]
  \includegraphics[width=8.7cm, viewport=30 155 590 687, clip]{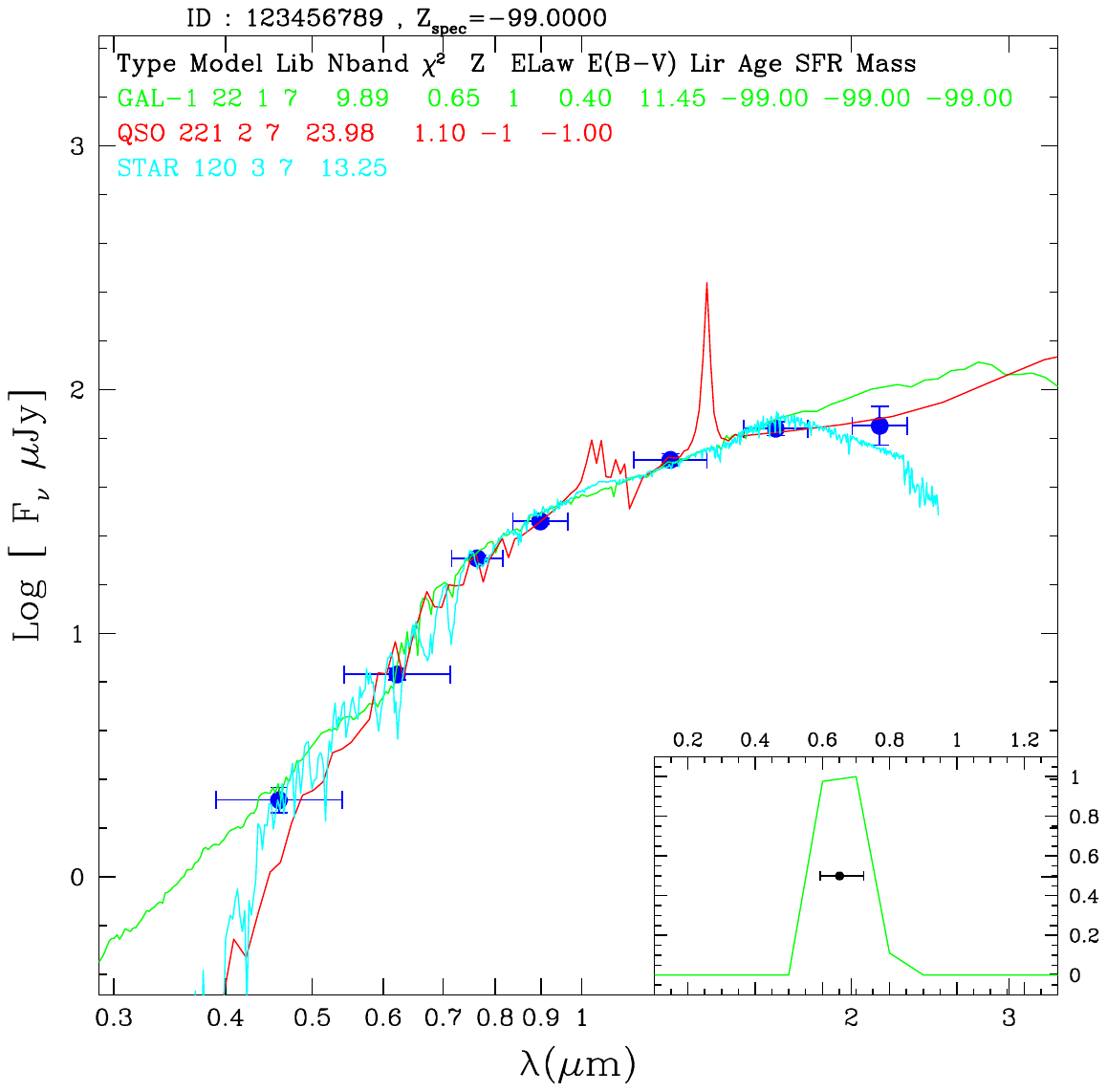}
  \vspace{-0.2cm}
  \caption[140331A]{A LePhare fit of the spectral energy distribution
    of the extended object within the Swift/XRT error circle of GRB 140331A,
    based on a stack of several GROND images with a total exposure of 1 hr.
   The data have been corrected for the foreground $A_{\rm V}$=0.15 mag.
  \label{140331A}}
\end{figure}

\subsection{GRB 140129A}
After the GROND \gp\rp\ip\zp-detection in the first night
(see Tab. \ref{unpub})
of the UVOT-detected afterglow of this Swift/BAT-triggered GRB, a second epoch
was performed two nights later, for which we measure
\gp = 24.30$\pm$0.24 mag,
\rp = 23.79$\pm$0.17 mag,
\ip $>$ 23.9 mag,
\zp $>$ 23.2 mag.
A final observation on July 21st, 2014, reveals no sign of host
emission within 5\asec\ radius, to upper limits of
\gp $>$ 24.5 mag,
\rp $>$ 24.3 mag,
\ip $>$ 23.6 mag,
\zp $>$ 22.9 mag,
J$>$20.8 mag,
H$>$20.2 mag, and
K$>$18.7 mag.

\subsection{GRB 140331A}

GROND observations of this Swift GRB started about
21.5 hours after the GRB trigger, and a second epoch was
done on  April 18, 2014.
The source, identified by \cite{Littlejohns+2014} 
within the enhanced Swift-XRT error circle \citep{Beardmore+2014},
is constant between the two epochs, and our magnitudes are consistent
with those of \cite{Littlejohns+2014} taken about 1 hr after the GRB,
as well as the SDSS measurements (SDSS J085927.51+024304.0).
The SED is very steep in the blue, and \cite{Littlejohns+2014}
suggested a photometric redshift of 4.65 (+0.34,-2.80).
However, the source is clearly extended, and a fit using LePhare
\citep{Arnouts+1999, Ilbert+2006}
(instead of a powerlaw) provides a decent fit for a galaxy at redshift
0.65$\pm$0.10, and E(B-V) = 0.4 mag (Fig. \ref{140331A}).
The SDSS DR16 query returns a photometric redshift of 0.676$\pm$0.100.
Whether or not this is the host galaxy remains open.
The upper limits in Tab. \ref{unpub} are for the region outside this
  galaxy within the XRT error circle.

\subsection{GRB 140719A}

For this Swift-detected GRB, the X-ray afterglow was rapidly localised
\citep{ Starling16606}. GROND observations started 16.8 hrs after the GRB
trigger and revealed a faint source in two filter bands \citep{Bolmer16611},
at coordinates RA(J2000.0) = 11:26:24.3, Decl.(J2000) = -50:08:04.8
($\pm$0\farcs2), 1\farcs7 from the X-ray position which itself has a
1\farcs6 error (enhanced position, \citep{Evans+2009a}). This is close to the
bright (8th mag) star HD 99481, which
hampered detection in other bands. No detections from other groups were
reported. A second epoch GROND observation performed 5 nights later
(mid-time 2014-07-24T23:34), finds \rp = 24.33$\pm$0.27,
\ip $>$ 23.8 and \zp = 23.01$\pm$0.32 (forced detection). 
The marginal fading prevents a secure statement about an optical afterglow
detection, but the second epoch magnitudes could likely resemble host emission.

\subsection{GRB 140815A}

For this INTEGRAL-IBIS detected burst with a positional accuracy of 2\farcm5
\citep{Mereghetti16690}, a Swift follow-up observation at 19.2--31.8 ks
after the GRB identified an X-ray source \citep{Mangano16691} within the
INTEGRAL error circle. No further Swift observation was done, so the afterglow
nature through fading could not be established. GROND observations
revealed an optical source within the X-ray error circle \citep{Graham16693}.
Additional GROND observations two months later (midtime 2014-10-17T09:18
and 2014-10-24T06:01) show this source at unchanged brightness. Thus, this
object is not the optical afterglow of GRB 140815A, but could be the
counterpart of the X-ray source in case it is not the X-ray afterglow.

\subsection{GRB 141022A}

For this Swift-detected GRB, a rapid X-ray afterglow was reported, but no
UVOT detection \citep{Vargas16938}. GROND observations started 11 min. after
the GRB trigger and revealed a single source inside the X-ray error circle
\citep{Kann16939} which did not vary substantially over the next hour.
Another GROND observation in the following night was too shallow for a
decisive answer, but a third GROND observing run half a year later
(mid-time 2015-04-24T08:14) showed this source still at the same brightness,
thus excluding its afterglow nature.

\subsection{GRB 150428A}

For this Swift-detected GRB, a rapid X-ray afterglow was reported, but no
UVOT detection \citep{Page17765}. The originally reported source from
GROND observations \citep{Knust17767} was not consistent with the enhanced
XRT position \citep{Osborne17768}, and also shown with several later
observations to remain constant in time. \cite{Andersen17771}
reported a marginally detected source with R = 22.5$\pm$0.43 mag at a
median time of 0.887 hrs after the GRB trigger. Checking our early GROND
images, we do not detect this source (or similarly at the slightly offset
radio source position \citet{Leung30324}), with 3$\sigma$ upper limits of
\rp $>$ 23.8 (mid-time 1.05 hrs post-burst) and
\rp $>$ 24.3 (mid-time 1.41 hrs post-burst). 
If the above marginal detection were true, this would imply a decay slope
of $-7$ between their and our GROND epoch. We therefore assume that
an optical/near-infrared afterglow of this GRB has not been detected.

\subsection{GRB 150915A}

\begin{figure}[ht]
  \includegraphics[width=8.8cm]{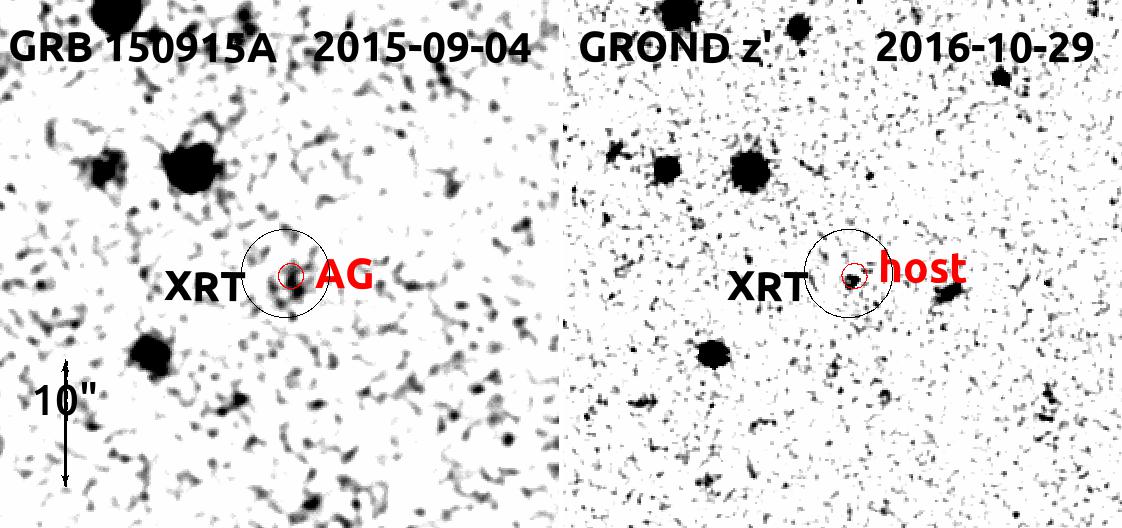}
  \vspace{-0.2cm}
  \caption[150915A]{GROND images of two epochs of the GRB 150915A afterglow
    on top of the spectroscopically identified host galaxy (right panel),
    with clear fading by up to 1.3 mag in the \zp-band.
    \label{150915A}}
\end{figure}

For this Swift-detected GRB a bright, fading X-ray afterglow was readily
detected \citep{Delia18315}, but no UVOT counterpart was identified.
With GROND-observations starting 2.2 hrs post-burst, a candidate afterglow
was found at the border of the X-ray error circle \citep{Yates18317}.
Spectroscopy with X-Shooter on the ESO/VLT derived a redshift of $z = 1.968$
of this source based on both, metal line absorption features as well as
emission lines from the putative host galaxy \cite{Delia18318}.
Contemporaneous NOT observations found the object to brighten during the
first three hours, but like for the initial GROND observation did not
find a fading behaviour \citep{Xu18323}. We obtained two further epochs
two nights later (mid-time Sep. 18, 01:47 UT) and about one year later
(mid-time 2016-10-29T01:03) which clearly identifies the fading nature,
but also reveals the host galaxy underneath the afterglow (Fig. \ref{150915A}),
at the following brightness:
\gp = 23.24$\pm$0.05 mag,
\rp = 23.45$\pm$0.05 mag,
\ip = 23.52$\pm$0.11 mag,
\zp = 23.89$\pm$0.21 mag,
$J$ = 22.31$\pm$0.41 mag.

\subsection{GRB 151023A}
 
GROND observations of this Swift/BAT-detected burst started about 10 hrs
after the BAT trigger, and a new source had been quickly reported in
\cite{Knust+2015}. However, further GROND imaging did not reveal fading,
and also \cite{Butler+2015} detected this source at similar brightness
2--3 hrs later. Also, after correction for the Galactic foreground
(E$_{B-V}$=1.38 mag) the SED of the reported source resembles the
stellar spectrum of a cool star. Thus, the object reported in
\cite{Knust+2015} is unrelated to GRB 151023A.

\subsection{GRB 151212A}

Swift follow-up observations of this MAXI-detected GRB at 8.4--8.6 ks
after the trigger revealed a likely X-ray afterglow \citep{Roegiers18687}.
GROND observations at $\approx$14.1 hrs after the GRB trigger
identified an optical source with a powerlaw-like SED, suggestive of the
afterglow \citep{Wiseman18688}. Follow-up GROND observations in
the second and third night after the GRB trigger found this object to
fade by $>$1.3 mag in the bluest filter bands, thus confirming the
afterglow identification.

\subsection{GRB 160804A}

\begin{figure}[ht]
  \includegraphics[width=5.98cm]{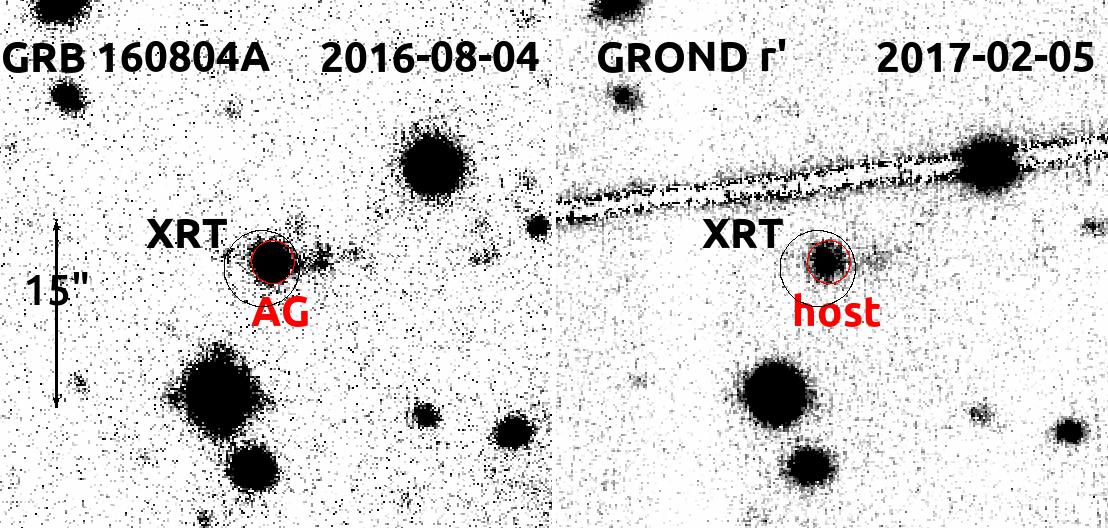}
  \hspace{-0.075cm}\includegraphics[width=2.99cm, viewport=560 0 1120 545, clip]{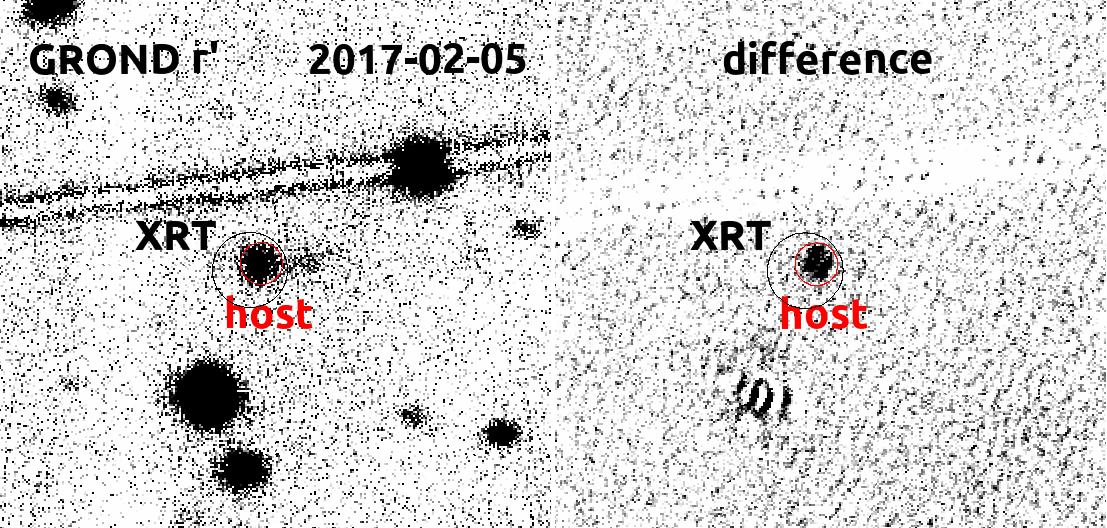}
  \vspace{-0.3cm}
  \caption[160804A]{GROND images of two epochs of the GRB 160804A afterglow
    on top of a relatively bright SDSS galaxy (middle), with the afterglow
    emission clearly showing up in the image subtraction (right panel).
    \label{160804A}}
\end{figure}

For this Swift-detected GRB, a fading optical source was identified
in the first UVOT exposures \citep{BreeveldMarshal19764}, at a position
consistent with the X-ray afterglow \citep{Osborne19762}. The position
of this optical source was readily noticed to be coincident with that
of a SDSS galaxy \citep{Tyurina2016}.
All ground-based observations started a couple hours (or later) after
the BAT trigger, returning brightness estimates consistent at first
glance with the SDSS catalogue magnitudes
\citep{Tyurina2016, Malesani19770, Moskvitin19771, Mazaeva19775, Watson19779},
including the first GROND epoch (mid-time Aug. 5, 00:15 UT)
\citep{BolmerGreiner2016}.
A second GROND epoch was taken 6 months later (mid-time 2017-02-05T08:17),
providing the following estimates (numbers in bracketts are from the
first epoch): \\
\gp = (21.50$\pm$0.02) 22.47$\pm$0.04 mag, \\
\rp = (21.21$\pm$0.02) 22.19$\pm$0.03 mag, \\
\ip = (20.85$\pm$0.02) 21.28$\pm$0.03 mag, \\
\zp = (20.66$\pm$0.02) 21.02$\pm$0.04 mag, \\
J = (20.20$\pm$0.03) 20.67$\pm$0.18 mag, \\
H = (19.88$\pm$0.04) 20.10$\pm$0.13 mag, \\
K = (19.69$\pm$0.08) $>$20.0 mag, \\
which are consistent with the SDSS-catalogue PSF-magnitudes.
This demonstrates clear fading in the blue bands (see Fig. \ref{160804A}),
suggesting that most
of the first-day observations indeed detected afterglow light. The implied
shallow brightness decline is matching that of the X-ray emission observed
with Swift-XRT. A powerlaw fit to the Galactic foreground (E(B-V)=0.03 mag)
corrected magnitudes gives a slope of 1.12$\pm$0.05; when omitting the $JHK$
bands due to their larger fraction of host light, the slope remains
unchanged, but the error towards flatter slopes increases to $-0.20$.

\end{document}